\documentclass[sn-basic]{sn-jnl}

\jyear{2021}%

\theoremstyle{thmstyleone}%
\theoremstyle{thmstyletwo}%

\theoremstyle{thmstylethree}%

\usepackage{framed}
\usepackage{soul}
\usepackage{amsmath}
\usepackage{comment}
\usepackage{subcaption}
\usepackage[labelformat=parens,labelsep=quad,skip=3pt]{caption}
\usepackage{graphicx}

\newcommand{\interviewquote}[2]{
 \def\FrameCommand{%
    \hspace{0pt}%
    {\color{blue}\vrule width 1.5pt}%
    {\color{white}\vrule width 4pt}%
    \colorbox{white}
  }%
  \MakeFramed{\advance\hsize-\width\FrameRestore}%
  \noindent\hspace{-4.55pt}
  {\small ``\emph{#1}'' - {#2}}\vspace{0.5pt}\endMakeFramed%
}

\begin{document}

\title[Adopting Automated Bug Assignment]{Adopting Automated Bug Assignment in Practice --- A Longitudinal Case Study at Ericsson}


\author*[1,3]{\fnm{Markus} \sur{Borg}}\email{markus.borg@cs.lth.se}

\author[2]{\fnm{Leif} \sur{Jonsson}}\email{leif.jonsson@ericsson.com}

\author[3]{\fnm{Emelie} \sur{Engstr\"om}}\email{emelie.engstrom@cs.lth.se}

\author[4]{\fnm{Béla} \sur{Bartalos}}\email{bela.bartalos@ericsson.com}

\author[4]{\fnm{Attila} \sur{Szab\'o}}\email{attila.szabo@ericsson.com}

\affil*[1]{\orgname{RISE Research Institutes of Sweden}, \orgaddress{\city{Lund}, \country{Sweden}}}

\affil[2]{\orgname{Ericsson}, \orgaddress{\city{Kista}, \country{Sweden}}}

\affil[3]{\orgdiv{Dept. of Computer Science}, \orgname{Lund University}, \orgaddress{\city{Lund}, \country{Sweden}}}

\affil[4]{\orgname{Ericsson}, \orgaddress{\city{Budapest}, \country{Hungary}}}

\abstract{[Context] The continuous inflow of bug reports is a considerable challenge in large development projects. Inspired by contemporary work on mining software repositories, we designed a prototype bug assignment solution based on machine learning in 2011-2016. The prototype evolved into an internal Ericsson product, TRR, in 2017-2018. TRR's first bug assignment without human intervention happened in April 2019. [Objective] Our study evaluates the adoption of TRR within its industrial context at Ericsson, i.e., we provide lessons learned related to productization of a research prototype within a company. Moreover, we investigate 1) how TRR performs in the field, 2) what value TRR provides to Ericsson, and 3) how TRR has influenced the ways of working. [Method] We conduct an industrial case study combining interviews with TRR stakeholders, minutes from sprint planning meetings, and bug tracking data. The data analysis includes thematic analysis, descriptive statistics, and Bayesian causal analysis. [Results] TRR is now an incorporated part of the bug assignment process. Considering the abstraction levels of the telecommunications stack, high-level modules are more positive while low-level modules experienced some drawbacks. On average, TRR automatically assigns 30\% of the incoming bug reports with an accuracy of 75\%. Auto-routed TRs are resolved around 21\% faster within Ericsson, and TRR has saved highly seasoned engineers many hours of work. Indirect effects of adopting TRR include process improvements, process awareness, increased communication, and higher job satisfaction. [Conclusions] TRR has saved time at Ericsson, but the adoption of automated bug assignment was more intricate compared to similar endeavors reported from other companies. We primarily attribute the difference to the very large size of the organization and the complex products. Key facilitators in the successful adoption include a gradual introduction, product champions, and careful stakeholder analysis.}



\maketitle

\section{Introduction} \label{sec:intro}

In large development projects, the continuous inflow of bug reports is a considerable challenge~\citep{bettenburg2008duplicate,just2008towards}. The Bug Tracking System (BTS) is a central repository in contemporary software development organizations. There are two archetypal bug assignment processes, i.e., approaches to distribute bug reports to developers. First, as is common in Open-Source Software (OSS) communities, individual developers can select bug reports to resolve in a \textit{pull-based process}. Second, a \textit{push-based process} can be used where a change control board or product manager assigns bug reports to either development teams or individual developers. In our research, we focus on a hybrid model, i.e., push-based bug assignment to development teams and pull-based assignment from the teams themselves.

Push-based bug assignment is normally done manually. However, several studies report that manual bug assignment is labor-intensive and error-prone~\citep{baysal_bug_2009,jeong2009improving}, resulting in ``bug tossing''~\citep{anvik2011reducing,bhattacharya2012automated,jonsson2012towards} and potentially slower bug resolution. Several researchers have proposed mitigating the challenges by automating bug assignment. The most common automation approach uses supervised Machine Learning (ML), i.e., a classifier is trained to find patterns in historical bug reports to make recommendations for new bugs. Early research on automated bug assignment focused on OSS development communities, especially the Eclipse and Mozilla projects \citep{sajedi2020guidelines}. However, the OSS context differs from proprietary development in several aspects, e.g., organizational structures and developer incentives. Recent studies from Türkiye İş Bankası~\citep{aktas2020automated} and LGE Brazil~\citep{oliveira2021issue} constitute rare examples of empirical studies in large companies.

In 2016, we presented a controlled experiment on ML-based bug assignment using five datasets from two companies in telecommunications and process automation~\citep{jonsson2016automated}. This study was the first step in an incremental design science research process~\citep{engstrom2020software}. Our findings in this controlled setting were positive and led to internal productization of a simplified version of the solution within Ericsson. Since 2017, a team in Hungary owns and maintains the solution, referred to as Trouble Report Routing (TRR). To align the terminology, we refer to bug reports as Trouble Reports (TR) in the remainder of this report. Furthermore, we use the terms ``bug assignment'' and ``bug routing'' synonymously in this paper, i.e., auto-routing refers to TRs assigned by TRR.

We have previously reported lessons learned from deploying TRR in an anecdotal manner~\citep{carver2018industry}. Furthermore, we conducted a quantitative analysis of the prediction accuracy of TRR's assignments~\citep{sarkar2019improving} on a subset of the modules in the systems. In the latter paper, we concluded that the confidence based approach of triaging TRs~\citep{jonsson2016automatic} eventually adopted by TRR is promising. We continued improving and customizing TRR and in 2019 activated the solution --- the very first TR assignment without human intervention happened on April 10, 2019. Since then, TRR has been in continuous operation and automatically routed roughly 30\% of the incoming TRs. In this study, our overall goal is to evaluate the adoption of TRR within its industrial context. We now investigate 1) how TRR evolved from a prototype to an internal Ericsson product, 2) the accuracy of TRR, 3) how much value TRR provides, and 4) how the TRR adoption influenced the way of working.

This paper presents an industrial case study to evaluate the adoption of TRR within its industrial context. The case study protocol, developed in line with guidelines by~\citet{runeson2012case}, was accepted as a registered report at the 15th International Symposium on Empirical Software Engineering and Measurement (ESEM) in 2021. As we now report our findings, we present new perspectives on automated bug assignment in proprietary contexts by moving beyond the prediction accuracy that has been in focus in previous work~\citep{sarkar2019improving,oliveira2021issue}. Our study provides insights regarding direct as well as indirect effects of deploying this research-based intervention in an operational setting. Thus, we add empirical support, and provide deeper insights, for a general \emph{technological rule}~\citep{runeson2020design}:

\begin{quote}{
To achieve more efficient and effective assignment of bug reports to teams in large scale industrial contexts, use machine learning to automate bug assignment.
}\end{quote}

The rest of this paper is organized as follows: Section~\ref{sec:rw} introduces previous work on automated bug assignment and tool adoption in general. Section~\ref{sec:method} describes the research method used in this case study. Sections~\ref{sec:collection}~and~\ref{sec:analysis} explain the data collection and analysis, respectively. In Sections~\ref{sec:rq1}--\ref{sec:rq4} we share our findings and discuss their implications. Finally, we discuss the validity of our research in Section~\ref{sec:quality} before concluding the paper in Section~\ref{sec:conc}.

\section{Related Work} \label{sec:rw}
Numerous studies report quantitative evaluations of issue assignment using ML-based tools. Most studies target assignment to individual developers in OSS projects~\citep{sajedi2020guidelines}, whereas we target team assignment in proprietary contexts~\citep{jonsson2016automated}. In this section, we focus the discussion on qualitative industrial experiences. Two recent studies match our interests, i.e., a case study at a full-service bank in Turkey and an experience report from a consumer electronics company in Brazil. Moreover, we discuss previous work on factors that influence adoption of software engineering tools in industry.

\subsection{Evaluations of ML-Based Bug Assignment in Industry}
Closest to our work is the case study by \citet{aktas2020automated} at Softtech, a subsidiary of the large Turkish bank Türkiye İş Bankası (IsBank). Since January 2018, the tool IssueTAG automatically assigns all incoming issues (on average 380 per day) to teams. The authors conducted an experiment on 13 months' worth of data to assess IssueTAG's accuracy. Moreover, they shared qualitative usability insights collected through informal meetings with stakeholders and a short questionnaire.

The deployment of IssueTAG was successful and \citet{aktas2020automated} report four main insights. First, \textit{deploying IssueTAG necessitated changes to the manual issue assignment process} at IsBank. In our study at Ericsson, we explain how the process co-evolved with the introduction of TRR. Second, \textit{IssueTAG does not need to match the manual assignment accuracy at IsBank to be useful}, i.e., slightly less accurate but more efficient issue assignment was reported as an improvement. We investigate this perspective at Ericsson and report a contrasting view. Third, \textit{successful adoption of IssueTAG required two features beyond automated assignment}: 1) accuracy monitoring and 2) an explainability solution for assignments rationales. We confirm that the same features were needed for TRR. Fourth, \citet{aktas2020automated} \textit{did not identify any objections at all regarding deployment of IssueTAG} at IsBank. At Ericsson, we present a richer analysis of skeptical stakeholders and thus complement the picture. Compared to the case study at IsBank, our work investigates issues collected during a longer time period and we focus more on qualitative insights from interviews.

\citet{oliveira2021issue} present an industrial experience report from ML-based issue assignment at LG~Electronics mobile division in Brazil. While the study is substantially less rigorous than both our work and the study on IssueTAG adoption, it contributes real-world insights from tool adoption. The quantitative analysis shows very accurate results ($>$90\%) based on 5,684 issues collected during 2.5 years. The researchers worked closely together with the practitioners by following the six phases of the established data mining process model CRISP-DM~\citep{wirth2000crisp}. Following CRISP-DM, the researchers organized regular meetings with LG~Electronics from which they collected qualitative data.   

\citet{oliveira2021issue} report four primary lessons learned. First, adoption of automated issue assignment \textit{requires an iterative process with effective communication and gradual trust development}. This finding resonates with best practice for industry-academia collaboration~\citep{garousi2019characterizing} and corresponds to the gradual introduction of IssueTAG done at IsBank~\citep{aktas2020automated}. Second, the \textit{researchers must be flexible and add new tool features when needed} --- which could be done as part of the mentioned iterative process. Third, ML model accuracy must be monitored over time. This insight is covered by the third insight reported by~\citep{aktas2020automated}. Fourth, automated assignment can provide value even when the accuracy is not so high. As this insight was also reported for IsBank, we note that misclassifications are considered a bigger problem at Ericsson. 

\subsection{Tool Adoption in Industry}
Numerous studies seek to understand factors that influence the adoption of novel innovations in software-intensive businesses. In this section, we present three papers from three different decades --- indicating that the factors appear stable over time.

\citet{premkumar1995adoption} studied characteristics impacting software engineering tool adoption in the 1990s. Based on a questionnaire survey of 90 managers in the US, inspired by research on innovation adoption, the authors developed a model of seven factors that are important for successful tool adoption. First, they discovered five technology variables: T1) \textit{Relative Advantage} (how superior the new tool is perceived to be  compared to the current solution) T2) \textit{Cost} (the initial investment cost as well as costs for operations and training) T3) \textit{Complexity} (the degree to which a tool is perceived as difficult to understand and use), T4) \textit{Technical Compatibility} (how compatible the new tool is with existing technology), and T5) \textit{Organizational Compatibility} (how consistent with the existing values, past experiences, and needs of the organization a tool is perceived to be). Second, they found three organizational variables: O1) \textit{Product Champion} (an internal person who actively facilitates the adoption), O2) \textit{Top Management Support} (active involvement to allocate adequate resources and sending signals about the importance of the tool), and O3) \textit{expertise} (existing capabilities and skills in the organization). Discriminant analysis showed that O1, O2, O3, T1, and T2 mattered the most.

\citet{favre2003tool} presented an experience report from a decade of collaborations with Dassault Systèmes in France. The development context presented is large (about 1,000 engineers working on the same product), although not as large and complex as the one we study at Ericsson. At Dassault Systèmes, three stakeholder groups must be convinced for successful adoption of a new tool: managers, the tool support team (orchestrating all internally used tools), and software engineers (the end users). The authors report 10 categories of issues that can hinder tool adoption: 1) \textit{scalability}, 2) \textit{usability}, 3) \textit{tool integration}, 4) \textit{process integration}, 5) \textit{customization}, 6) \textit{deployment}, 7) \textit{administration}, 8) \textit{evolution and continuity}, 9) \textit{training}, and 10) \textit{strategical}. The lists represent different abstraction levels, but, we find that \citet{favre2003tool}'s specific issues can be mapped to \citet{premkumar1995adoption}'s seven factors.

\citet{hameed2012conceptual} developed a model for the process of IT innovation adoption in organizations. The authors integrated theories from innovation research and user acceptance models into a comprehensive model. The model consists of five categories of factors, extracted from the literature, that influence adoption: 1) \textit{innovation characteristics} (20 factors, e.g., relative advantage, cost, complexity, and compatibility), 2) \textit{organizational characteristics} (41 factors, e.g., top management support, organization size, expertise, and product champion), 3) \textit{environmental characteristics} (16 factors, e.g., competitive pressure and government support), 4) \textit{individual (decision makers') characteristics} (8 factors, e.g., CEO attitude and innovativeness, and 5) \textit{user acceptance attributes} (22 factors, e.g., perceived usefulness and ease of use, user attitude, and experience). While \citet{hameed2012conceptual} extracted many individual factors from the literature, the ones reported as the most significant resemble the conclusions by~\citet{premkumar1995adoption} from the 1990s, i.e., the factors appear to be stable.

Looking specifically at adoption of ML-based products, \citet{paleyes2020challenges} recently conducted a survey of case studies and experience reports. In their synthesis, they present 44 issues mapped to 13 ML deployment steps and four cross-cutting aspects. Furthermore, the deployment steps are organized into the deployment stages 1) \textit{data management}, 2) \textit{model learning}, 3) \textit{model verification}, and 4) \textit{model deployment}. We found that issues related to data management are not a major concern in the TRR adoption since the data is internal at Ericsson and mostly well-structured. Similarly, the model learning and model verification issues were not major obstacles, although Ericsson's R\&D put considerable effort into model selection and definition of evaluation metrics. Regarding model deployment, our study confirms issues of monitoring and updating in the Ericsson context. Finally, in relation to the cross-cutting aspects, we discovered no issues related to \textit{ethics}, \textit{law}, and \textit{security} in the TRR adoption. The category \textit{end users' trust}, however, was vital in the Ericsson context, including \citet{paleyes2020challenges}'s listed issues of end user involvement, user experience, and explainability. 

In an experience report from Atlassian, \citet{flaounas2017beyond} specifically discussed challenges in a software engineering context. The author organized eight challenges into three phases of building an ML-based product feature. The ideation phase involves challenges of 1) \textit{data availability}, 2) \textit{privacy concerns}, and 3) \textit{project risk estimation}. In the execution phase, they report 4) \textit{build vs. rent} (ML engineers are a scarce resource, thus online service providers might be needed), 5) \textit{scalability}, and 6) \textit{productionization}. Finally, the operation phase brings challenges to 7) monitor and \textit{maintain accuracy} over time and 8) \textit{stability of data sources}. In our work at Ericsson, we found none of these eight challenges to be major impediments to TRR adoption.



\section{Overview of the Research Method} \label{sec:method}
\begin{figure}
    \centering
    \includegraphics[width=0.7\textwidth]{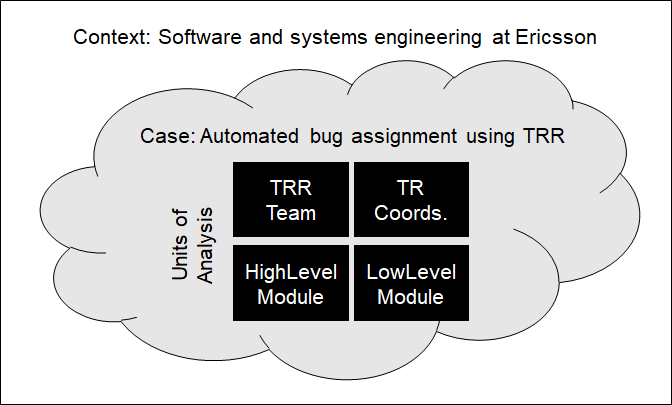}
    \caption{The context, case, and units of analysis.}
    \label{fig:contextcaseunits}
\end{figure}

The case study protocol is available as a peer-reviewed registered report from ESEM 2021~\citep{borg2021adopting}. We conduct interpretivist research as the methods of natural science are insufficient for understanding the case in its social reality context~\citep{baltes2020sampling}. Figure~\ref{fig:contextcaseunits} illustrates the context, the case under study, and the units of analysis. As defined by~\citet{runeson2012case}:

\begin{quote}{
case study in software engineering is an empirical enquiry that draws on multiple sources of evidence to investigate one instance (or a small number of instances) of a contemporary software engineering phenomenon within its real-life context, especially when the boundary between phenomenon and context cannot be clearly specified.
}\end{quote}

Since the adoption of TRR cannot be isolated from the development context at Ericsson, we design an industrial case study. Our study relies on a flexible design, i.e., the sampling, data collection as well as the data analysis involve components relying on our evolving knowledge about the phenomenon.

\subsection{Rationale and Purpose} \label{sec:ratpurp}
Our overall goal is to evaluate the adoption of TRR within its industrial context at Ericsson (cf. Figure~\ref{fig:rqs}). Several aspects motivate us to pursue this goal. First, we want to follow up on research that was initiated 10 years ago~\citep{jonsson2012towards}. How does the automated bug assignment solution actually perform in the field? Are the assignments provided by TRR sufficiently accurate to provide value in the industrial context? Do engineers at Ericsson appreciate the support provided by TRR? How has the introduction of TRR influenced the ways of working? Are there any surprising indirect effects that should be reported? As discussed in Section~\ref{sec:intro}, there is a lack of industrial case studies sharing these types of insights. 

Second, we seek to provide insights regarding the industrial adoption of a research prototype. By conducting this study, we highlight an example of \textit{industry-academia collaboration} \citep{rico2021case} and successful technology transfer of \textit{practically relevant research}~\citep{garousi2020practical}. The study contains a retrospective analysis of the evolution from prototype to internal product. We explore obstacles experienced in the productization and share lessons learned on how they were tackled in the industrial context. We expect our findings to be highly relevant for other software engineering researchers proposing new tools for use in proprietary contexts.

\subsection{Context} \label{sec:context}
As illustrated in Figure~\ref{fig:contextcaseunits}, the context is software and systems engineering at Ericsson. Ericsson is a global actor in telecommunications. We characterize the context inspired by the facets proposed by~\citet{petersen_context_2009}, focusing on the factors that we believe are the most relevant for our study.

\textbf{Product} The products in the analysis consists of two large systems in the Information and Communications Technology (ICT) domain. Various programming languages are used in the products, but a majority of the code is developed in C++ and Java. Other languages such as hardware description languages and tailored domain-specific languages are also used. The two systems are mature with old code bases.

\textbf{Processes} The project model used to develop both systems is an adapted agile development process. Development in the ICT domain is heavily standardized, and adheres to standards by regulatory bodies such as 3GPP, 3GPP2, ETSI, IEEE, IETF, ITU, and OMA. Moreover, Ericsson is ISO~9001 and TL~9000 certified. 

\textbf{Practices and Techniques} The development projects use agile practices that have been customized for the organization, e.g., sprint planning meetings, retrospectives, self-organization, and test automation. The development projects are organized into two-week sprints followed by releases. 

\textbf{People} Staff turnover is very low in the development organization. Many of the engineers are seniors developers who have been working on the same, or similar, products for many years.

\textbf{Organization} Thousands of engineers are distributed over several countries, e.g., Sweden, Hungary, China, and Canada. In total, Ericsson has 100,000 global employees. The BTS is the central point for organizing the bug handling process. Tracking of analysis, implementation proposals, and verification are all coordinated through the BTS.

\textbf{Market} Both systems are deployed at customer sites world-wide in the ICT market. The telecommunications market is currently in a transition from the last generation of 4G networks to 5G. Software-oriented technology improvements are increasingly flexible high-speed connectivity at ultra-low latency.

\subsection{Case and Units of Analysis} \label{sec:caseunits}
\begin{figure}
    \centering
    \includegraphics[width=0.70\textwidth]{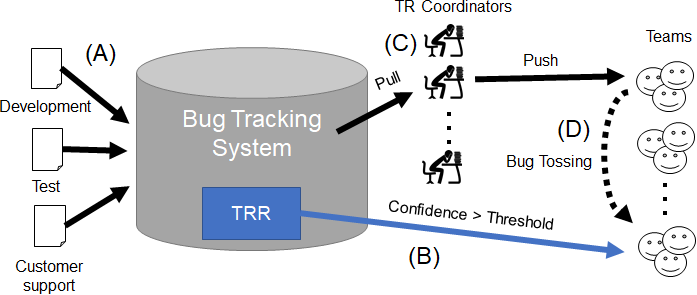}
    \caption{Preunderstanding of how TRR operates in the industrial context.}
    \label{fig:preunderstanding}
\end{figure}

The case under study is automated bug assignment using TRR in its industrial context. Figure~\ref{fig:preunderstanding} shows our preunderstanding of how TRR had been integrated in the BTS at the company before initiating the case study. As our understanding evolved during the study, the complete picture of how TR assignment is conducted at Ericsson is now substantially more complex, which we elaborate on in Section~\ref{sec:rq1_process}.

Different organizational units submit TRs to the BTS (A). TRR, operating as a BTS add-on, predicts which development team would be the most likely to resolve the bug and appends this information to the TR. If the prediction has a high confidence value, i.e., above a configurable threshold, the TR is automatically assigned to the corresponding team (B). If the confidence value is lower than the threshold, the assignment process relies on the normal manual approach by one of the TR coordinators (C). The manual approach encompasses a TR coordinator pulling a TR from the BTS, analyzing it --- possibly perceiving TRR as a recommendation system for issue reports~\citep{borg2014changes} --- and pushing the TR to one of the development teams. Bug tossing entails reassignment of a TR to another team (D). Note that the phenomenon of bug tossing is not necessarily caused by an incorrect initial team assignment. On the contrary, it can be a required step when resolving complex bugs that necessitate changes by multiple teams.

TRR is currently in operation in a BTS used for development of two major systems (4G and 5G) consisting of more than 20 high-level modules. The modules corresponds roughly to the level of abstraction of the telecommunications technology stack, with applications on top, via modules responsible for traffic control and user equipment handling, to low-level modules such as radio technology and hardware in the bottom. Each module is maintained by several development teams. The largest module encompasses 1,000+ engineers in teams distributed globally. In this study, we refer to modules on the higher and lower levels of the technology stack \textit{HighLevel} and \textit{LowLevel}, respectively. 


Since 2017, TRR is maintained by a team in Hungary, see \textbf{TRR Team} listed as a unit of analysis in Figure~\ref{fig:contextcaseunits}. Furthermore, we define three additional units of analysis. First, a HighLevel development team that opted in as early adopters of TRR, heavily involved in the transition from research prototype to operational tool (cf. \textbf{HighLevel Module}). Second, a LowLevel development team that initially opted out from automatic routing, i.e., the \textbf{LowLevel Module}. Third, senior engineers that act as TR coordinators for the 4G/5G systems (cf. \textbf{TR Coords.}). TR coordinators have different roles within Ericsson, but perform TR assignment as part of their routine work.

\subsection{The Issue Assignment Process at Ericsson} \label{sec:rq1_process}
This section provides a detailed description of the current issue assignment process at Ericsson. While this is part of our results, i.e., it shows how our understanding of the process evolved from the preunderstanding in Figure~\ref{fig:preunderstanding} by interviewing key stakeholders, we present the content here. This piece of the Ericsson context is important for readers to interpret all findings we present in Sections~\ref{sec:rq1}--\ref{sec:rq4}.

\begin{figure}
    \centering
    \includegraphics[width=0.8\textwidth]{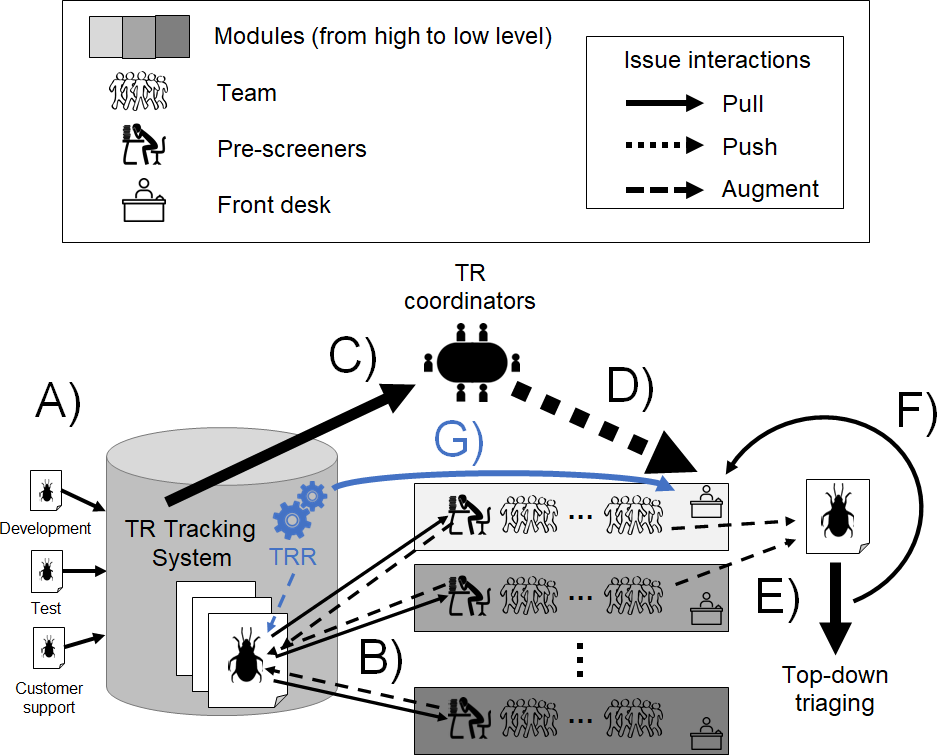}
    \caption{The TR assignment process at Ericsson.}
    \label{fig:complex_process}
\end{figure}

4G/5G product development at Ericsson is a highly complex endeavor involving thousands of globally distributed engineers. At this scale, issue management and TR triaging inevitably becomes complex activities. Figure~\ref{fig:complex_process} shows an overall picture of the process --- without TRR --- at Ericsson. The telecommunications technology stack is hierarchical and deep, spanning from higher application levels via layers such as traffic control and baseband down to the bottom radio layer. At Ericsson, the overall architecture is based on modules corresponding to the technology stack. Several modules are very large, e.g., the radio system with 1,000+ engineers organized into several sub-modules in different countries with a many development teams. In the figure, we depict lower layers with increasingly dark shades of gray. At this scale of software development, with thousands of engineers in hundreds of teams across the planet, finding the development team corresponding to a particular responsibility is hard. To support navigating the modules' internal team landscapes, each module has a \textit{front desk}, i.e., engineers working on locating internal expertise. As indicated by thicker arrows, the large flow of TRs through the TR coordinators in the upper part of the figure is what TRR is intended to support.

\textbf{A)} in Figure~\ref{fig:complex_process} shows the inflow of TRs into the tracking system. The main sources of TRs are the development organization, the internal test organization, and TRs that have been processed by the customer support organization. TRs originating in customer issues shall always be given a high priority. Every early morning, an assigned group of \textit{pre-screeners} from each module analyzes the newly registered TRs. The duration of the daily pre-screening meetings varies based on the TR inflow the last day, but are typically concluded within an hour. In this meeting, pre-screeners augment TRs based on an initial analysis, e.g., based on analyses of extracted logs, of the issue to support the subsequent issue assignment process, or they immediately pull TRs to their module (cf. \textbf{B)} in Figure~\ref{fig:complex_process}) and assign them to one of the teams within the module. Modules rotate the pre-screeners to support knowledge sharing within the organization. 

Later in the morning, the TR Coords. (cf. \textbf{C)} in Figure~\ref{fig:complex_process}) meet to analyze the newly arrived TRs that have not yet been assigned to any module. The TR coordinators are highly senior engineers with significant Ericsson experience. Based on their analysis, potentially advised by TR augmentations added by the pre-screeners, the TR Coords assign each TR to the module most suitable to initiate an investigation into the issue (cf. \textbf{C)} in Figure~\ref{fig:complex_process}). Note that the TR Coords.' meetings are not only related to supporting TR assignment, as other equally important tasks are completed such as severity assignment and impact analysis related to the high-variability systems in Ericsson's product lines.

In many cases, TR triaging starts at a higher layer of the technology stack. If a development team cannot resolve the issue at their level, teams augment the TR with their analysis results before passing them down to the front desk of a lower level module for further triaging (cf. \textbf{E)} in Figure~\ref{fig:complex_process}). The phenomenon of ``bug tossing'' is present as development teams can reassign TRs both within their own module or to front desks of other modules (cf. \textbf{F)} in Figure~\ref{fig:complex_process}). As explained in Section~\ref{sec:quan_anal}, we measure the average length of bug tossing chains.

The blue cogwheels in Figure~\ref{fig:complex_process} depict the TRR add-on in the TR tracking system. For each incoming TR, TRR predicts how likely it is for each module to resolve it. If the prediction for a single module has a very high confidence level, TRR bypasses the TR coordinators and immediately sends the TR to the corresponding front desk. If the confidence level is lower, TRR only augments the TR with its predictions. Moreover, modules can opt-in to get email notifications when TRR has provided relevant medium confidence predictions, i.e., an early heads-up for their next prescreening meeting.

\subsection{Theory} \label{sec:theory}
The general \emph{problem} of inefficient and ineffective bug assignment was observed in the literature~\citep{bettenburg2008duplicate,just2008towards,oliveira2021issue} as well as in the specific industrial contexts where this research was conducted~\citep{jonsson2012towards,sarkar2019improving}. With the solution in mind (to use ML techniques to assign TRs to modules), the characteristics of the targeted \emph{problem instance} were identified, i.e., we explored the nature of the TRs, the BTS, and the organizational context within a subset of the development at Ericsson. Related work on bug classification as well as on ML techniques was identified and carefully compared~\citep{jonsson2016automated}, which underpinned the \emph{design decisions} for the proposed solution. The ML solutions were implemented and trained using the Weka framework~\citep{hall_weka_2009}. Several alternative solution instances were \emph{validated} on real data (about 50,000 TRs) from five projects across two companies/domains. A design artifact was produced specifically for Ericsson, namely a prototype ensemble-based bug assignment tool built on top of Weka.

In our 2016 paper, we stated that the translation from TRR's prediction accuracy to the practical value of the solution might not be linear. Furthermore, we discussed this aspect in terms of the QUPER model~\citep{regnell_supporting_2008}, a theoretical construct describing the perceived benefits of different degrees of quality as continuous and non-linear. The QUPER model suggests three quality breakpoints for TRR:
\begin{itemize}
    \item \textbf{Utility} Engineers start considering TRR as a useful addition to manual bug assignment.
    \item \textbf{Differentiation} Engineers recognize that TRR provides a competitive advantage compared to fully manual work.
    \item \textbf{Saturation} Increasing the quality of TRR beyond this points adds no practical value.
\end{itemize}

We base our evaluation of the adoption of TRR on three theoretical models. First, we revisit the QUPER model to assess where TRR belongs on the sliding quality scale. Second, as we also did in the original paper, we discuss the direct and indirect effects of increased levels of automation using the model by~\citet{parasuraman2000model} (AUTO). The latter model opens up for an analysis of both direct and indirect effects of increased automation. Third, we study the Ericsson engineers' impressions of working with TRR. 

\subsection{Research Questions} \label{sec:rqs}
\begin{figure}
    \centering
    \includegraphics[width=0.8\textwidth]{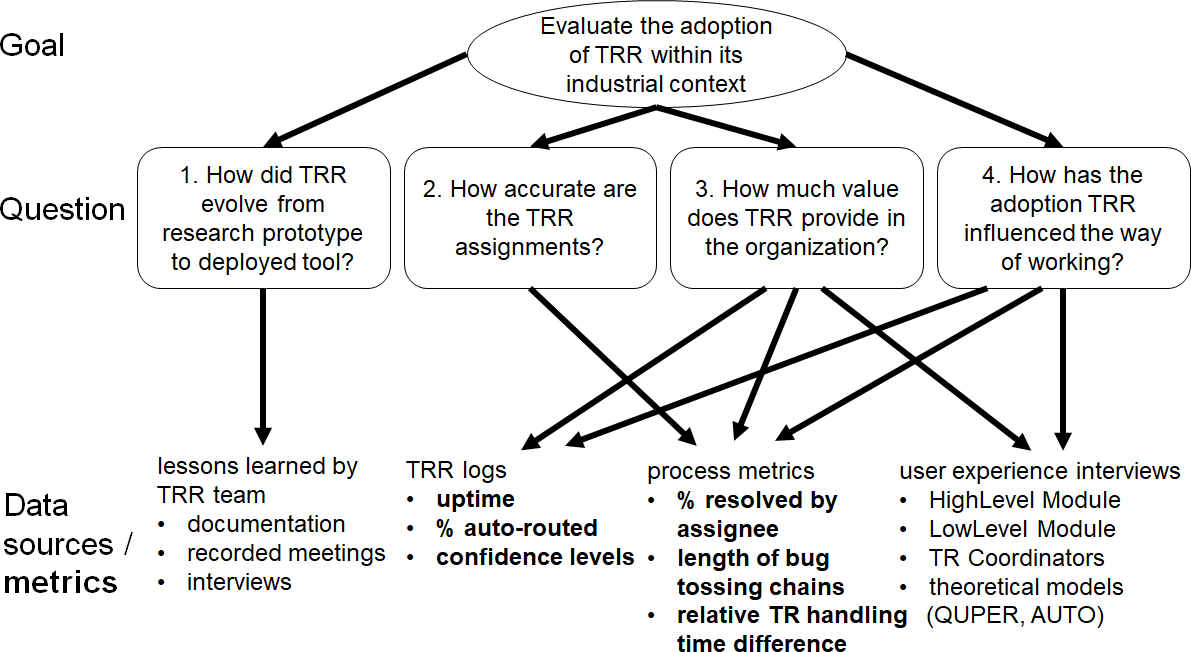}
    \caption{Breakdown of our goal to research questions, data sources, and metrics.}
    \label{fig:rqs}
\end{figure}

As visualized in Figure~\ref{fig:rqs}, the aim of the study is to evaluate the adoption of TRR within its industrial context. We have defined four main research questions, which may all be answered by applying both qualitative and quantitative methods. The lower part of the figure presents data sources and metrics, where the latter are indicated in bold font.
\begin{itemize}
\item[RQ1] How did TRR evolve from prototype to deployed tool?
\item[RQ2] How accurate are the TRR assignments?
\item[RQ3] How much value does TRR provide in the organization?
\item[RQ4] How has the adoption of TRR influenced the way of working?
\end{itemize}

We answer RQ1 by studying the design decisions Ericsson engineers made along the way. How and why were potential adaptations to the original solution made? What were the major challenges during the tool introduction, including processes, technology, organizational issues, and human factors? The TRR team, one of four units of analysis, shared a collection of recorded virtual sprint meetings and internal documentation. Furthermore, we conducted interviews to collect lessons learned. After the analysis, the interviewees from the TRR team joined the research team (authors 1--3) as coauthors of this paper --- the value of their contributions go far beyond the credits a mention in the acknowledgements section could offer.

RQ2 involves a quantitative analysis of TRR's prediction accuracy in the light of our previous work~\citep{sarkar2019improving}. Previously, we studied the feasibility of increasing the accuracy of TRR by augmenting the ML input data with logs and alarms. The study was performed on a set of roughly 10,000 TRs originating in nine of the modules, i.e., a subset of the full systems. Relying on easily accessible textual and categorical features, we obtained precision and recall values around 80~\% on the subset of modules. Adding the alarm and log data however, did not improve upon the standard TRR implementation. As the overall accuracy of TRR on the full system at the time (around 66\%) was reported as insufficiently accurate for regular use, we proposed to only assign TRs for which the ML classifier was confident. In this study, we revisit the accuracy RQ to evaluate how TRR performed in the field using historical data since deployment in April 2019. As shown in Figure~\ref{fig:rqs}, we measure the fraction of TRs resolved by the first assigned module, the length of bug tossing chains, and the relative TR handling time difference between TRs assigned by humans and TRR. To mitigate confounding factors, the cycle time covers the implementation time but not the verification by the test organization and subsequent deployment activities.

RQ3 targets the utility of TRR and its added value in the organization. We will complement the insights provided by RQ2 with an analysis of the TRR utilization, i.e., whether it has been available (uptime) and sufficiently confident to be effective (fraction of automatic TR assignments and distribution of confidence levels). Moreover, we will complement the analysis with qualitative insights from interviews with members of the HighLevel and LowLevel Modules and a sample of TR coordinators (cf. Figure~\ref{fig:contextcaseunits}). Section~\ref{sec:collQual} presents how we designed an interview guide supported by the theoretical models QUPER and AUTO~\citep{regnell_supporting_2008,parasuraman2000model}. We have previously used QUPER to discuss the relation between tool accuracy and perceived value in the context of automated change impact analysis~\cite{borg2016supporting}. 

RQ4 explores the direct and indirect effects of introducing TRR in the organization. A tool never exists in isolation, i.e., the introduction of tool-oriented interventions ought to be studied through a holistic perspective. Among other things, we seek to understand what made certain teams quickly opt-in to an increasing level of automation whereas others remained skeptical. Analogous to RQ3, RQ4 will be answered using a combination of quantitative metrics and rich information from interviews.


\section{Data Collection} \label{sec:collection}
The study relies on non-probability sampling~\citep{baltes2020sampling}, i.e., there is no element of randomness when selecting items in the sampling frame. Instead, we used a combination of purposive and referral-chain sampling to select interviewees. To mitigate selection bias, our initial set of interviewees included engineers from different levels of the organization as well as with different perceptions of TRR. Furthermore, we included questions in the interview guide with the purpose of identifying people having complementary insights. For our artifact analyses, i.e., document analysis and mining software repositories, we used whole-frame sampling. We collected all relevant meeting protocols and mined all available TRs during the time period, as opposed to our previous work~\citep{sarkar2019improving}.

\subsection{Quantitative Data Collection}
The BTS is an important source of data that constitutes a valuable target for mining of software repositories~\citep{borg2014changes}. The BTS data contain details of TRs, e.g., assignments, submitters, severity levels, and time stamps. The data in the BTS does not directly contain the name of the modules, but rather lower level designations that are then mapped to the 20+ modules in the data pre-processing stage of the analysis pipeline. This pre-processing can sometimes introduce a lag, in the sense that low level designations can be added by the design organization without this being reflected in the mapping to modules. This phenomenon, sometimes called \emph{data drift}, is usually corrected in time when it is discovered in the ML model monitoring process that erroneous predictions are made by TRR due to low level designations missing in the mapping process. This lag can also during shorter periods skew the accuracy of TRR downwards to this missing mapping. 

We collected all data available in the BTS related to the development of the 20+ modules, resulting in 21 months' worth of data, i.e., 2019-04-10--2022-02-28. Thus, we apply \textit{whole-frame sampling} by selecting all items in the sampling frame~\citep{baltes2020sampling}. TRR logs all its actions and output in the BTS. The logs primarily show the TRR predictions, i.e., the bug assignment output provided by the tool and the confidence levels accompanying the individual predictions. Finally, all TRR actions have individual time stamps.

\subsection{Qualitative Data Collection} \label{sec:collQual}
We selected interviewees from the four units of analysis based on \textit{purposive sampling}. Our goal was to identify the candidate interviewees that can provide the richest information, while also complementing the perspectives of the previous interviewees from a heterogeneity perspective, e.g., roles, background, site, age, and gender. As case study research allows a flexible design, we complemented the interviewee selection with \textit{referral-chain sampling}. In practice, each interview session concluded by asking the interviewee to refer other members of the population whom they believe would provide valuable perspectives on the adoption of TRR. 

We developed an interview guide with some variation points for the four units of analysis. The interview sessions with TRR Team, HighLevel Module, and LowLevel Module focused on challenges, solutions, and opportunities related to the evolution of TRR from a research prototype to an internal Ericsson tool. On the other hand, the interview sessions with the TR Coordinators primarily focused on the user experience and perceived value of TRR (corresponding to perceived ease of use and usefulness in TAM~\citep{davis1989perceived}). The interview questions were intermixed in several interview sessions, i.e., we performed semi-structured interviews. The complete interview guide is available in Appendix~\ref{app:int_guide}, whereas an initial overview is presented below:

\begin{enumerate}
    \item A formal introduction including overall purpose, non-disclosure agreements, integrity, security, and research ethics.
    \item A brief description of the interviewee's current role and engineering background.
    \item Open questions related to TRR's evolution from a research prototype to an internal tool.
    \item Closed questions on TR assignment and TRR. Perceived TRR value and ease of use.
    \item An open discussion on the value of TRR and its direct and indirect effects on the related work tasks.
    \item Perceived value of TRR in relation to its prediction accuracy.
    \item Final comments and suggestions for additional interviewees.
\end{enumerate}

We conducted six individual interview sessions and a group interview with three TR coordinators between Oct 2021 and Jan 2022. The interview sessions lasted 60--90 minutes and were conducted by at least two interviewers, i.e., authors 1 and 3 were always present. Due to the Covid-19 pandemic, all sessions were done remotely using MS Teams. All interviewees provided consent for recording of both audio and video. 

Table~\ref{tab:interviewees} shows an overview of the nine interviewees. [TRR1] and [TRR2] are the main developers of TRR. Both are experienced developers with roughly ten years employment time at Ericssson. They are primarily Java developers, but they acquired practical skills in deploying and operating ML models while developing TRR. [CO1--3] are highly seasoned Ericsson engineers with substantial system and domain expertise. [CO1] has an overarching TR management role, whereas [CO2] and [CO3] are responsible for TRs for the 5G and 4G systems, respectively. [HL1] and [HL2] represent two separate high-level modules for which they have acted as single points of contact for TRs. Finally, [LL1] and [LL2] belong to the same (very large) low level module. [LL1] is currently a line manager, whereas [LL2] is a maintenance leader responsible for orchestrating TRs for the module.

\begin{table}[]
\caption{Overview of the interviewees. (TC=Technical Coordinator, SPoC=Single Point of Contact.)}
\label{tab:interviewees}
\begin{tabular}{|l|p{1.7cm}|p{1.5cm}|p{0.7cm}|p{3.5cm}|l|}
\hline
\textbf{IntID} & \textbf{Unit of Analysis} & \textbf{Role(s)}      & \textbf{Years at Ericsson} & \textbf{Experience}                                                  & \textbf{Gender} \\ \hline
[TRR1]   & TRR Team         & Developer          & 11                & Developer and product owner.                                         & Male   \\ \hline
[TRR2]   & TRR Team         & Developer          & 8                 & Developer and product owner.                                         & Male   \\ \hline
[CO1] & TR Coords.       & Main TC             & 20                  & Developer, quality management, PhD.                                                                  & Female \\ \hline
[CO2] & TR Coords.       & Main TC 5G         & 14                  & Systems engineer, quality management.                                                                     & Male   \\ \hline
[CO3] & TR Coords.       & Main TC 4G         & 27                  & Systems engineer, quality management.                                                                     & Male   \\ \hline
[HL1]  & HighLevel Module A (HL-ModA) & Line manager       & 23                & Various engineering and management roles. TR SPoC for a module.      & Female \\ \hline
[HL2]  & HighLevel Module B (HL-ModB) & Project manager   & 16                  & Developer. Technical project manager. TR SPoC for different modules. & Male   \\ \hline
[LL1]   & LowLevel Module A (LL-ModA) & Line manager      & 8                 & Engineering and management.                                         & Male   \\ \hline
[LL2]   & LowLevel Module A (LL-ModA) & Maintenance leader, Module TC & 15                & Developer, project manager, line manager.                            & Male   \\ \hline
\end{tabular}
\end{table}

In the context of this paper, we use the term Technical Coordinator (TC) for engineers that assign TRs. TR SPoCs for specific modules are also TCs, i.e., [HL1] and [HL2]. [CO1--3] are TCs on the highest product level, whereas [HL1], [HL2], and [LL2] are TCs on the module level.

\section{Data Analysis} \label{sec:analysis}
This section describes how we analyzed the collected BTS/TRR data and our approach to qualitative analysis.

\subsection{Quantitative Data Analysis} \label{sec:quan_anal}
We open the discussion on quantitative data analysis with an important disclaimer. Bug data is highly sensitive to any development organization. As a result, we are not allowed to report any absolute numbers related to TRs at Ericsson. Instead, bug counts will mostly be presented in relative numbers.

We used the extracted data to calculate simple descriptive statistics for the HighLevel Modules (High-A and High-B) and the LowLevel Module (Low-A). The descriptive statistics were used as input to the interview sessions (cf. subsection 6 in Appendix~\ref{app:int_guide}), to tailor figures for the individual interviewees when applicable.

As explained in the registered report~\citep{borg2021adopting}, we worked iteratively with the quantitative data. Data collection and analysis were intertwined, and we found new research angles as we got more familiar with the data. Our final list of metrics (M1--M6), also presented in Figure~\ref{fig:rqs}, are: 

\begin{enumerate}
    \item[M1] \textbf{Uptime} was estimated by the TRR maintenance team.
    \item[M2] \textbf{Fraction automatically routed} is calculated from the TRR logs. 
    \item[M3] \textbf{Distribution of confidence levels} for the TRR predictions is collected from the TRR logs. The confidence level is fundamental in TRR as it must surpass a certain threshold to allow automated assignments.
    \item[M4] \textbf{Fraction of TRs resolved by the assignee} was calculated by combining BTS data and TRR logs. This represents an ideal case, i.e., the team assigned the TR also resolved it.
    \item[M5] \textbf{Length of bug tossing chains} shows the number of TR reassignments as recorded in the BTS. This measure is commonly reported in studies on automated bug assignment~\citep{jeong2009improving,wu2018empirical}.
    \item[M6] \textbf{Relative TR handling time difference} is the relative difference in handling time between human-routed and auto-routed TRs based on the BTS data.
\end{enumerate}

We present what we call a Bayesian Causal Analysis (BCA) were we combine causal analysis~\citep{pearl2009causal,hernan2020causal} with Bayesian statistics \citep{mcelreath2020statistical,gelmanbda13} to estimate full posterior distributions of quantitative causal effects of interest. These are then used to investigate how automatic TR assignments impact the cycle time of TRs within Ericsson. Responding to calls for Bayesian data analysis in empirical software engineering~\citep{furia2019bayesian}, we present the first causal graph on the impact of automatic assignment of bug reports in large proprietary contexts where we estimate the effects using Bayesian analysis. Our visual model can be openly scrutinized by the community, as we quantify the confounding factors and measure the sensitivity to model noise and model misclassification as part of the BCA workflow.

\subsection{Qualitative Data Analysis}
\label{sec:qualitative_analysis}
To answer RQ1, RQ3, and RQ4 we primarily analyzed interview transcriptions. We iterated over the five steps of thematic analysis as described by~\citet{cruzes_recommended_2011}: 1) extract relevant data, 2) code the extracted data, 3) translate codes into themes, 4) create a model based on the themes, and 5) validate the synthesis. Moreover, we performed method triangulation by analyzing meeting minutes and recorded sprint meetings to validate claims from the interviews. Finally, we assessed our interpretations through member-checking. 

\subsubsection{Thematic analysis of interviews}
Thematic coding was carried out by the first and third author of this manuscript. We combined inductive and deductive coding, i.e., new codes were suggested based on the data, however, identified and organized guided by the current coding scheme. Data extraction, coding and interpretation were carried out in iterations of two interviews at a time. The current coding scheme evolved after each iteration, see Appendix~\ref{app:coding}. 

Our starting themes, and input to the first iteration, were the high-level RQs and a general description of the case under study. For RQ1, our starting point was to identify and code information regarding design decisions when implementing and deploying TRR, while for RQ3 and RQ4 our starting point was to code effects (direct and indirect) of adopting TRR. 
Codes and themes that were more context-specific (and thus not good candidates for a general theory) were used to provide a more in-depth description of the problem instance (our case under study) to support analytical generalization. 


In each iteration the two authors independently coded one interview transcript each, based on the current coding scheme. Then they met to discuss interpretations of new and previous codes, and to identify themes among the codes. After each iteration new codes and themes were derived, which in turn were used as input for the next iteration. When all interviews had been coded once, the coding scheme was reworked, i.e., the codes were restructured and new themes identified. In the final iteration, all transcripts were revisited to align the coding according to the final coding scheme. Sec~\ref{app:coding} in Appendix shows how the coding scheme evolved in each iteration. 

In Section~\ref{sec:rq1}--\ref{sec:rq4}, we use the following conventions when reporting the results of our qualitative analysis. To maintain a chain of evidence, we provide references in brackets to the interview IDs in Table~\ref{tab:interviewees}, e.g., [HL2] and [TRR1]. Most raw data is presented as inline quotes, but longer snippets appear in separate paragraphs to support readability. When words in quotes have been replaced for clarity, they appear in brackets, e.g., [TRR] and [augmented TRs].

\subsubsection{Method triangulation and member checking}
We triangulated findings from the interviews using evidence from the TRR team's sprint meetings. This was especially useful when building a timeline of the TRR introduction as events remembered by interviewees' often could be confirmed. 

The second, fourth, and fifth authors had access to the sensitive evidence from the sprint meetings and conducted the analysis. The TRR Team has a mature meeting culture and all sprint meetings result in detailed minutes-of-meeting (MoM) for the archive. Furthermore, as most meetings in the Covid-19 period were entirely virtual, almost all instances were available as recorded MS Teams meetings. In most cases the minutes contained the information needed, but the video recordings were available when even more details were needed. In total, we had access to evidence from 56 sprint meetings that typically lasted 30 minutes. 

Finally, we validated our interpretations by testing the coding scheme, the main takeaways, and the causal effect model on the study participants. All interviewees also received a draft version of the manuscript for internal review before we finalized the article. In the takeaways boxes, presented last throughout Sections~\ref{sec:rq1}--\ref{sec:rq4}, we generalize from the Ericsson case to practical implications that should apply to other organizations adopting automated bug assignment. On the other hand, for highly specific findings, we explicitly mention TRR and leave it to the reader to attempt analytical generalization based on our rich context description.

\section{RQ1-A: Evolution From Prototype to Tool} \label{sec:rq1}
In this section, we present how TRR evolved from prototype to deployed tool in the Ericsson context. Note that the overall issue assignment process also evolved during the last years, and while the preunderstanding presented in Figure~\ref{fig:preunderstanding} used to be a valid abstraction, the current process depicted in Figure~\ref{fig:complex_process} is considerably more complex. For the adoption of TRR, we present a longitudinal perspective from the conceptual ideas in the early 2010s to an operational internal Ericsson product roughly a decade later. As we investigated the TRR adoption process, we discovered several obstacles and facilitators. While these belong to RQ1, we present these separately in Section~\ref{sec:obst} to support the presentation of our results.

\begin{figure}
    \centering
    \includegraphics[width=1\textwidth]{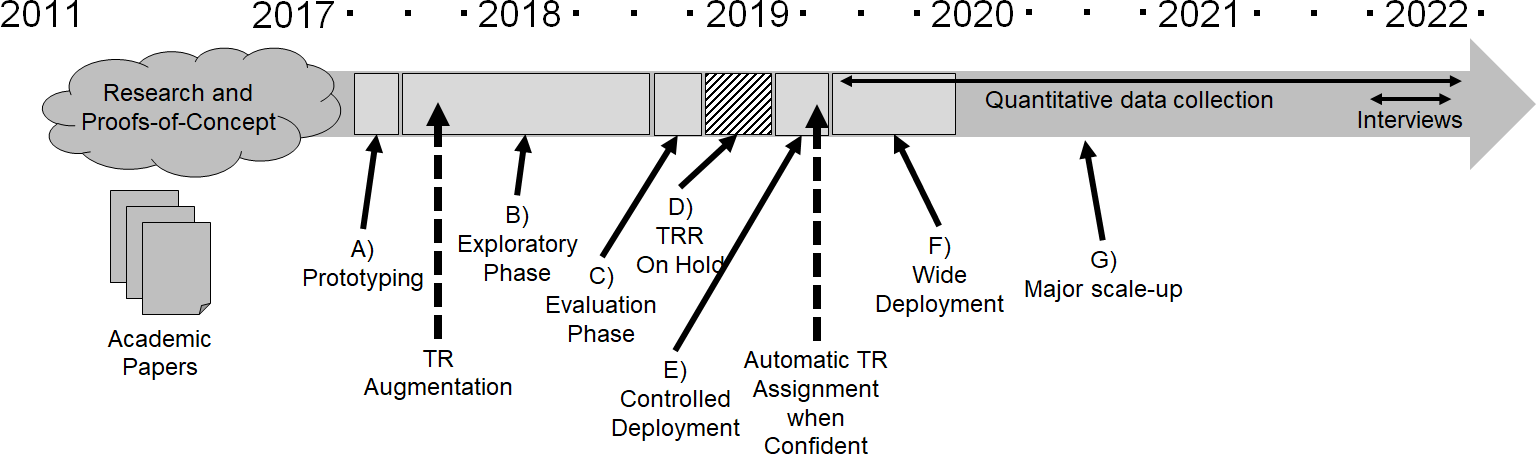}
    \caption{Timeline of the TRR evolution and adoption.}
    \label{fig:timeline}
\end{figure}

Figure~\ref{fig:timeline} shows a timeline of the TRR evolution from prototype to deployed tool. The cloud on the left side illustrate the early research and proofs-of-concept described in Section~\ref{sec:rq1_poc}. The large horizontal arrow depicts the TRR evolution until today, including six identified key phases A)-F). Dashed vertical arrows indicate when key TRR features were released. The rightmost part of the figure shows interviewees' suggestions for future TRR improvements. Finally, the horizontal black arrows show the time interval during which quantitative data was collected for this study and the time frame of the individual interviews, respectively.

Table~\ref{tab:codes_timeline} presents the 10 codes that emerged during the qualitative analysis of the interviews. The evolution of the codes is presented in the upper part of Figure~\ref{fig:code_evol_rq1} in Appendix~\ref{app:coding}.

\begin{table}[]
\caption{Codes used to describe the TRR adoption from the timeline dimension.}
\begin{tabular}{|p{2.3cm}|p{8cm}|}
\hline
\textbf{Code}            & \textbf{Description} \\ \hline
Key step                 & Important events in the TRR introduction, e.g., a meeting, a decision or a feature release.                 \\ \hline
Growing need for tool    & The need for tool support grew as issue assignment turned into a bottleneck.                 \\ \hline
Evolutionary development & TRR's accuracy increased gradually and new features were added over time.               \\ \hline
Co-evolving process      & Ericsson's issue management process changed concurrently with the TRR development.                 \\ \hline
Interruption             & Putting the development of TRR on hold.                 \\ \hline
Evolving use case        & The envisioned TRR use case evolved over time as stakeholders learned about the potential.               \\ \hline
Gradual introduction     & A step-wise approach to TRR roll-out.                 \\ \hline
Gradual acceptance       & Users' trust in TRR increased over time.                 \\ \hline
Maintenance              & Aspects related to TRR's lifecycle, e.g., retraining of ML model and ML pipelines. \\ \hline
Improvement suggestion   & Users' ideas about how TRR could evolve in the future.                 \\ \hline
\end{tabular}
\label{tab:codes_timeline}
\end{table}

\subsection{Research and Proofs-of-Concept (2011--2017)} \label{sec:rq1_poc}
The ambition to increase the level of the automation in the TR assignment originated in a process improvement initiative at Ericsson. The TR inflow in any large software organization is considerable. Given the size of Ericsson and the increasing software complexity of modern telecommunications systems, the amount of manual work needed to analyze and route incoming TRs was identified as an important improvement candidate. Moreover, further motivating research on the topic, is that the only people who know enough of the overall product family to successfully assigning incoming TRs to the right modules are highly senior engineers --- an obvious bottleneck in the process. As [HL1] expressed it: \textit{``It originated in lots of manual work and lots of that work was on our top players, the main TR Coords. Who had a lot to do with just deciding who should start a TR and they became blockers because they had so many other technical issues.''} Guided by Ericsson's ambitions to improve TR handling, we initiated corresponding solution-oriented research in 2011. 

Before diving into the details about the TRR introduction, we emphasize that improved tool support was not the only approach to improve TR handling at Ericsson. As [HL1-247] elaborated: \textit{``TRs are a cost that no one wants. So there is a constant hunt for better ways of handling TRs. Faster ways of handling TRs, routing them to low-cost countries and such things.''}. Note that the TR process also evolved at this time. [HL1-58] described the introduction of the prescreeners' meetings (cf. Figure~\ref{fig:complex_process}) as follows: \textit{``It was clear that a change was needed. What we did fast was this manual task where a couple of guys locked [themselves] into a room and looked at [all incoming TRs].''}  

Before 2018, we explored the concept, developed several proofs-of-concept, and published academic papers containing evaluations. Using supervised ML was found as the most appropriate way forward, as manually encoding rules for TR assignment would not scale to the problem. Furthermore, it would not be sustainable for Ericsson to maintain such a solution over time. [TRR1] elaborated: 

\interviewquote{The software product family has like a zillion things that can go wrong and lead to a TR. /.../ A rule-based thing that would be a huge product in itself. And it would be an overkill, because just a fraction of things that can go wrong will actually go wrong. Most of your rules won't ever be used. And it's not easy [to identify rules]. We don't have the domain knowledge for that, it's a really complex product family.}{TRR1-249}

The research leading to TRR resulted in several academic publications. The first paper was published in 2012~\citep{jonsson2012towards} and focused on using stacked generalization, an ensemble technique, to train classifiers for TR classification. The second author presented a discussion on the applicability of the approach at the ICSE~2013 doctoral symposium~\citep{jonsson2013increasing} --- and met the first author. They jointly conducted controlled experiments in 2013--2014 to find useful ensemble models, including an analysis of the generalization to another set of TRs in another company in a different domain~\citep{jonsson2016automated}. Concurrently, the second author explored Bayesian classification for TR assignment~\citep{jonsson2016automatic}. Although published somewhat later, i.e., after the first release of TRR, the fourth author did another quantitatively-oriented study trying to find improved ML models in 2019~\citep{sarkar2019improving}.

Despite several studies on improved ML model accuracy, Ericsson eventually preferred a less complicated ML model. The TRR implementation eventually deployed neither uses ensemble models nor Bayesian learning. Instead, a conventional logistic regression model is trained for TR classification. The features used for training the model are nominal features characterizing individual TRs in the BTS, e.g., submitting site, country, and severity, complemented by terms that frequently appear in the free-text TR descriptions. The natural language processing is limited to the established TF-IDF normalization, often used in software engineering research prototypes~\citep{borg2014changes}, i.e., each word count is divided by the number of documents this word appears in. TF-IDF effectively discriminates common words and makes rare words more prominent. 

Ericsson found the advantages of implementing a simpler and more maintainable ML model more important than having the most accurate ML solution. In the Ericsson case, it was evident that the more sophisticated models did not attain considerably higher accuracy --- an observation in line with observations in other fields adopting ML-based solutions~\citep{hansen2020virtue}. Maintainability is a prioritized TRR quality requirement, and both [TRR1] and [TRR2] are involved in automating ML training and deployment steps in a pipeline. [TRR2] explains that model retraining is currently triggered more frequently, i.e., it has changed from a quarterly to a weekly process --- which appears to be the right frequency for TRR. Ericsson's focus on maintainability of internal ML products resonates with previous work~\citep{flaounas2017beyond} and developing a customized pipeline is a recommended solution~\citep{john2021towards}.\\

\noindent\fbox{\begin{minipage}{0.97\textwidth}
Takeaways: 
\begin{enumerate}
    \item Cost reductions and the possibly to free up key people strongly motivate automated TR assignment.
    \item It is hard to select the right level of automation and appropriate enabling technology. Influencing factors include scalability, maintainability, and usability.
    \item A simple ML model with a mature pipeline might be the best technology option.
\end{enumerate}
\end{minipage}}

\subsection{Productization Phase I: Recommendations for all TRs (2017--2018)} \label{sec:rq1_phase1}

In March 2017, [TRR1] was the first developer tasked with turning the the various proofs-of-concept into an internal Ericsson product (cf. A) in Figure~\ref{fig:timeline}). He was told that \textit{``it can be productized within two weeks''} (this initial optimism was confirmed by [TRR2]) but TRR's first product owner assigned a development budget of two months. Adapting TRR for the operational context was non-trivial, but, in May 2017 [TRR1] presented the first prototype: \textit{``We had the team demo at 10:00 AM and I was sent the final fix that really did the real thing and not just a fake thing at 9:57.''} A few meetings with the TR Coords. followed, and they were very supportive --- TR assignment can be sensitive and trigger blame games, and [TRR1] remembered that the TR Coords. were happy to see that additional tool support was on the roadmap.

After the demo meeting, the TRR team integrated the tool into the BTS and implemented automatic augmentation of all incoming TRs. From this point in time, all new TRs got an attached note with the module predictions from TRR. The scheme for recommending modules was based on the cumulative confidence score reported by TRR, i.e., as long as the cumulative confidence was less than 80\%, additional modules were appended to the recommendation list in order of decreasing confidence levels. That is, if TRR predicted only one module with 100\% accuracy, only that single module would be in the recommendation list. On the other hand, if TRR's prediction was more uncertain, for instance, one module with 50\% confidence, one with 20\% confidence, one with 12\% confidence and two with 9\% confidence, the recommendation list would contain three modules (50\%+20\%+12\%=82\% $>$ 80\%). [TRR1-64] explained: \textit{``First, it was just putting a prediction, some information, into the [BTS] and it was not routing [TRs] at all.''} This type of decision support corresponds to \citet{parasuraman2000model}'s third level of automation, i.e., ``the system narrows the selection and presents these to the human.'' Data were collected during roughly a year to enable an internal evaluation of TRR's prediction accuracy (cf. B) in Figure~\ref{fig:timeline}).

Unfortunately, the evaluations showed that TRR's prediction accuracy remained mediocre. [TRR2] said that while the manual process corresponded to 75-78\% accuracy, TRR obtained roughly 50\% (verified in Sprint meeting MoM 2018-06-04). [TRR1] explained the consequences: \interviewquote{The line manager on our side in [the] summer of 2018 stated that if we couldn't reach 70\% until November the project would be shut down. And we didn't reach it. So the project was put on hold.}{TRR1} 

The evaluation phase and the period on hold are shown as C) and D) in Figure~\ref{fig:timeline}, respectively. [TRR2] remembered that internal ML engineering pushed the accuracy toward 65\% (verified in Sprint meeting MoM 2018-07-17), but the goal was not reached. Internal negotiations with several stakeholders followed, including TR coordinators, senior developers, the TRR team, and line managers. [TRR1] continued: \textit{``we saw that there is a module where we had a really good accuracy''} --- this was HL-ModB for which [HL2] was the TR Single-Point-of-Contact.\\

\noindent\fbox{\begin{minipage}{0.97\textwidth}
Takeaways: 
\begin{enumerate}
    \item The accuracy of the automated issue assignment is important for acceptance within organizations.
    \item How an organization uses the tool, including for which system modules it is activated, may need to be adapted to the accuracy level.
\end{enumerate}
\end{minipage}}

\subsection{Productization Phase II: Automatic Assignment for a Subset of TRs (2019)} \label{sec:rq1_phase2}
The TRR team met with representatives from HL-ModB during a physical two-days workshop in the end of 2018 to find a way forward. During the workshop, they decided to proceed with a less ambitious use case for TRR. Instead of automatically assigning all incoming TRs, only the ones for which TRR's predictions were particularly confident should be automatically assigned --- or as [TRR2-101] put it '\textit{``we got a chance to route part of the TRR inflow, not all''}. [TRR2-105] reported from evaluations that showed a correlation between the confidence and accuracy of predictions --- thus, the confidence threshold turned into a vital parameter that could be customized for different modules or types of TR. Several interviewees remembered this as a critical turning point in the TRR productization, e.g., \textit{``That was a game changer /.../ We knew that we couldn't route every TR.''} [TRR1] and \textit{``we had a different idea in the beginning, that we needed to automate everything. But /.../ we couldn't do that. The [accuracy] was not enough.''} [TRR2-384].

The HL-ModB turned into an early TRR adopter and the TRR team worked on tailored controlled releases. We identify [HL2] as an internal champion facilitating the TRR deployment and successfully navigating the internal corporate politics. HL-ModB proactively approached the TRR team [TRR2-153] and [HL2] remembered the setup as somewhat unorthodox: \interviewquote{``There was some sort of budgetary thing or project continuation thing, and I don't know all the details on that. But somewhere along the line, /.../ we could essentially take the [TRR] development team under our wings in [HL-ModB].}{HL2-140}

[HL2] hypothesized that the setup was possible since HL-ModB is a \textit{``slush bucket''} module where miscellaneous components often end up. From the TRR team's perspective, [HL2] was perceived as very supportive as he promised to smoothen the introduction of automatic bug assignment for HL-ModB: \textit{``Because the big question was... what happens if [TRR] would start routing incorrectly. And he said that we have this chance lets start it. Two or three hiccups he could manage.''} (TRR1-220). Furthermore, [HL2] proposed a set of other suitable modules (beyond HL-ModB) that could be the next targets in the gradual introduction of TRR at Ericsson during 2019. We recognize that [HL2] was the type of product champion reported by \citet{premkumar1995adoption} and \citet{hameed2012conceptual} to strongly influence the success of industrial tool adoption.

In early 2019, the TRR team initiated a controlled deployment of the re-engineered tool (cf. E) in Figure~\ref{fig:timeline} (verified in Sprint meeting MoM 2019-02-27). In April 2019, HL-ModB and two additional high-level modules were the first modules to receive automatic TR assignments from TRR, but, only for predictions with a confidence above a specific threshold. When confident, TRR's action corresponded to \citet{parasuraman2000model}'s seventh level of automation, i.e., ``the system executes tasks automatically, and informs the human.'' When less confident, TRR just augmented TRs the same way as before. [TRR2-236] confirmed that TRR matched the accuracy of the manual process for HL-ModB at this time. [TRR1-87] remembered that enabling automatic TRs assignments was a major step that involved a top management decision. 
 
Later during April 2019, three additional modules were included in the controlled deployment of TRR's higher level of automation (verified in Sprint meeting 2019-05-09). [TRR1-223] shared: \textit{``We learned a lot, we received a lot of feedback, but, basically the the predictions and the routings made sense.''} One of the learning outcomes was the tuning of the confidence thresholds for automatic TR assignment --- stakeholders had contrasting preferences as will be described in Section~\ref{sec:obst}. [HL2], the champion from HL-ModB, remained involved in the TRR adoption process, and explained the importance of managing the internal politics: \textit{``We had to flex the diplomatic muscles, ensure that we didn't piss people off. Ensure that we had support from the organizations and so on''} [HL2-158]. Moreover, [HL2-147] elaborated: \textit{``We very, very well established diplomatically in the organization. Present in all the relevant meetings. We knew all the stakeholders for essentially anything concerning processes. And we had established a fairly decent reputation as knowledgeable in the area and reasonably trustworthy.''} Some modules valued the increased level of automation from the start, such as [HL1-70]: \textit{``the real gain came when we started to use [TRR] as a routing tool and actually allowed the tool to route TRs.''} who considered this as \textit{``just a natural step.''} [HL1-281]. 

Several TRR users actively provided feedback during the tool adoption. One of [HL1]'s feature requests ended up in TRR: email notifications. The feature allows users to opt-in to email notifications if TRR produces moderately confident predictions (roughly 50\%) for an assignment to the user's module, i.e., a ``heads-up'' for the next pre-screening meeting. This was a major improvement according to her: \textit{``That was a large improvement to our work, because as soon as we got the emails, we started to look at the TR and could say if it was ours. Or we could put a comment next to the comment from TRR. We could say `[HL-ModA] front desk saying: We suggest that [anonymous module] starts with this [TR]. We think it's something with the throughput.'''} [HL1-146]. Another feature that was introduced was an explicit feedback button in connection to individual TRR predictions, i.e., users can now easily provide free-text feedback to the TRR developers [HL1-170].

Later in 2019, modules were incrementally added for auto-routing with TRR (cf. F) in Figure~\ref{fig:timeline}). The new modules were large, both in terms of people and number of TRs handled. By the end of the year, TRR encompassed all 4G/5G modules. TRR gradually auto-routed a larger fraction of TRs and [TRR1-223] reported the progress as\footnote{Values normalized for confidentiality.}: \textit{``In the summer of 2020 we really scaled it up. We went from routing 0.16-0.24 TRs [per time unit] to 0.6 plus TRs. /.../ and now we are between 0.8 and 1 TRs''} This statement (cf. G) in Figure~\ref{fig:timeline}) is supported by quantitative data that will be discussed in relation to Figure~\ref{fig:routing_ratio}. Two factors explain the scale-up in numbers of routed TRs. First, Ericsson rolled out many 5G systems at this time. Second, the TRR team decreased confidence thresholds to auto-route more TRs.

TRR was seen as a natural part of the TR process by several interviewees in the end of 2021. [HL1-242] stated \textit{``[TRR] is for us the new normal already, so I think we have incorporated it''}. Interestingly, we found that not all TR Coords. were aware that TRR currently performs automatic assignments. \textit{``We had a long discussion about whether [TRR] would actually route or just give a recommendation in the text and that was the big thing at that point. /.../ I don't know whether it's actually auto-routing now.''} [CO1-103] [CO2] believed that TRR provided only recommendations, while [CO3] perceived that automatic assignments actually are in use. This suggests that the TR Coords. do not need to know on what automation level TRR operates. Automated assignments limit the number of TRs they need to manually process, but there has been no drastic difference in the work task.\\

\noindent\fbox{\begin{minipage}{0.97\textwidth}
Takeaways: 
\begin{enumerate}
    \item The product champion, diplomats, and early adopters are essential for gradual tool introduction in large organizations.
    \item After deploying an early version of a tool, it can evolve incrementally, resulting in gradually improved accuracy and new features.
    \item TRR is now an incorporated and invisible part of the issue assignment process at Ericsson.
\end{enumerate}
\end{minipage}}

\subsection{Potential Future Direct and Indirect Improvements} \label{sec:rq1_improve}
TRR will inevitably need to keep evolving to match the needs of its users within the organization.
While discussing TRR and the ways of working at Ericsson, we explicitly asked the interviewees to propose new tool features and related process improvements.

Some interviewees discussed \textit{increasing TRR's accuracy}. Perhaps TRR's accuracy would increase if TR attachments, especially log files, were considered by the tool. Indeed, only a subset of the available information is currently used as features when training the ML model. [LL1] was particularly vocal about the importance of extracting information from the TRs' attached logs, and repeatedly stressed the importance of extending TRR in this direction during the interview. It is obvious that [LL1] is used to finding key information in the logs when processing TRs manually. [TRR1] also speculated about the possibility to increase TRR's by extracting information from attached logs, thus confirming that the TRR team are aware of the potential. 

Another approach to increasing TRR's accuracy would be to \textit{make the predictions more fine-grained}, which would be beneficial for large modules that are organized into sub-modules. [LL2-297] proposed the following improvement: \textit{``[Predict] which radio type. And which [communication] band /.../ That would be so nice not to search in the logs for this.''} Again, the interviewee indicated the value of extracting information from the logs.

In the same vein, \citet{aktas2020exploratory} explored adding information from screenshots in IssueTAG. They found that many of IssueTAG's incorrect assignments at IsBank contained screenshots rather than detailed textual descriptions. While screenshots are less relevant in the embedded telecommunications context at Ericsson, it is possible that TRs with logs contain less descriptive text and thus are misclassified more frequently.

A subset of the interviewees proposed enhancing TRR to \textit{predict TR fields beyond the team assignment}. During the issue management process, technical coordinators and pre-screeners assign values to several fields in the BTS, e.g., severity, priority, and target versions for patches. Both [HL1] and [LL2] proposed this, indicating a broad potential within the organization as they represent contrasting viewpoints, i.e., an optimistic high-level module and a skeptical low-level module. On the other hand, the two TC Coords. [CO1] and [CO2] expressed worry toward automating such sensitive aspects as the decisions require deep understanding of the market and different customers. [HL2] shared the same concern but from a developer's perspective: 

\interviewquote{Some of [the TR fields] would be complete suicide to try to automate /.../ for instance criticality which is a field that essentially translates into priority with very specific working conditions connected to it. For instance, if you have something classified as a stopping issue, it essentially means mandatory weekend work for the organization. And so you can imagine what would happen if we let loose something with good, but not perfect accuracy on that field.}{[HL2-285]}

Three interviewees imagined how \textit{TRR could trigger other changes related to both processes and products}. First, [HL1] proposed the introduction of an online forum where TR specifics could be discussed instead of stacking attached notes in the BTS. Second, [LL2] suggested that TRR should operate in two phases. For newly submitted TRs, he would prefer TRR to only route if highly confident. Then, after the pre-screening meeting  TRR could get a second chance to auto-route based on the additional information is available in the TR. Third, [TRR1] envisioned that TRR could impact the Ericsson products themselves to generate and augment TRs that more easily could be auto-routed. From this innovative perspective, using ML-based issue assignment could lead to TRs that are easier for TRR to route:

\interviewquote{That would be the real deal. If 100\% of the TRs could be auto-routed then you could really switch from one day to another, freeing up hundreds of people from pre-screening and save a lot of senior staff on this tedious job.}{[TRR1-382]}

Finally, based on our interviews, we posit that a short internal training of how TRR works could be useful to remove some incorrect assumptions. In the interviews, we found that [LL2] had the wrong impression of when TRR acts, [CO2] was not aware that TRR currently does auto-routing of TRs, and also [CO1] was unsure. All stakeholders do not need to know the implementation details, but at which level of automation TRR operates should be a shared understanding within the organization. The internal TRR introduction appears to have been insufficient, as indicated by [LL2-200]: \textit{``I had just one introduction presentation from some of your colleagues in one hour, and then we knew that this was the way forward.''} Maybe the fact that TRR is a ``background tool'' with limited user interaction (discussed next in Section~\ref{sec:obst}) made the organization provide only minimal training.\\

\noindent\fbox{\begin{minipage}{0.97\textwidth}
Takeaways: 
\begin{enumerate}
    \item TRR's accuracy could be increased by 1) considering attached logs in the ML model and 2) providing finer-grained module predictions.
    \item Automated bug assignment might catalyze both process and product improvements in organizations.
    \item Internal training could align all stakeholders' understanding of when and how automated bug assignment operates. 
\end{enumerate}
\end{minipage}}

\section{RQ1-B: Obstacles and Facilitators in the TRR Adoption} \label{sec:obst}
In this section, also addressing RQ1, we present the most important obstacles and facilitators in the TRR adoption at Ericsson. Although some overlaps are inevitable, the discussion is organized into: acceptance, ML technology obstacles, organizational factors and character traits, and facilitators. Tables~\ref{tab:codes_obst} and~\ref{tab:codes_fac} present the 25 codes that emerged in the analysis. The evolution of the codes is presented in Figure~\ref{fig:code_evol_rq1} in Appendix~\ref{app:coding}.

Several challenges related to the transition from research prototypes to industrial tools have been reported in the literature, e.g., limited trust, scalability, usability, process integration, staff training, and maintenance~\citep{favre2003tool,lee2004trust}. The SE community is also aware that industrial tool adoption often takes years~\citep{garousi2019characterizing}. We discuss our RQ1-B results in light of such previous work.

\begin{table}[]
\caption{Codes used to describe obstacles and facilitators in the TRR adoption related to the high-level (HL) codes acceptance, technology, character traits, and internal politics.}
\begin{tabular}{|p{1.5cm}|p{2cm}|p{7cm}|}
\hline
\multicolumn{1}{|c|}{\textbf{HL Code}}                                                          & \textbf{Code}             & \textbf{Description}                                                                                                                                                                                \\ \hline
\multirow{4}{*}{Acceptance}                                                     & Motivation                & Factors and incentives motivating indivuduals or teams .to use TRR, e.g., reducing mundane work tasks                                                                                               \\ \cline{2-3} 
                                                                                & Risk-reward               & Views on the perceived benefit of using TRR vs. the involved risks, e.g., the cost of misclassifications.                                                                                           \\ \cline{2-3} 
                                                                                & Cost-benefit misalignment & Concerns about who benefits from TRR and who pays the price for misclassifications. E.g., Person A might not accept regularly spending 30 min extra work to save Person B hours of work on average. \\ \cline{2-3} 
                                                                                & Rigor vs. relevance       & Views on rigor vs. relevance when evaluating TRR.                                                                                                                                                   \\ \hline
\multirow{7}{*}{\begin{tabular}[c]{@{}l@{}}Technology\\ obstacles\end{tabular}} & Accuracy                  & TRR's accuracy is perceived as an obstacle.                                                                                                                                                         \\ \cline{2-3} 
                                                                                & Biased classification     & Perceptions of biased TRR output, e.g., too many TRs go to top-level modules or the bottom Radio level.                                                                                             \\ \cline{2-3} 
                                                                                & Information entities      & Concerns about what information is available in TRs, e.g., log extracts, screenshots, and product information.                                                                                      \\ \cline{2-3} 
                                                                                & Data quality              & Concerns about the quality of TR information and how it affects TRR's predictive power.                                                                                                             \\ \cline{2-3} 
                                                                                & Tool integration          & Concerns about integrating TRR with other Ericsson tools.                                                                                                                                           \\ \cline{2-3} 
                                                                                & Sensitivity to changes    & Changes to Ericsson's products, e.g., replaced modules, could lead to systematic errors in the routing.                                                                                             \\ \cline{2-3} 
                                                                                & Explainability            & Understanding the rationale of TRR's output, i.e., what influenced the TR assignments. \\ 
\hline
\multirow{6}{*}{\begin{tabular}[c]{@{}l@{}}Character\\ traits\end{tabular}} & Diplomats            & Employees with considerable diplomatic skills.                                                                             \\ \cline{2-3} 
                                  & Early adopters       & Employees who embraced TRR early by opting in to use higher levels of automation earlier.                                  \\ \cline{2-3} 
                                  & Enthusiasts          & TRR stakeholders who are very optimistic about what what TRR and ML can solve.                                             \\ \cline{2-3} 
                                  & Skeptics             & Employees with negative impressions on increasing automation levels, e.g., perceptions on TRR's use of ML.                 \\ \cline{2-3} 
                                  & Limited awareness    & TRR stakeholders are not fully aware of the complete issue assignment process or the internals of TRR, e.g., its ML model. \\ \cline{2-3} 
                                  & Skilled engineers    & TRR users have different needs depending on how skilled and tech-savvy they are.                                           \\ \hline
\multirow{2}{*}{Politics}         & Convincing           & Concerns about how individuals and teams were convinced to accept TRR in the issue assignment process.                     \\ \cline{2-3} 
                                  & Stakeholder analysis & Stakeholder analysis and initial requirements engineering, e.g., setting accuracy target for the ML model.                 \\ \hline
\end{tabular}
\label{tab:codes_obst}
\end{table}

\begin{table}[]
\caption{Codes used to describe key facilitators in the TRR adoption.}
\begin{tabular}{|p{2.3cm}|p{8cm}|}
\hline
\textbf{Code}            & \textbf{Description} \\ \hline
Established adoption process                 & Following a structured tool adoption process in the organization.                 \\ \hline
Tool-process alignment    & The TRR use case is streamlined to be aligned with the issue assignment process.                \\ \hline
Confidence thresholds & Customizing module-specific confidence thresholds for when to let TRR make automatic assignments.      \\ \hline
Stakeholder feedback       & Direct access to TRR users facilitated evolutionary development.                 \\ \hline
Evaluation method and results            & Reliable methods to measure TRR's output and effects in the organization, and promising measurement results.                 \\ \hline
Background tool        & TRR operates in the background without needing additional user actions, i.e., users don't need much training.               \\ \hline
\end{tabular}
\label{tab:codes_fac}
\end{table}

\subsection{Accepting a Higher Level of Automation} \label{sec:rq1_acceptance}
Ericsson engineers must develop trust in TRR to accept its adoption in the organization. Many researchers have tried explaining the multi-faceted concept of trust. A systematic review by \citet{hoff2015trust} reports that most explanations contain three constituents. First, there must be 1) a truster to give trust, 2) a trustee to accept the trust, and 3) something must be at stake. Second, the trustee must have an incentive to perform some task. Third, there must be a risk that the trustee will fail to perform the task. In the TRR case, 1) TRR is the truster, 2) Ericsson engineers are trustees, and 3) additional work is at stake if TRR does not perform adequately. Ericsson engineers are incentivized to delegate bug assignments to TRR, but, there is a risk that TRR will misclassify TRs. [HL2] elaborated on the risk: \interviewquote{One of the reasons [TRR] wasn't done before was quite simply because it was perceived as risky. To actively interject actions into the process flow of something as important as the TRs.}{[HL2-198]}

Trust in increased levels of automation evolves as users interact with tools. \citet{lee2004trust} discuss how trust dynamically increases or decreases as users process information about the capabilities of automation. They distinguish three parallel processes, i.e., the \textit{affective process} (emotional responses to violations and confirmations of implicit expectancies), the \textit{analytical process} (rational evaluation), and the \textit{analogical process} (comparing personal trust to opinions of others). Our study identified elements of all three, and [HL2] summarized it as: \interviewquote{It took a little while for everyone to get the sense of what TRR was good at, what it was bad at, and how it actually interacted with the organization.}{[HL2-109]}

The interviewees reported several instances of automatic TR assignments that decreased engineers' trust in TRR. [LL1-280] emphasized TRR's problem of assigning new TRs that do not resemble anything in the training data. If a batch of such similar TRs appear in a short time frame, it can quickly deteriorate trust: 

\interviewquote{When something new appears, a new issue that TRR doesn't actually recognize very well. And then you get five very similar to yours in in in the same week or in a few days. And they all get routed in the exactly same way, and it turns out to be wrong. /.../ It causes the people to have a reaction of `Why do we have this thing? Why?'}{[LL1-283]}

[LL1] explained that part of the problem was that TRR could not distinguish between Radio as a product, Radio software, and the Radio hardware box itself. Apparently, TRR assigned everything to Radio software --- during one period, Radio even requested the TRR team to disable automatic routing to their module. [LL2] mentioned several mails from Radio engineers who were unhappy about TRR assignments --- the existence of such heated communication was confirmed also by [TRR1].

While there have been setbacks, our interviews confirm the development of trust in TRR at Ericsson --- even the skeptical trustee [LL2-164], who personally suffered from TRR's downsides, clearly showed signs of increasing trust: \textit{``But of course, machines are getting more and more intelligent, more inputs, better decisions, more accurate decisions.''} The similar trust development was reported by [LL1] who initially did not at all believe in the approach for [LL-ModA] but gradually recognized the benefits over time.

In comparison, \citet{aktas2020automated} reported trust development of IssueTAG at IsBank as easier. In fact, the authors strikingly identified no objections at all and hypothesize that \textit{``this was because all the stakeholders believed that they would benefit from the new system and none of them felt threatened by it.''} It appears that the downsides of roughly 15\% misclassifications at IsBank is a negligible problem compared to the benefits of savings time for most of the issues. \citet{aktas2020automated} also found that IssueTAG was accepted because no stakeholders were afraid to loose their jobs. At Ericsson, we found no indications that any engineers were afraid to be made obsolete by TRR --- on the contrary, several interviewees welcomed a future involving less effort manual work in issue assignment.\\

\noindent\fbox{\begin{minipage}{0.97\textwidth}
Takeaways: 
\begin{enumerate}
    \item Existing theoretical models for trust in automation can largely explain the adoption of automated bug assignment.
    \item Despite some setbacks, users gradually developed trust in TRR.
\end{enumerate}  
\end{minipage}}

\subsection{Machine Learning Technology Obstacles} \label{sec:rq1_tech}
As for any engineering effort, the design and implementation of TRR necessitated overcoming some obstacles related to technology. Several implementation hindrances would apply to any tool adoption in a large organization, e.g., integrating TRR with the existing tool suite and developing proper web services for TRR operations. None of those aspects stood out in any way, thus we focus the rest of this section on novel considerations originating in the ML aspects of TRR. 

\textit{Reaching useful accuracy levels} is often a challenge when developing predictive ML models~\citet{vogelsang2019requirements}. Despite not being trained data scientists, the TRR team needed to perform ML engineering activities to reach the accuracy goals, e.g., devising proper evaluation methods and feature engineering. The accuracy challenge is described in relation to the timeline of the TRR adoption in Section~\ref{sec:rq1_phase1}. 

The \textit{availability and quality of training data} are essential for the success of supervised ML. TRR's predictions largely rely on the textual content of TRs. Consequently, for some TRs, TRR barely has any input with predictive power as explained by [TRR2-236] \textit{``sometimes it is impossible for [TRR] when the TR is saying `OK, you can download the logs here.' Or `some images is a attached and you can find the relevant information there.' It is invisible for [TRR] or hard to retrieve.''} [LL1] confirms this issue, and strongly calls for future versions of TRR to consider attached logs. Moreover, [LL1-90] complained that the quality of the textual content in TRs sometimes is limited: \textit{``[TRs] sometimes has wrong information and in most cases it doesn't contain all the necessary information to determine where the problem actually is.''}.

One intrinsic challenge with ML-based systems is \textit{explainability}. Explainable AI is a trending research topic, and initiatives exist also in software engineering~\citet{xai4sebook}. In this study, we explicitly asked interviewees about their thoughts in relation to TRR explainability (cf. Part 5 of the interview guide in Appendix~\ref{app:int_guide}). While none of the interviewees considered it to be a major issue, most remembered automatically assigned TRs that appeared strange, e.g., \textit{``Why on Earth did we get this one?''} [HL1-230]. [HL1], [TRR1] and [LL2] brought up anecdotes of when TRR assignements triggered angry emails within the organization, e.g., \textit{``This was routed to us. This is unheard of /.../ This machine is always wrong!''} [TRR1-335].

To support explainability, the TRR team developed a browser plugin to extract the rationale behind individual TRR predictions. The tool is not distributed to all users, but has been used on a case by case basis when needed. [TRR1-433] explained how this tool has resolved arguments when TR assignees were unhappy: \textit{``When someone's finger is in [the TRR Team's] eye because how stupid the prediction was. More often than not, there was a quite plausible explanation why [TRR] said that /.../ What features were extracted and what led to the prediction.''} [TRR-1] clearly described how convincing it can be to extract a list of domain specific keywords that motivate certain assignments. [LL1] agrees that rationales could be interesting, especially in cases where TRR makes predictions with low confidence levels.

On the other hand, presenting explanations when not requested can result in an information overload. [CO1] and [LL2] were very clear about this, and stress that they do not want to know any details about how TRR predicts assignment destinations, for example: \interviewquote{As far as I can trust the information [TRR] is generating, we are happy. As far as we have accuracy, which is at least 85\% /.../ Then the intelligence behind that is not so interesting for us to know.}{[LL2-285]}

Contradictory to advice for recommendation system output~\citet{murphy-hill_recommendation_2014}, the TRR team deliberately chose not to present any motivation of TRR's predictions along with its output. In practice, finding the balance between explainability and output conciseness is non-trivial --- it is evident that different types of users have contrasting preferences. \\

\noindent\fbox{\begin{minipage}{0.97\textwidth}
Takeaways: 
\begin{enumerate}
    \item A mature internal tools team decreases engineering risks, e.g., tool integration and web development, when adopting automated bug assignment,
    \item Substantial ML engineering might be required to make automated bug assignment accurate enough.
    \item Finding the balance between explainability and output conciseness is non-trivial.
\end{enumerate}
\end{minipage}}

\subsection{Organizational Characteristics and Personality Traits} \label{sec:rq1_org_traits}

Numerous studies have targeted the characteristics of organizations adopting novel technology. A review by  \citet{hameed2012conceptual} identified 41 factors in the innovation adoption literature. In this section, we report the most prominent related findings from the interviews, including the political and social dynamics that always are at play in large companies~\citep{premkumar1995adoption}. Furthermore, we present findings collected to personality traits, e.g., the tendency to champion new solutions and conservative points of view. The TRR team presented the internal inertia as more difficult overcome than the technology challenges reported in Section~\ref{sec:rq1_tech}. [TRR1-297] set the tone as follows:

\interviewquote{Implementation wise, that was the minority. We wanted to turn a huge organization with many, many people. Of course there were voices who were really supportive. There were voices who were sceptical and we didn't have one single point to talk to. We talked to many people and you can imagine there were always some debate.}{TRR1-297}

Working toward acceptance within a large organization requires a trust development process as discussed in Section~\ref{sec:rq1_acceptance}. Various aspects of negotiation and finding convincing ways forward were presented in relation to the TRR timeline in Sections~\ref{sec:rq1_phase1}--\ref{sec:rq1_phase2}. [HL2], previously identified as a champion with noteworthy diplomatic skills, stressed the importance of stakeholder identification and analysis during tool adoption in general: 
\interviewquote{We often worked the grassroots and tried to find the users that own the end-to-end problem. So those tool [adoptions] that fit that niche were quite successful. But the tools for which Person A spends 15 minutes to save Person B a full day, those were a little bit trickier.}{[HL2-245]}


[LL2-117] vocalized several skeptical concerns, e.g., \textit{``I have seen this auto-routing that I really don't like. I have also stated that this is not in my interest to see TRs auto-routed to [LL-ModA].''} His primary concern was that too many TRs are incorrectly assigned to modules related to LL-ModA and reassigning them to other modules is a very costly activity --- once a TR has been initially assigned, they tend to stick. His top goal as a maintenance leader is to increase the accuracy of TRs assigned to LL-ModA as he clearly found that the previous manual process (corresponding to 75\% accuracy) generated too much waste in the organization. He reported that increasing the level of automation before reaching high accuracy levels would lead to additional costs that the maintenance organization cannot bare. The problem of rerouting TRR's initial assignments is confirmed on the high-level modules: \textit{If you get an incorrectly assigned TR from [TRR], you have no human to reach out to. It came from a machine. /.../ You just cannot fight a machine.} [HL2-296]. 

[LL2]'s solution to minimize the number of misrouted TRs was to first let human pre-screeners add comments based on their analyses before TRR makes any predictions. He also wants TRR to use the prescreeners' comments as input, also as a sign of respect: \textit{``There is a prescreening behind every TR and these guys are really putting a lot of efforts at the moment. /.../ 
And we think that they should be respected because they are the guys who know the system and make the best assessments.} [LL2-149] On the other hand, [LL2] agrees that for very confident TRR predictions, LL-ModA could accept bypassing the human analysis to reduce manual effort.

[TRR1] expressed two statements that illustrated the opposite view of TRR's automatic assignments to LL-ModA. First, he remembered an occasion when LL-ModA rejected a TR automatically assigned by TRR with the motivation that it did not come from the human TR Coords. --- which was seen as provocative by the TRR team who responded by providing extensive statistics of manual and automatic accuracy levels. Second, he raised the point that it is hard for anyone to assign TRs to LL-ModA: \interviewquote{If [LL-ModA] would accept TRs more easily. Let it be machine routed. Let it be coming from other modules. Then that would be a better place for everyone because it's really hard for other modules to do handover of TRs to them.}{ [TRR1-499]}

On the other hand, [TRR1] also stressed that he was aware of how difficult the maintenance activities are for LL-ModA as \textit{``their module is always on fire''} and must \textit{``support an insane product portfolio''} of legacy systems. He understood that TRR added to the pressure: \interviewquote{The existence of [TRR] itself challenges the organization. And you have these fortresses in organizations, as in this case LL-ModA that has a habit to be defensive and just push back everything. Because now there is this robot [TRR] who keeps them bombarded with TRs.}{[TRR1-514]} Despite their compassion for LL-ModA, the TRR team confirmed that it would have been detrimental to the TRR adoption if LL-ModA got an exemption to not receive automatically assigned TRs --- it could have set a bad example for other modules within Ericsson.


\noindent\fbox{\begin{minipage}{0.97\textwidth}
Takeaways: 
\begin{enumerate}
    \item Organizational inertia can be harder to overcome than technology challenges.
    \item Stakeholder identification and analysis are vital activities for successful tool adoption.
    \item The initial perception of TRR contrasted substantially between users from different modules.
\end{enumerate}
\end{minipage}}

\subsection{Key Facilitators in the TRR Adoption}
This section concludes the discussion on RQ1 by presenting key facilitators for the TRR adoption identified at Ericsson. The findings are both of organizational and technical nature.

The TRR adoption was facilitated by the existence of \textit{explicit stakeholders}. As described in Section~\ref{sec:rq1_poc}, the development of TRR originated in a real need at Ericsson, i.e., the need to improve the costly TR handling process. The TR Coords. were stakeholders from the beginning, and additional stakeholders in the modules were gradually identified by the TRR Team: \textit{``In the beginning it was harder. But later, /.../ it went smoother because we already had the [communication] channels, the representatives to talk to.''} [TRR2-167] [TRR2-184] also stressed that it is easier to convince engineers if the tool development originated in Ericsson's internal improvement initiatives rather than external sources. This resonates with the well-known ``not-invented-here'' phenomenon in corporate cultures~\citep{stefi2015developers}.

Another key TRR facilitator was the approach of \textit{gradual tool introduction}. The gradual introduction encompasses both 1) the two steps of increased automation levels and 2) the controlled deployment of TRR to carefully selected modules. [CO1] explained it as follows: \interviewquote{The way we took [TRR] in by [letting modules] opt in or opt out and and doing trial runs with only recommendations made it fairly straightforward to see `Yes, [TRR] is doing OK. Yes, we can adopt it.' /.../ So I think that made it a fairly smooth ride.}{[CO1-133]} [CO1] also commended the TRR team for carefully aligning the TRR deployment with the current TR handling process, i.e., there was no extra work required by the TRR Coords. post-deployment. In the IssueTAG adoption at IsBank, \citet{aktas2020automated} emphasized the importance of a gradual tool introduction to build confidence and subsequently support acceptance of a higher automation level.

Decreasing the TRR ambition to only operate on the higher level of automation when above a \textit{configurable confidence threshold} was reported as a game changer by [TRR1] in Section~\ref{sec:rq1_phase2}. Also [HL2] emphasized the importance of this decision, as it gave modules a greater sense of control. [HL1] recognized how her preferred confidence configuration stood out in the organization: \interviewquote{I know that many of the other modules were really conservative and said `OK, [TRR] has to guess 99\% ours to send it.' Or 98 maybe. I mean that was at a very different mindset. Whilst we compared it a lot to what would happen if [TRR] didn't send it to us. /.../ and when I said about 50\% is good. They were like... `ooh?' really surprised.}{HL1-96} [HL1]'s viewpoint was that as long as her module had the highest confidence level among the candidates, the TR would most likely anyway reach her --- she perceived her top-level module as the default target for most TR assignments.

Finally, we found that a key facilitator of the TRR adoption was that it is \textit{a background tool}. Users do not have to operate TRR in anyway, they only need to act on its output. TRR automatically augments a TR: \textit{``[The TRR adoption] went really smooth because it is really simple. On our end it is a note, a row in the notebook of the TR.''} [HL1-278] Only if confident, the TR gets automatically assigned to the most likely module. Also [HL2] stressed that TRR is a tool from which no explicit user input needed: \interviewquote{The automatic routing is something that just disappears from peoples' backlogs if you do it with sufficient accuracy. It's very uncontroversial. People just get less to do and that they will always be happy about.}{[HL2-242]}

\noindent\fbox{\begin{minipage}{0.97\textwidth}
Takeaways: 
\begin{enumerate}
    \item Explicit stakeholders and a gradual introduction facilitated the TRR adoption.
    \item Restricting TRR to only operate on the higher level of automation when above a configurable confidence threshold was a key decision.
    \item As TRR operates as a background tool with very limited user action, Ericsson could more easily adopt it.
\end{enumerate}
\end{minipage}}

\section{RQ2: Accuracy of TRR's Assignments} \label{sec:rq2}
This section presents our quantitative analysis of TRR's accuracy. The analysis focuses on two metrics introduced in Section~\ref{sec:quan_anal}, i.e., Fraction of TRs resolved by the assignee (M4) and Length of bug tossing chains (M5).

\begin{figure}
    \centering
    \includegraphics[width=1\textwidth]{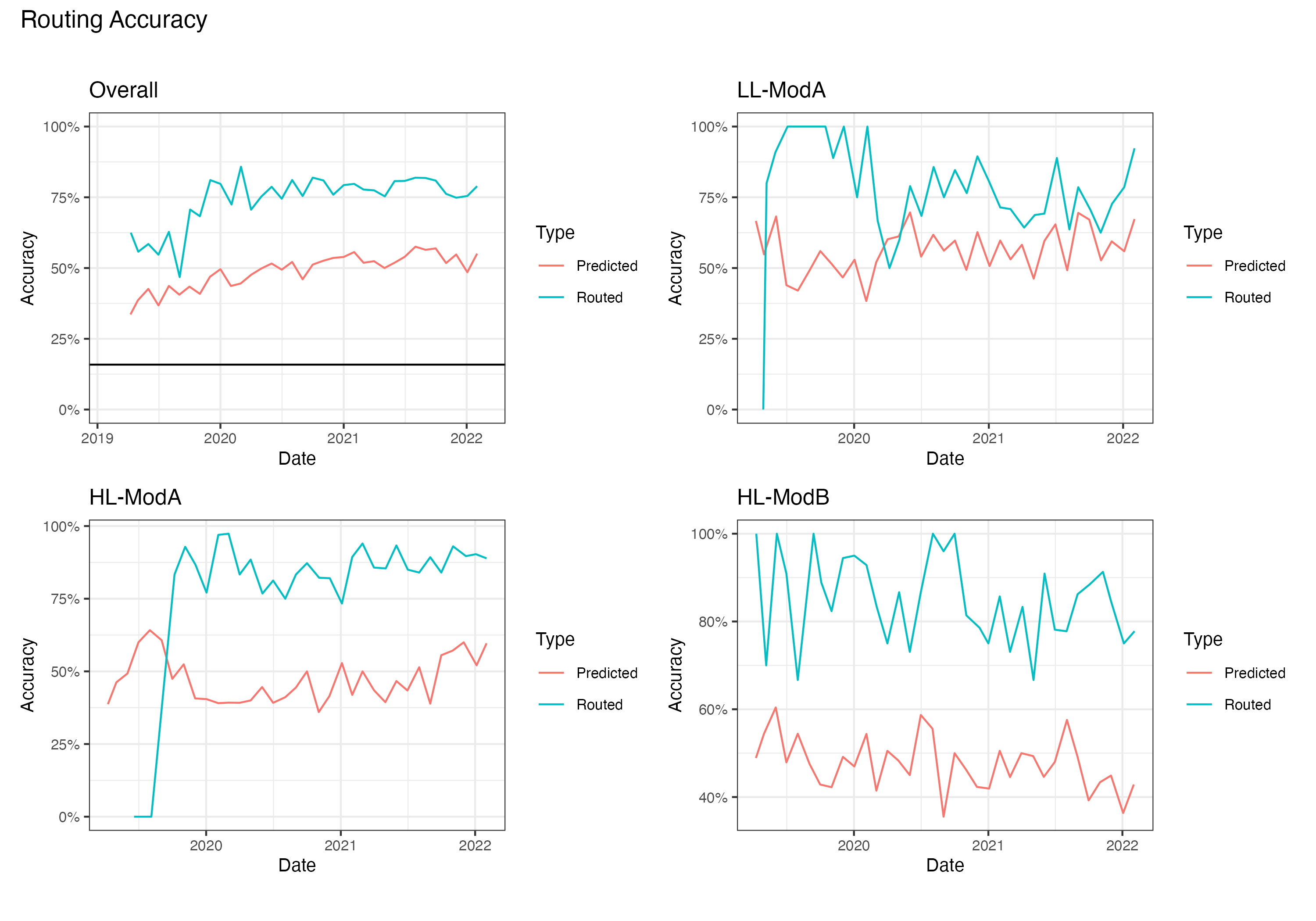}
        \caption{TRR's accuracy over time. The turquoise lines represents auto-routed TRs whereas the red lines depict the accuracy of all predictions.}
    \label{fig:accuracy_over_time}
\end{figure}

Figure~\ref{fig:accuracy_over_time} shows TRR's routing accuracy over time plotted per month. The upper left subplot depicts an overall view across all modules. The turquoise line represents the accuracy of the TRs that were auto-routed by TRR. This subset of TRs corresponds to predictions for which TRR's confidence level was above the specified threshold. The red line, on the other hand, shows the accuracy of all predictions no matter the confidence level, i.e., it encompasses both the auto-routed TRs and the TRs that were only augmented in the BTS. Finally, the horizontal black line shows the accuracy of the ZeroR baseline, i.e., always predicting the majority class.

Since the summer of 2020, TRR has consistently auto-routed TRs with an accuracy within the range 75\%--80\%. Considering also the less confident TRR predictions, i.e., both auto-routed and augmented TRs, TRR's accuracy has fluctuated around 50\% in the same time frame. Both the turquoise and red lines show increases since TRR was deployed in 2019, but, we find that TRR's accuracy appears to now be stable. Finally, it is clear that TRR outperforms the ZeroR baseline at 14\%.

The bottom subplots in Figure~\ref{fig:accuracy_over_time} show TRR's accuracy for HL-ModA and HL-ModB, respectively. As the total numbers of TRs are fewer on the module level, the values fluctuate more and there is no apparent increasing trend. TRR's accuracy for HL-ModA is within the range 75\%--100\% since the last quarter of 2019, whereas the accuracy has been around 80\% on average for HL-ModB. We find that TRR's predictions for the two high-level modules under study have been more accurate than the average module. 

The upper right subplot in Figure~\ref{fig:accuracy_over_time} depicts TRR's accuracy for LL-ModA. This module displays maximum initial variation just after TRR was deployed. Since the summer of 2020, TRR's accuracy has stabilized within the range 65\%--85\%. While TRR is slightly less accurate for LL-ModA than HL-ModA and HL-ModB, it is not the sole reason why we found more skeptical views on the lower level --- a deeper understanding is presented in RQ1 and RQ4.

\begin{figure}
    \centering
    \includegraphics[width=1\textwidth]{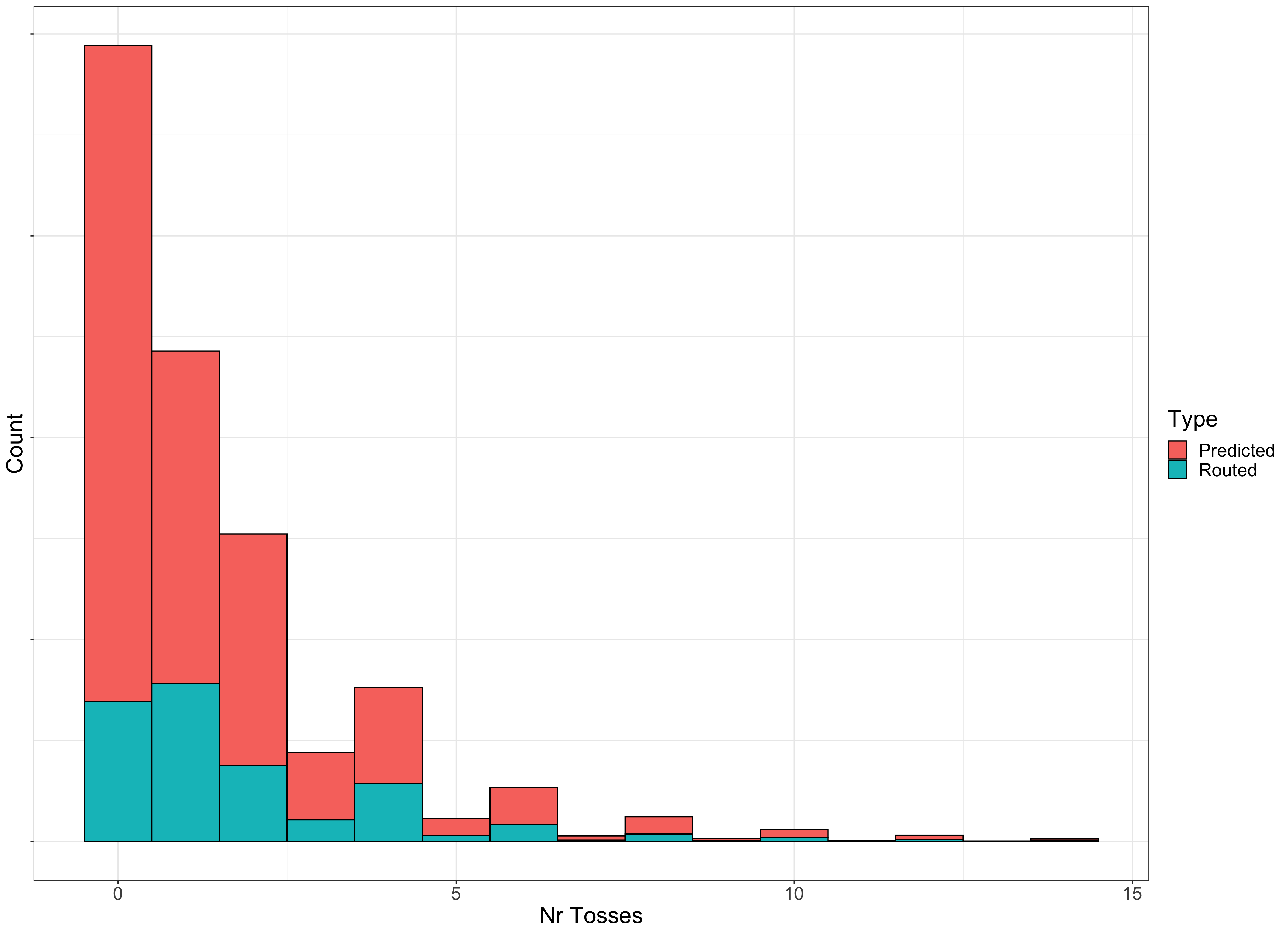}
    \caption{Distribution of bug tossing chain lengths.}
    \label{fig:tossing_histogram}
\end{figure}

Figure~\ref{fig:tossing_histogram} shows the distribution of bug tossing chain lengths for the same time period. The data shows reassignment for both auto-routed only augmented TRs. We find that a majority (81\%) of TRs are either never reassigned or reassigned once or twice. Only rarely, is a TR tossed five or more times (7\%). Through an example, [TRR1] explained that some tossing is sometimes inevitable as the module that should analyze a TR first might not be the same as the closing module: 
\interviewquote{There is a significant chunk of TRs that has to visit more than one module. The textbook example is a simulator issue. No matter what, the [organization] has to rule out that there is no product issue first, then it can go to the simulator organization for fixing. The pre-screening job is to tell which module should start the investigation. So there is a gap between the two because in some cases these are not the same .}{[TRR1-563]}

Bug tossing does not appear to be a major concern at Ericsson. Still, shorter tossing chains was mentioned as a direct effect in RQ4 discussed in Section~\ref{sec:rq4_direct}. We believe that although bug tossing does not occur frequently, a TR turning into a `hot potato' can be costly within the organization. Several interviewees remebered such noteworthy cases and thus brought them up in the interviews.

\noindent\fbox{\begin{minipage}{0.97\textwidth}
Takeaways: 
\begin{enumerate}
    \item TRR's overall prediction accuracy at Ericsson has increased and is currently about 62.5\%.
    \item The accuracy of TRR's auto-routing, i.e., when TR predictions are confident, is on average 75\%.
    \item Bug tossing chains at Ericsson are mostly short, i.e., 81\% of TRs are subject to 0--2 TR reassignments.
\end{enumerate}
\end{minipage}}



\section{RQ3: The Value of TRR at Ericsson} \label{sec:rq3}
This section is organized into descriptive statistics, the BCA, and a qualitative analysis based on the QUPER model. The section reports the metrics Uptime (M1), Fraction automatically routed (M2), Distribution of confidence levels (M3), and Relative TR handling time difference (M6) --- all introduced in Section~\ref{sec:quan_anal}.

\subsection{Descriptive Statistics of TRR's Auto-Routing}
For TRR to provide value at Ericsson, it must have a high availability. We did not find a way to measure this for the time period under study, thus we rely on a careful estimation by the TRR team: the estimated uptime of TRR since the tool was deployed is 97\%. We consider this value sufficiently high to discuss the other value dimensions.

\begin{figure}
    \centering
    \includegraphics[width=1\textwidth]{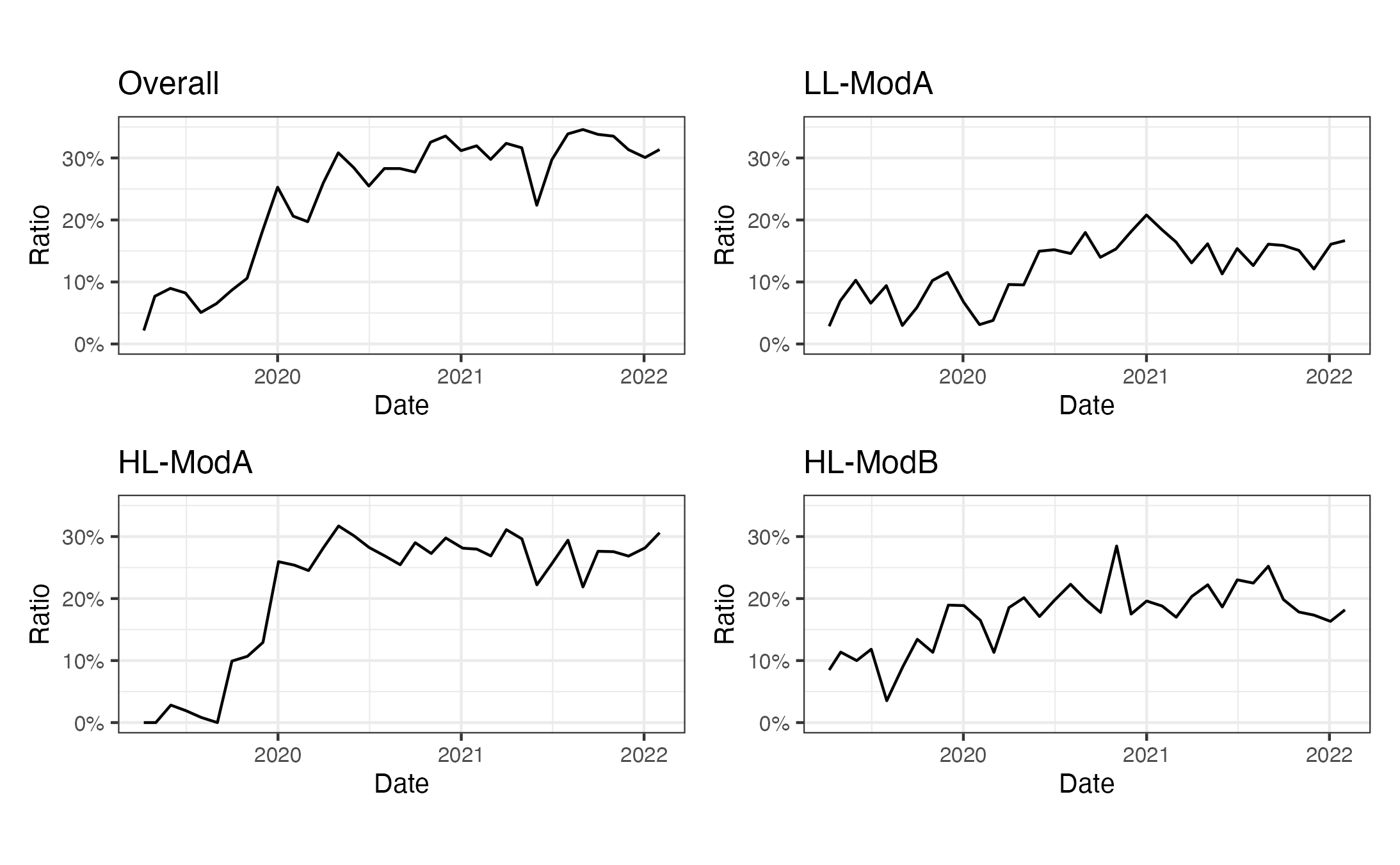}
    \caption{Fraction of auto-routed TR's over time.}
    \label{fig:routing_ratio}
\end{figure}

Figure~\ref{fig:routing_ratio} shows the fraction of auto-routed TRs over time. The more TRs that are automatically assigned, the more effort TRR can save. The top left subplot depicts the overall fraction for all modules. Starting with a low fraction in 2019, TRR has since 2020 consistently auto-routed 20--35\% of all 4G/5G TRs at Ericsson. This translates into fewer TRs to manually process, which saves time at Ericsson:  \interviewquote{I've attended a lot of TR pre-screening meetings, and it's a difference when there are 15 TRs to discuss within 10 minutes. Or three TRs to discuss. So sometimes the difference is just that the meeting ends after two minutes, but sometimes [the pre-screeners] have to go into details and they don't have time /.../ Then they have to push it to later that day.}{[TRR1-161]}

The upper right and bottom subplots in Figure~\ref{fig:routing_ratio} show the fraction of auto-routed TRs on module level. HL-ModA has been hovering between 25--30\% since the start of 2020. For HL-ModB, the fraction has been around 20\% since the summer of 2020 with a slight upward trend. The fraction is lower for LL-ModA, with an average below 20\% for the same time period. As [LL2] stressed the cost of misclassifications for LL-ModA, a lower fraction of auto-routed TRs appears logical.

Based on the fraction of auto-routed TRs by TRR, and the tool's current accuracy level, Ericsson can make a rough estimate of how much manual effort has been saved. As presented in the registered report, Ericsson estimates that a manual TR assignment by the TR Coords. takes three senior engineers 2~min on average~\citep{borg2021adopting}. As we are not permitted to disclose anything about the volume of TRs managed by Ericsson, we can only provide hypothetical numbers: 1~000, 10~000, and 100~000 auto-routed TRs would then translate to time savings for the TR Coords. of 99~h, 999~h, and 9 999~h, respectively. As the TR Coords. are very expensive resources, and since TRR roughly matches their manual accuracy, we find these results very valuable. While the interviewees representing LL-ModA shared that TRR resulted in extra work for them, [HL1]'s time saving estimates for HL-ModA were very positive: 
\interviewquote{Say that [pre-screening] involved two persons two hours every morning. 4 times 5, 20 hours per week in just analyzing TRs. Whilst as that compared to today. I don't know, maybe it's two hours per per week instead. /.../ Every wrongly routed TR costs 5 to 10 hours, and [TRR] has reduced the amount in half. [TRR] makes up to 50 hours per week time savings. That's quite amazing when you put numbers on it.}{[HL1-207]}
        
\begin{figure}
    \centering
    \includegraphics[width=1\textwidth]{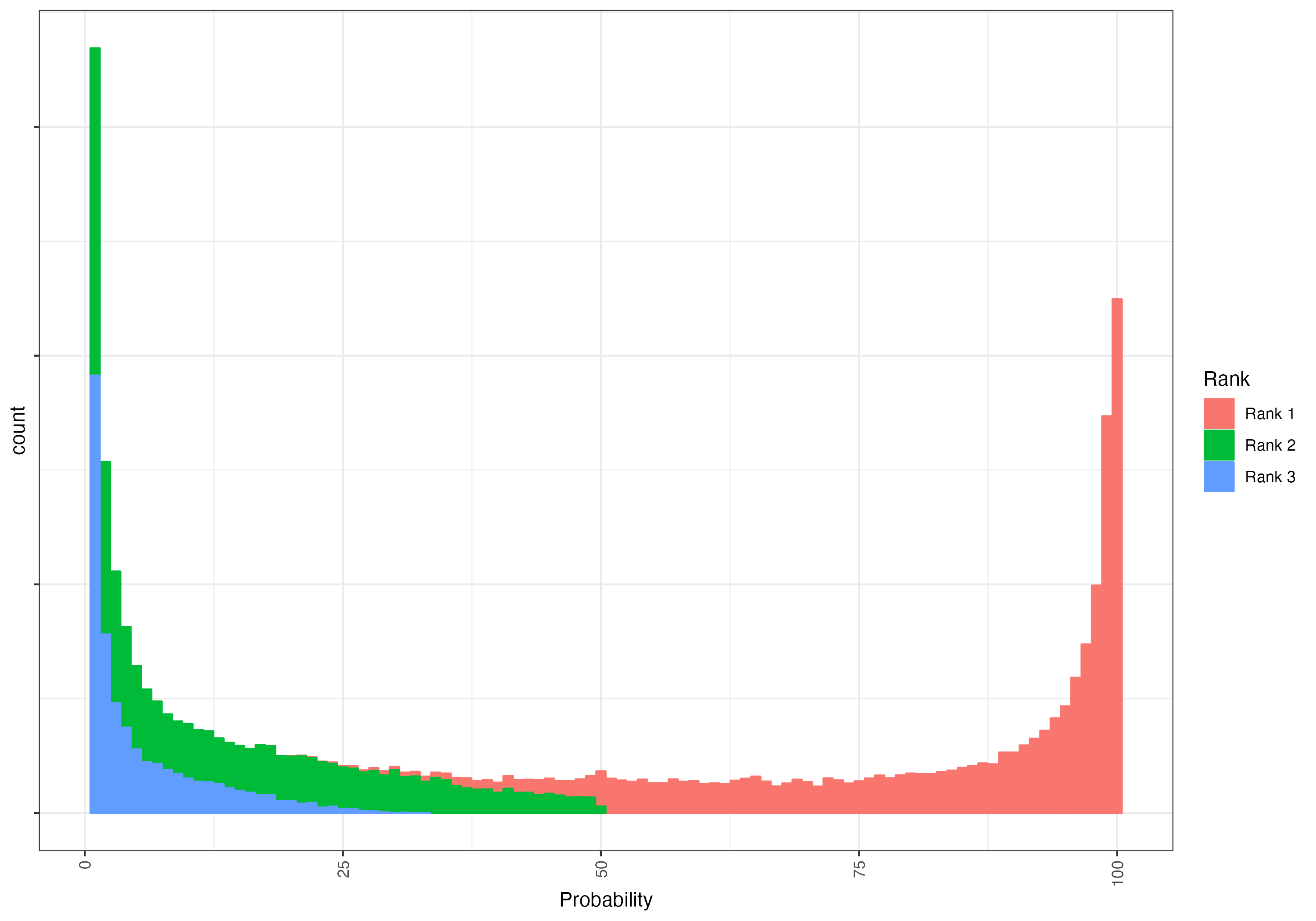}
    \caption{Distribution of confidence levels for the top-3 ranked TRR predictions.}
    \label{fig:confidence_levels}
\end{figure}

Figure~\ref{fig:confidence_levels} depicts the distribution of confidence levels for TRR's top three predictions. Red, green, and blue  bars show confidence levels for the first, second, and third TRR prediction. We find that the confidence of the first TRR predictions are left skewed with 29\% of the TRs predicted with a confidence level above 95\%. For the second and third predictions, which only are generated if needed to reach a cumulative confidence score of 80\% as described in Section~\ref{sec:rq1_phase1}, the scores are instead skewed toward 0. With only 32\% of all top-3 TR predictions in the range between 0.1 and 0.9, we conclude that TRR's logical regression model tends to produce either very confident or non-confident predictions. 

In some cases, especially due to the effect of time zones, TRR saves many hours and sometimes even entire workdays in routing lead-times. This can happen when a TR is detected during the night in the Central European Time and it would not have been routed until the next TC meeting in Sweden. TRR, on the other hand, automatically reacts as soon as a TR is registered in the BTS and (if confident) immediately routes the TR a module. In fortunate cases the module can immediately start working on the TR, rather than waiting for an assignment from the next TC meeting in the manual routing case. This phenomenon has been used by the TRR team when motivating their work on the tool. [TRR1] confirmed this and we share his illustrative success story under direct effects of TRR in Secion~\ref{sec:rq4_direct}.  

We were not able to find data to exactly calculate the number of days saved, since for each routed TR we would need to know where the team that handled the TR was located in the world and how soon they would be able to start working on the particular TR. We instead choose to calculate how often there was a potential for saving one day of work. This was done by calculating the time from when a TR was auto-routed until the next TC meeting. If this time exceeded eight hours we assumed there was a potential to save one day of work (eight hours). This occurred in roughly half of the cases of auto-routed TRs. This is \textbf{not} to say that it would happen that often since many TRs are pulled by the design teams themselves.


\noindent\fbox{\begin{minipage}{0.97\textwidth}
Takeaways: 
\begin{enumerate}
    \item TRR's uptime since deployment is estimated to 97\%.
    \item The fraction of TRs auto-routed by TRR has increased and is currently on average 30\%.
    \item Confidence scores for TRR's predictions are mostly either very high or very low.
\end{enumerate}
\end{minipage}}

\subsection{Bayesian Causal Analysis}
Both efficiency and effectiveness are important when introducing tool support for bug assignment. TRR was initially started when the efficiency in the TR handling was too low, i.e., costly senior human TR Coords. were struggling to keep up with the TR inflow. Pre-screening was introduced to help the TR Coords, but inevitably some TRs were assigned to modules that did not expect them. A substantial amount of work is involved in the process as described by [TRR1-107]: \textit{``You wouldn't believe how many hours are spent on meetings when [engineers] go through the TRs'' }. In this section we present a BCA in which we investigate the quantitative effects of using TRR for automatic assignment of TRs. We begin with a short introduction of the techniques of BCA, causal analysis, and Bayesian estimation techniques.

\subsubsection{Causal Analysis}

In the interest of space we can only give a short introduction to the topic of causal analysis, but for excellent introductions to the topic we recommend \citet{pearl18why} and \citet{pearl2016causal} and Chapters 5-6 in \citet{mcelreath2020statistical}. For an in depth technical discussion \citet{pearl2009causal} is the reference work. 

The first component of the causal analysis is a \emph{Directed Acyclic Graph} (DAG) that is imbued with causal properties. Figure~\ref{fig:causal_dag} visualizes the causal DAG that we use in our BCA. The DAG describes the causal relationships we model in our analysis. The DAG consists of \emph{nodes} and \emph{directed edges}. The nodes represent statistical variables that we study in our analysis as well as their corresponding real world phenomena. 

The directed edges represent the causal relationships between the real world phenomena. A directed edge from one node to another, signifies that the first node affects the second, but not vice versa. Thus, causal relationships (contrary to statistical) \emph{only goes one way}, in the direction of the arrows in the DAG. DAGs do not tell us anything about the concrete functional relationship between variables, only that a relationship exists and in which direction it goes. The functional relationship is later modeled in the estimation procedure which is separated from the modeling of the causal relationships. 

Grey nodes represent unobserved or \emph{latent} variables that we cannot measure in our context. The blue node, called the \emph{exposure} node, represents the variable we are particularly interested in measuring the effect of on the red node, the \emph{outcome} node. With this DAG, we communicate our interest in measuring the effect of machine vs. human routing (the blue \texttt{Routing Entity} node) on the TR handling time (M6), i.e., the time from TR submission to a finished implementation of a resolution (we do not include final handling and the release procedure to customers in this analysis). The handling time of the TR is represented by the red node denoted \texttt{Total Time} in the graph.

\subsubsection{Causal Graph of TR Routing Flow}

\begin{figure}
    \centering
    \includegraphics[width=1\textwidth]{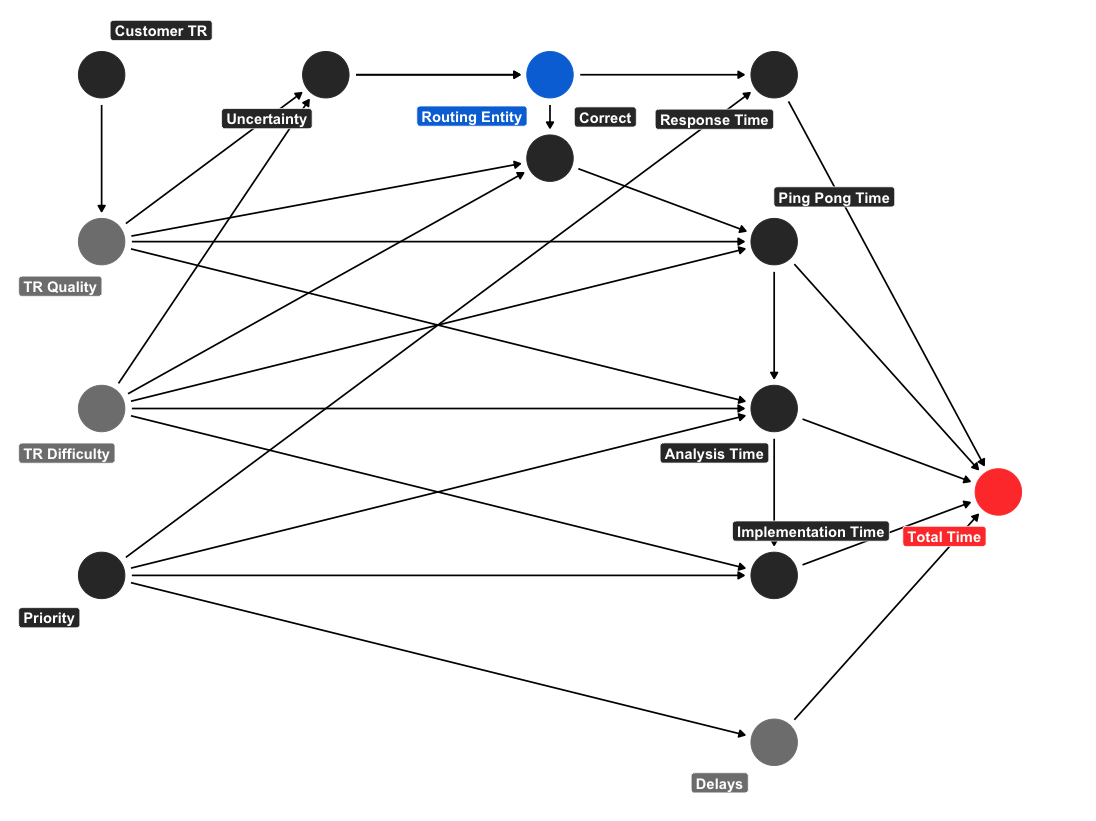}
    \caption{Causal DAG of the TR routing at Ericsson.}
    \label{fig:causal_dag}
\end{figure}

In this subsection, we motivate the causal relationships in Figure~\ref{fig:causal_dag} that reflect the TR handling process at Ericsson. Starting from the top left we conclude that \texttt{Customer TRs} are typically of higher quality than internal TRs in terms of documentation, so it affects the \texttt{TR Quality} node in our causal DAG. We have no quantitative measure of the TR quality in our data, hence this node is marked as a latent or unobserved node. From there on we claim that the quality of the TR (\texttt{TR Quality} --- how well it is written, how detailed it is, how many logs are included, etc.) affects several aspects of the TR flow.

Specifically in our modelling, we claim that it affects the \texttt{Uncertainty} of the machine prediction of where the TR should be routed (more quality data --- less uncertainty and vice versa), as well as if it will be correctly routed (\texttt{Correct}, better written TRs are more likely to be correctly routed) as well as the \texttt{Ping Pong Time} (the same as ``tossing time'', i.e., the time until the correct module starts working with the TR --- with better data, less tossing should be needed) and \texttt{Analysis Time} (better TR data should lead to less analysis time). In the DAG we separate between a ping-pong stage and an analysis stage although you can argue, and it is certainly true, that some part of the analysis happens already during the ping-pong stage. 

Next, the \texttt{TR Difficulty} (how hard it is to implement a solution for the TR, once it is understood what needs to be done), likely affects the uncertainty in the machine prediction as well as if it is correctly predicted. It also likely affects the \texttt{Ping Pong Time} (the more difficult it is to agree on how it should be implemented the longer a TR can be tossed around) and the \texttt{Analysis Time} (since a harder TR to implement, is likely also harder to analyze) as well as the \texttt{Implementation Time} (a TR that is hard to implement will likely also take longer time to implement. Perhaps code needs to be updated in several modules and extensive unit testing in several modules needs to be implemented). We have no direct data on \texttt{TR Difficulty}, hence this node is tagged as a latent variable.

Further, \texttt{Priority} affects the \texttt{Response Time} (a low priority TR can lie around for a longer time) as well as the \texttt{Analysis Time}, since low priority TRs may be assigned to less experienced engineers causing longer time for analysis, and work on it may be swapped out for work on more important TRs. \texttt{Priority} also affects \texttt{Implementation Time} (same argument as for \texttt{Analysis Time}) and eventual \texttt{Delays} (shortage of staff causes low priority TRs to lie in waiting for longer times). \texttt{Delays} are again a latent variable in the model since we have no data on delays.

The \texttt{Uncertainty} node represents the uncertainty of the ML algorithm in its classification of where to route the TR. \texttt{Uncertainty} will affect if a human or the TRR machinery (the \texttt{Routing Entity}) will route the TR by checking the uncertainty against pre-defined confidence thresholds as described in Section~\ref{sec:rq1_phase2}. In our context, the relation between \texttt{Uncertainty} and \texttt{Routing Entity} is in principle a deterministic relation since the confidence thresholds deterministically decides if a TR will be routed by the machine or not, but we will treat it as a random variable anyway since the thresholds have varied over time and that variation is not reflected in the data (but the uncertainty is). The \texttt{Routing Entity} in turn affects to which degree a TR is correctly routed. If a TR is correctly routed or not, it in turn affects the \texttt{Ping Pong Time}. Wrongly routed TRs must per definition be reassigned, while correctly routed TRs do not ping-pong at all. Finally, the \texttt{Routing Entity} affects the \texttt{Response Time} (i.e., how long a TR is waiting to be routed from arriving to the routing inbox), since the machine can immediately route the TR, while if a human routes the TR it must wait until a human is available to route it. 

As a final reflection on the causal model, we conclude that the \texttt{Routing Entity} has no \emph{direct effect}, i.e., an edge directly from \texttt{Routing Entity} to \texttt{Total Time}, which means that the \texttt{Routing Entity} only has an \emph{indirect effect} on \texttt{Total Time}, and this indirect effect is transmitted through \texttt{Response Time} and \texttt{Correct} (through \texttt{Ping Pong Time} and its descendants).

\subsubsection{Developing the Causal Graph} \label{sec:critique}

The causal graph in Figure~\ref{fig:causal_dag} has been iteratively developed in collaboration with TR experts in the organization. Version have been proposed, discussed, and critiqued --- leading to new versions developed based on the feedback. Our target estimand is the causal effect of ML-based routing on the Total Time so we will focus the discussion on this variable. What can mainly get us into trouble when we want to estimate causal effects are confounding variables, i.e., variables that introduce bias into our estimates of the effects of interest. We use causal inference libraries implemented in the R package `dagitty' by~\citet{textor2016robust} which implements graph algorithms~\citep{pearl2009causal} to extract the variables we need to include in our estimation procedure to ensure that we get unbiased estimates of the entities of interest. The algorithms implemented in 'daggity' will give us the so called \emph{adjustment set} (i.e., the variables needed in our estimation procedure to ensure no bias due to confounders) from a given graph, but these algorithms assume that the graph is correct. Considerable time and discussions have been spent on the development of the DAG but of course there is always the possibility that aspects have been missed. In general, each node in the DAG can have incoming arrows with so called \emph{unexplained} effects, but it is standard in the literature to omit these arrows. The effect of such unexplained effects will be visible in the estimation of the effects in terms of variation of the estimates. Next we move on to estimating the causal effects in the DAG.

\subsubsection{Estimating Causal Effects in the TR Routing flow}

With the causal graph in Figure~\ref{fig:causal_dag} and its motivation we can now proceed by estimating the causal effects of auto-routing. `dagitty' allows us to find the adjustment set, i.e., the set of variables to include in the estimation procedure to estimate the variables of interest. Running ‘dagitty’ on the graph with \texttt{Routing Entity} as the exposure node and \texttt{Total Time} as the outcome node, gives us the adjustment set \{\texttt{Uncertainty}\}. This means that to estimate the causal effect of \texttt{Routing Entity} on \texttt{Total Time} we must also \emph{control for} \texttt{Uncertainty}. In our case we do this by adding \texttt{Uncertainty} as a variable in the regression we use to estimate the effect of the routing entity on the total handling time.

\subsubsection{Selecting a Model}\label{sec:select_model}

\begin{figure}[!htb]
    \centering
    \includegraphics[width=.9\textwidth]{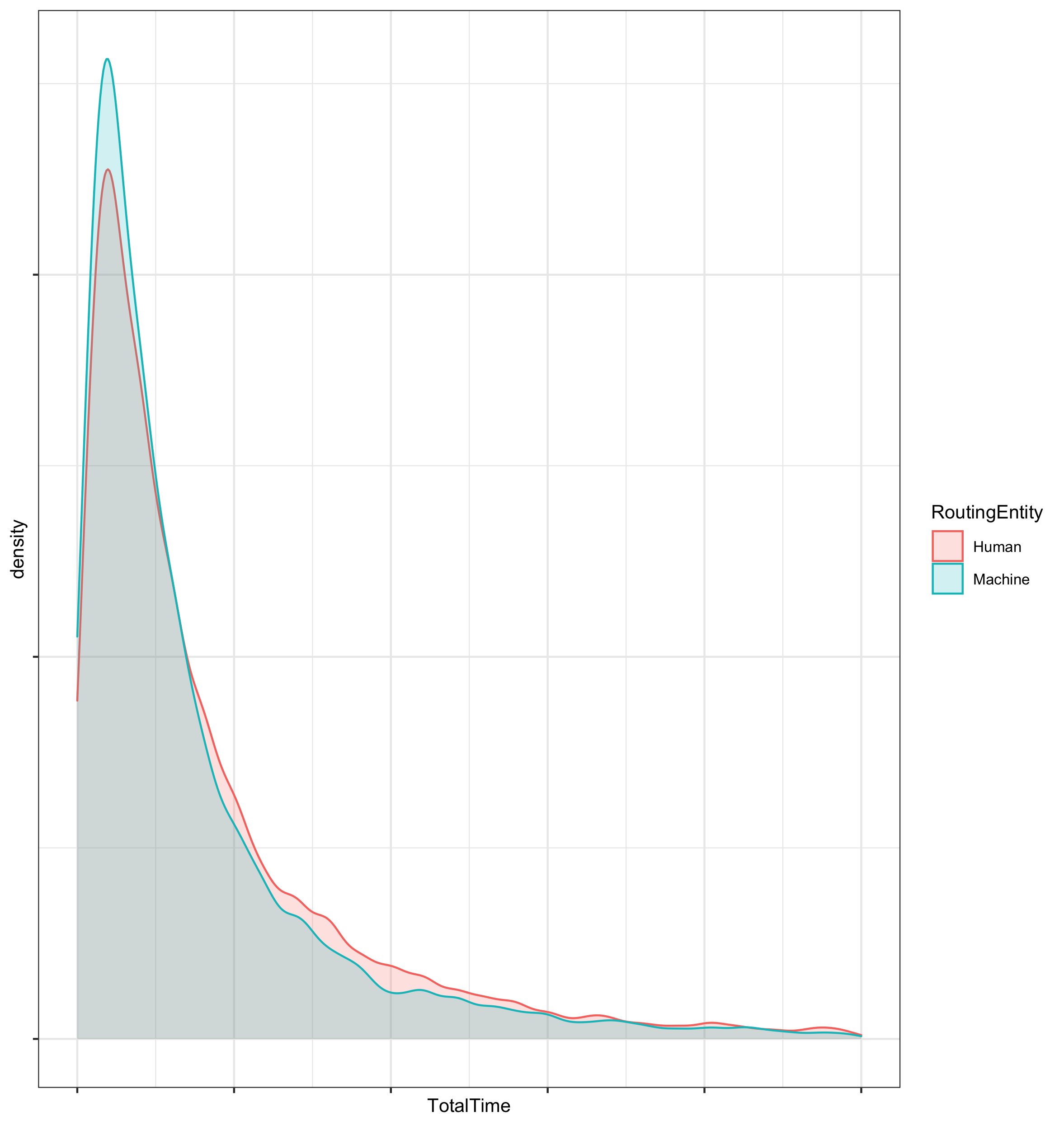}
    \caption{Distribution of total handling time by routing entity.}
    \label{fig:TotalTime_RoutingEntity}
\end{figure}

We base the model selection by reasoning about the \emph{data generating process}. Our data are the handling times of TRs, i.e., the data will be zero (some special case TRs) or positive, and continuous. We could arguably transform the data to integer data by rounding the handling time to full minutes, but we measure the time in minutes with second fractions. In practice there is an upper bound on how long a TR can be handled, but since there are some TRs that have been hanging in the system for a long time and we measure the time in minutes, there will be TRs with very large values of \texttt{Total Time} so we will model the data with an infinite upper bound. The distribution we select need to have support in (0,$\infty$). There are several models that could be used just based on the positive continuous data with support in (0,$\infty$) for instance the exponential distribution, the chi-square distribution, the truncated normal distribution and the gamma distribution, and inverse versions of these. 

To narrow the choice down, we think about more details of the data generating process. As can be seen from the causal DAG in Figure~\ref{fig:causal_dag}, the total time is a sum of a set of waiting times, and the gamma distribution is often used to model waiting times and the sum of several gamma distributions is again a gamma distribution~\citep{chattamvelli2021continuous}. By visual inspection of the distribution of the data in Figure~\ref{fig:TotalTime_RoutingEntity} we can also see that the data seems gamma distributed. It has a mean and mode that is $> 0$, and it is right skewed. With the gamma model we directly model the Total Time with a gamma distribution.

\begin{samepage}
The gamma regression model is formulated as (Model. \ref{eq:gamma_model}):
\begin{align} 
\mathsf{TotalTime} &\sim \mathsf{Gamma}(shape,rate) \label{eq:gamma_model}\\ 
\mu &= \beta_1 + \beta_2 \mathsf{Uncertainty} + \beta_3 \mathsf{RoutingEntity} \nonumber
\\
\beta_1 &\sim \mathcal{N}(mean\_hyper,beta1\_hyper) \nonumber \\
\beta_{2,3} &\sim \mathcal{N}(0,1) \nonumber \\
shape &\sim \phi^{-1} \nonumber \\
rate &= \frac{\phi^{-1}}{\mu} \nonumber \\
\phi^{-1} &\sim \mathsf{Exponential}(1) \nonumber
\end{align}
\end{samepage}

Another possible model that is frequently used when analyzing data with the mentioned properties and time-to-event data is the lognormal model (Model. \ref{eq:lognormal_model}) \citep{1988lognormal}. Specifically in the software engineering field, the lognormal model has been used by \citet{schroeder2009large} to model software repair times. They concluded that software repair times are better modelled by a lognormal rather than an exponential or gamma model. This is in line with our findings (see Section~\ref{sec:pp_checks} and Figure~\ref{fig:pp_checks}). \citet{zhang2013predicting} use the Weibull and lognormal models to fit repair times for three projects, where the Weibull gives the best fit for one project and the lognormal for the other two projects.

\begin{figure} [!htb]
    \centering
    \includegraphics[width=0.75\textwidth]{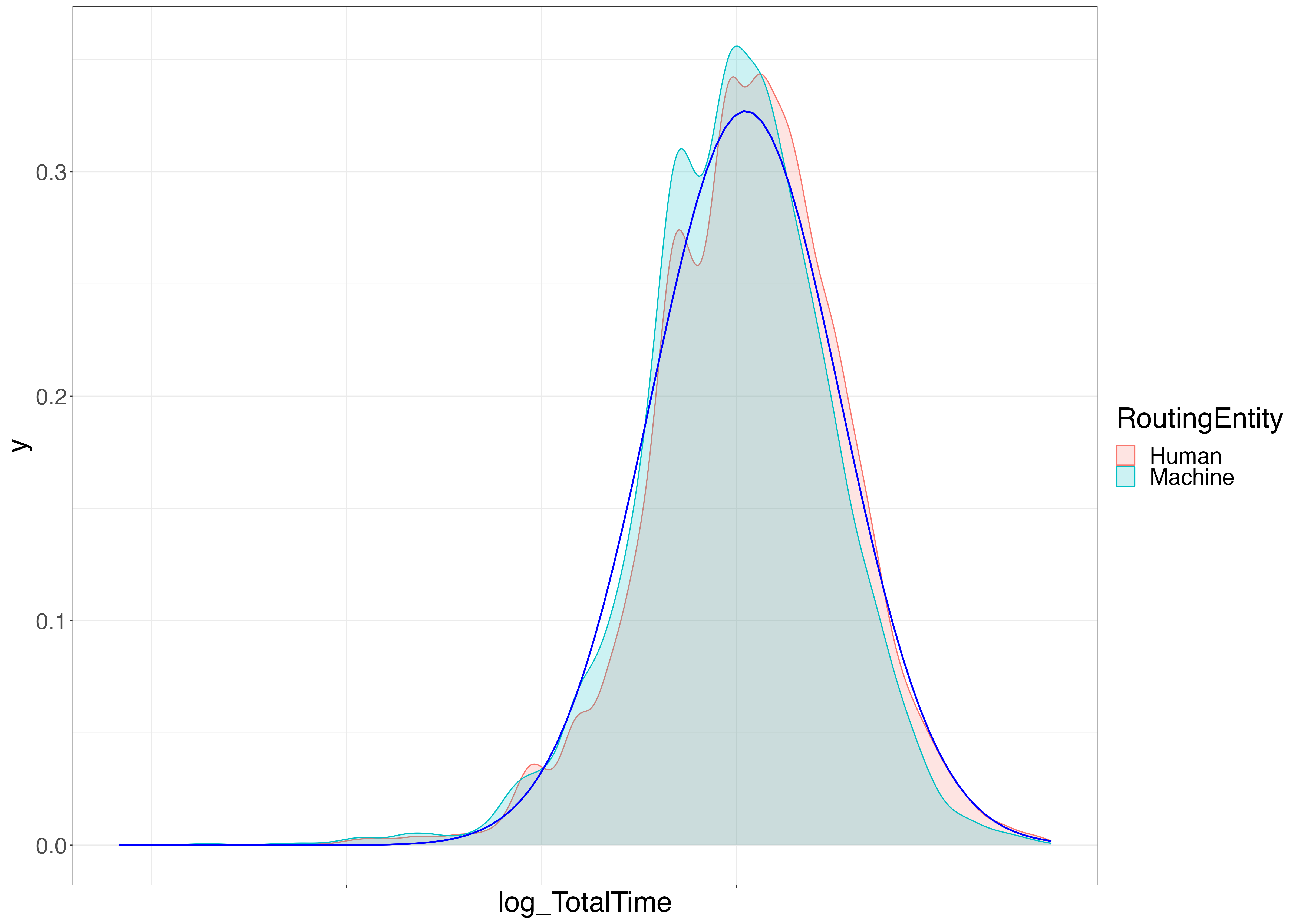}
    \caption{Log of distribution of total handling time by routing entity, with normal distribution overlayed in blue.}
    \label{fig:log_TotalTime_RoutingEntity}
\end{figure}

In the lognormal (Model. \ref{eq:lognormal_model}) model we claim that the log($TotalTime$) is normally distributed. We can check if this seems reasonable by plotting the incurred distribution. Figure~\ref{fig:log_TotalTime_RoutingEntity} shows the distribution of the log of the outcome with a normal distribution overlayed. We see that it is a very close fit. 

\begin{samepage}
\begin{align} 
log(\mathsf{TotalTime}) &\sim  \mathcal{N}(\mu_i,\sigma) \label{eq:lognormal_model}\\ 
\mu_i &= \beta_1 + \beta_2 \mathsf{Uncertainty} + \beta_3 \mathsf{RoutingEntity} \nonumber \\ 
\beta_1 &\sim \mathcal{N}(mean\_hyper,beta1\_hyper) \nonumber \\
\beta_{2,3} &\sim \mathcal{N}(0,1) \nonumber \\
\sigma &\sim \mathsf{Exponential(1)} \nonumber
\end{align}
\end{samepage}

We use Bayesian estimation techniques \citep{mcelreath2020statistical, gelmanbda13} to estimate the effects of interest. The effects will be estimated using standard Bayesian linear regression by implementing Models~\ref{eq:gamma_model} and~\ref{eq:lognormal_model} in the \texttt{Stan} programming language \citep{stan2022}. The \texttt{Stan} code for the models are available online.\footnote{\url{https://github.com/lejon/TRR_Evaluation}}

\subsection{Prior simulation}\label{sec:prior_sim}

The first step in a Bayesian workflow is deciding on priors for parameters in the model. Ideally we can think about what we are modeling and make sure that our selected prior distributions accommodate typical ranges of the data in question. We are working with data on software repair, i.e., the time it takes to resolve bug reports. We can then think about properties and typical ranges for such data. 
\begin{figure}[!htb]
    \centering
    \begin{subfigure}{4cm}
    \centering\includegraphics[width=5cm]{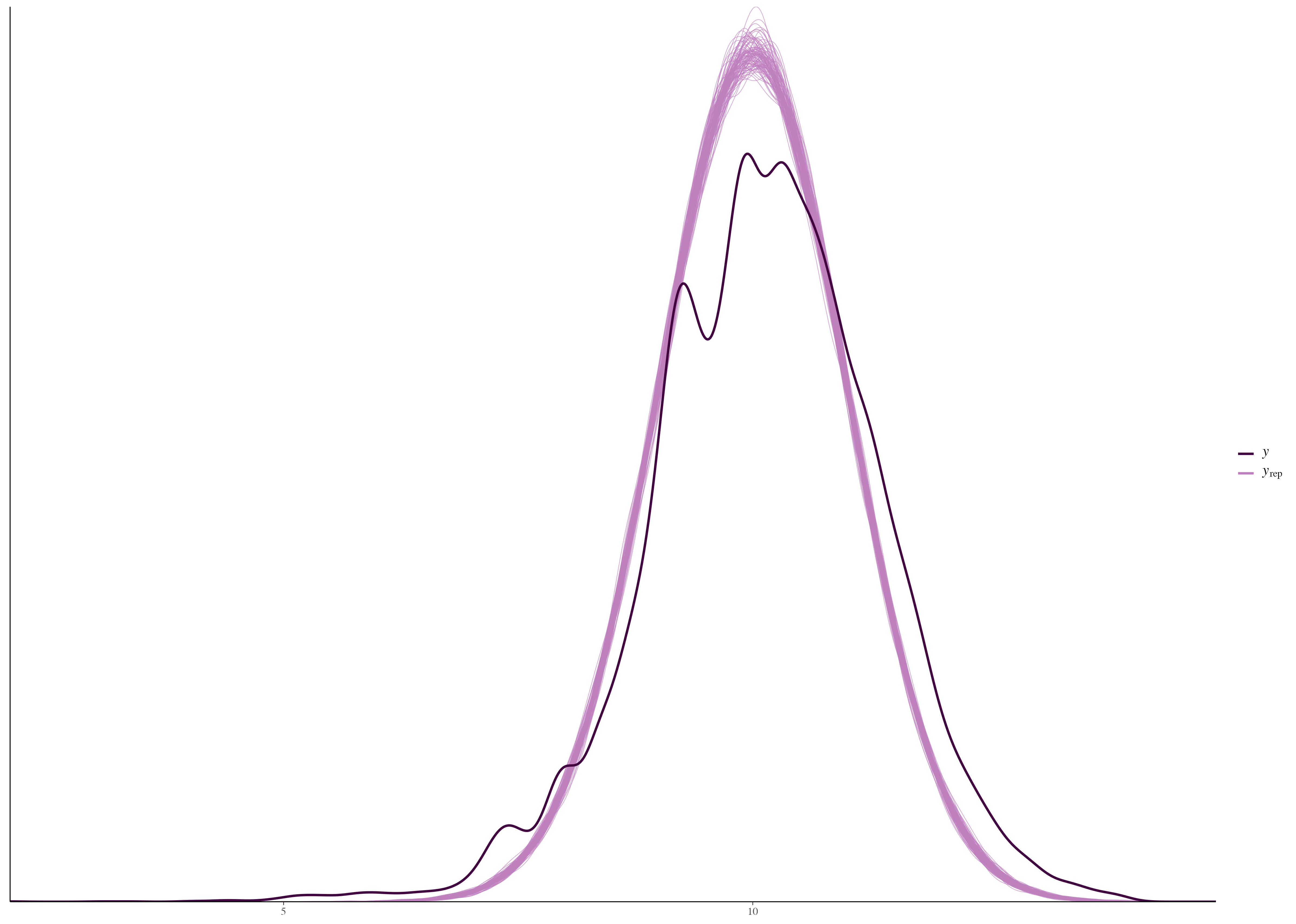}
    \caption{Narrow prior}\label{fig:prior1}
    \end{subfigure}%
\begin{subfigure}{4cm}
    \centering\includegraphics[width=5cm]{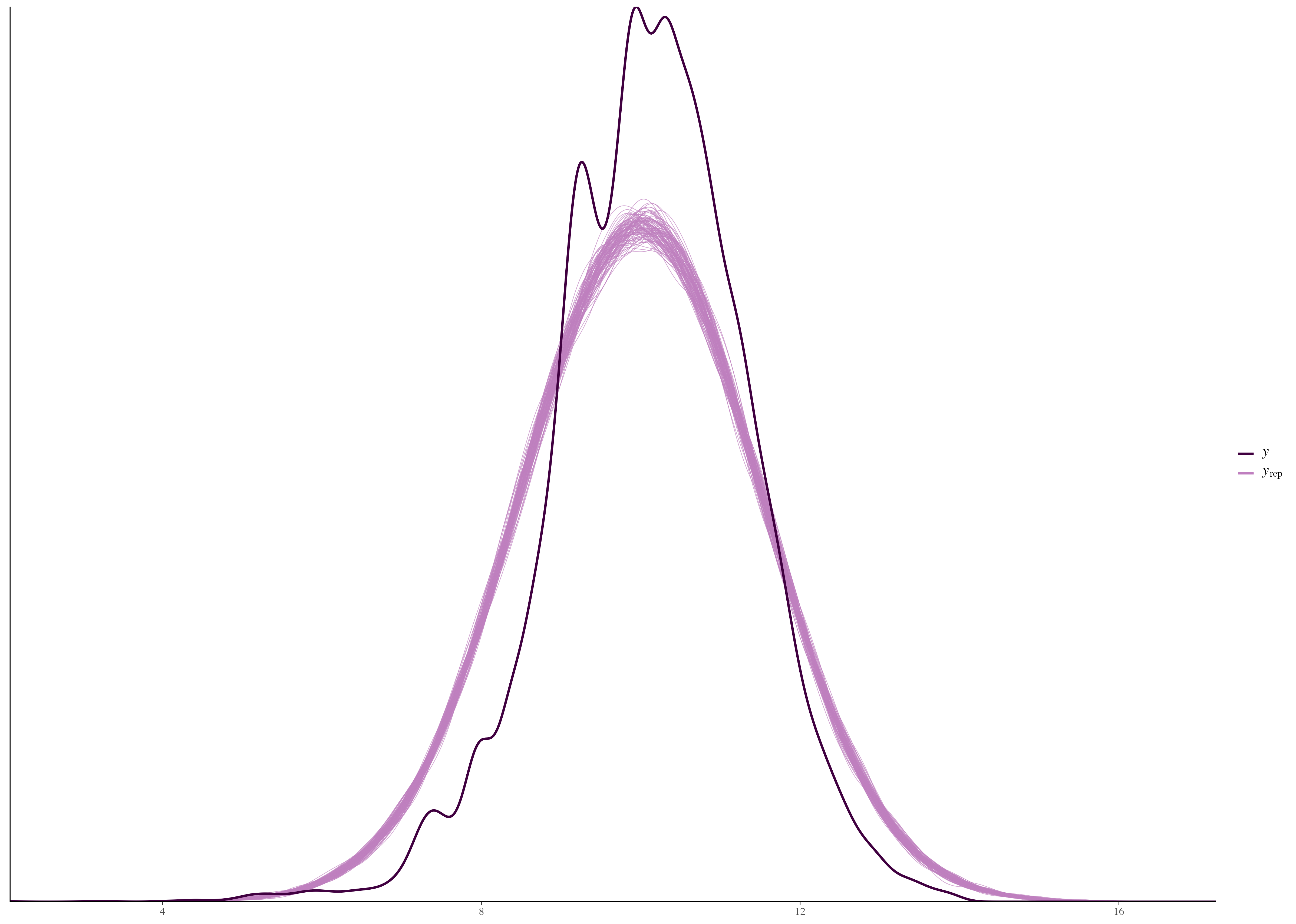}
    \caption{Medium prior}\label{fig:prior1.5}
    \end{subfigure}%
    \begin{subfigure}{4cm}
    \centering\includegraphics[width=5cm]{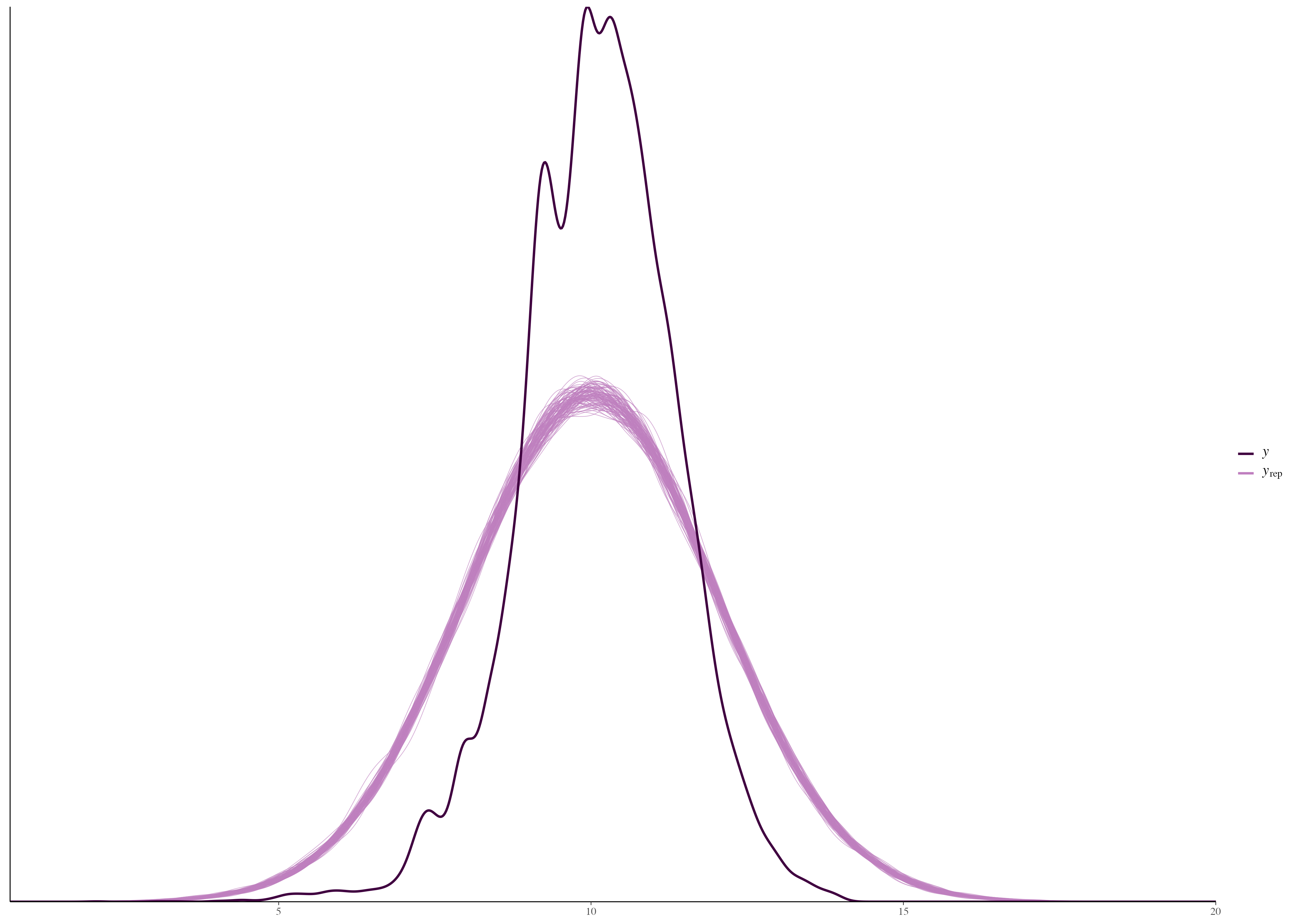}
    \caption{Wide prior}\label{fig:prior2}
    \end{subfigure}\vspace{10pt}
    \caption{Three different prior simulations for the variance of the lognormal}\label{fig:prior_sim}
\end{figure}

We have already discussed some aspects of the data in Section~\ref{sec:select_model}. We further conclude that in our system, there are some special cases where a TR goes directly through the system and thus the handling time is close to zero. Zero is obviously the minimum possible time. We can further think about the maximum time it can take to solve a TR. Sometimes unimportant TRs can be postponed for a long time, even several years, so we would like our model to be able to cater for such data. We can then think about the typical time it takes to handle a TR. This will of course depend on project specifics, but as the goal in the prior specification is not to be too precise, we do not need to exactly match our data. We just need to ensure that the prior places most of the probability on typical values. Figure~\ref{fig:prior_sim} shows simulations using three different values for the prior on the variance for the intercept in the lognormal model ($beta1\_ hyper$). These prior simulations are then used to select the so called \emph{hyperparameters} in the model, i.e., parameters that are selected using simulation and reasoning about the domain and then hard-coded into the model. 

In this case we select the `medium prior' since the `narrow' one did not cover the extreme values in the data, and the `wide' generated too many extreme values. We select the mean of the intercept to center on typical values on the log scale. For the other $beta$-coefficients we select a weakly informative standard normal prior following \citet{Gelman15stan:a}. We give $\sigma$ an exponential(1) prior following \citet{mcelreath2020statistical}.

\subsection{Fit and Posterior Predictive Checks}\label{sec:pp_checks}

\begin{figure}
    \centering
    \begin{subfigure}{6cm}
    \centering\includegraphics[width=5cm]{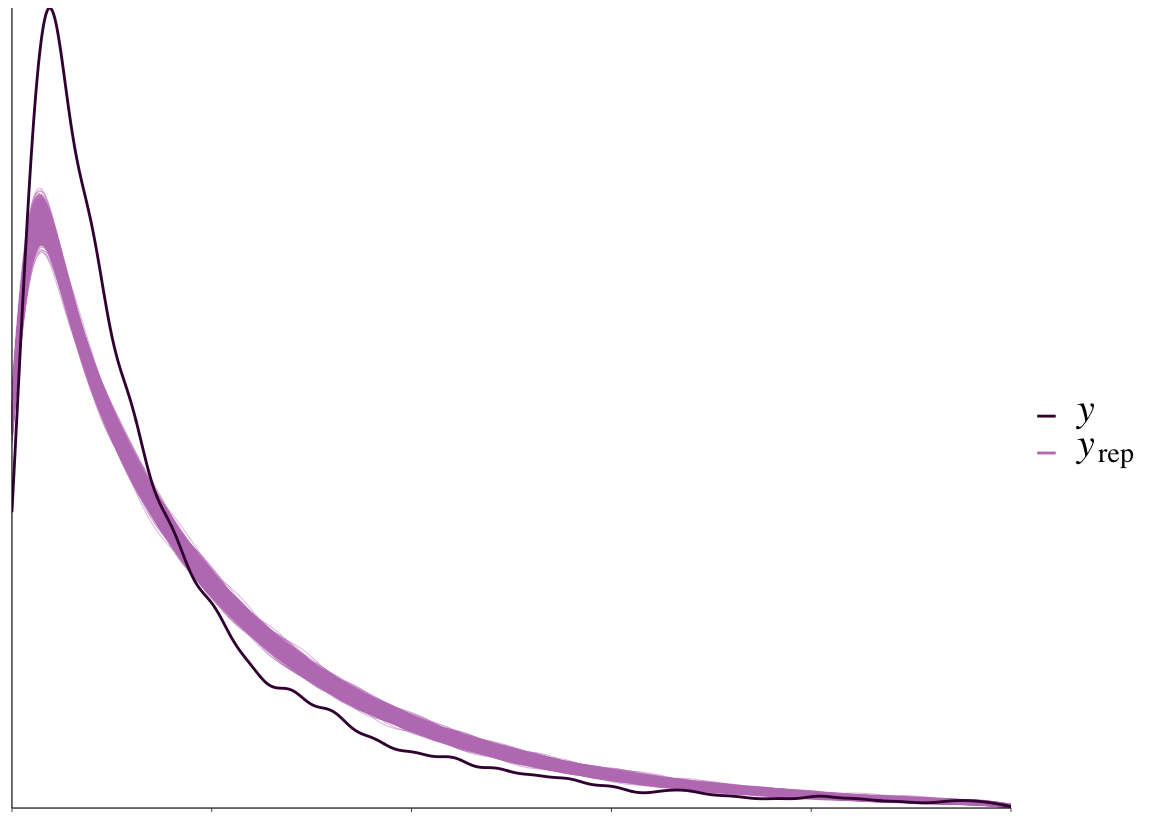}
    \caption{Gamma model}\label{fig:posterior_predictive_gamma}
    \end{subfigure}%
    \begin{subfigure}{6cm}
    \centering\includegraphics[width=5cm]{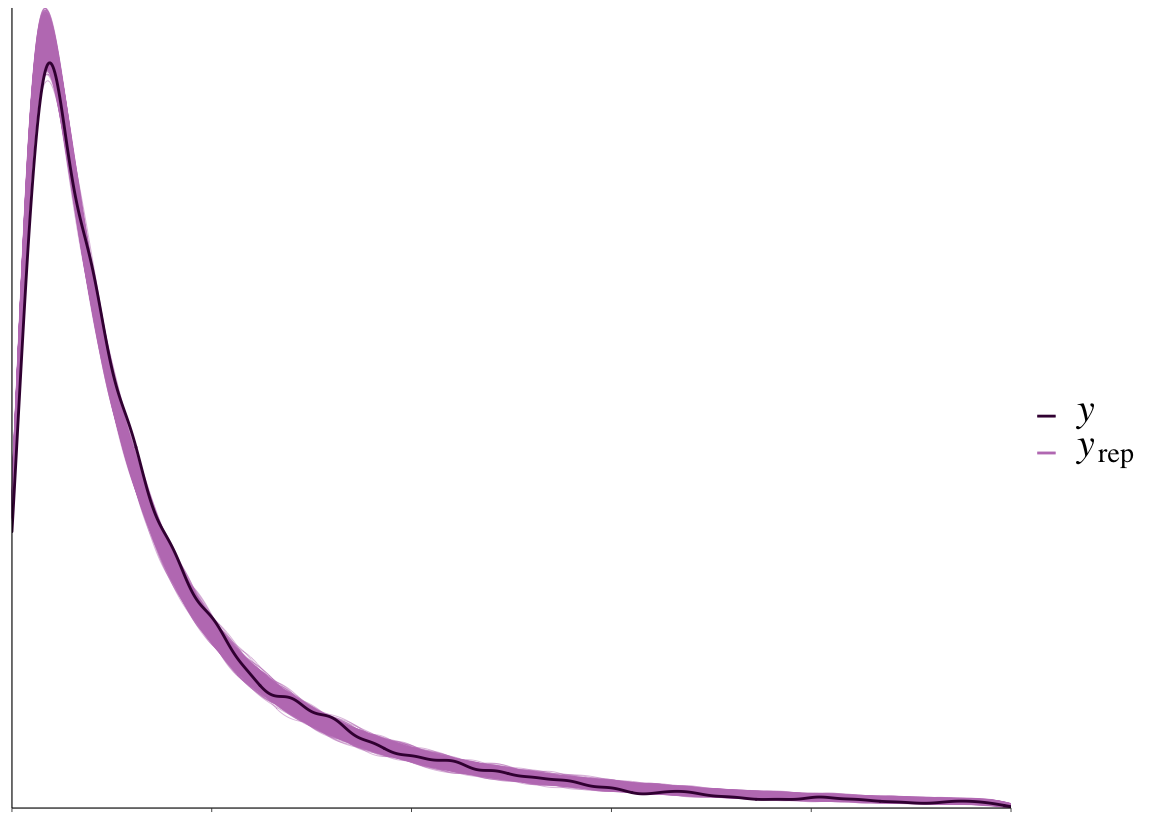}
    \caption{Lognormal model}\label{fig:posterior_predictive_lognormal}
    \end{subfigure}\vspace{10pt}
    \caption{Posterior predictive checks of the gamma and lognormal models.}\label{fig:pp_checks}
\end{figure}

The posterior predictive checks of the Bayesian workflow is intended to show that we have managed to fit the parameters satisfactorily. In Figure~\ref{fig:pp_checks} the dark purple line is the density of the observed data, while the light purple multiple lines (they are very tight in the graph so it can be hard to see the many simulations) are data simulated from the estimated model using the fitted parameters. Figure~\ref{fig:pp_checks} shows the posterior predictive checks of the two models we have analyzed.


We see in Figure~\ref{fig:posterior_predictive_gamma} that the posterior predictive check shows that the fitted gamma model does not properly generate simulated data similar to the data we have, so we instead turn to the lognormal model. Figure~\ref{fig:posterior_predictive_lognormal} shows that the lognormal model gives an excellent fit.



Having verified that we get a good model fit with the lognormal model, we move on to inspecting the fitted parameter of interest. This is the effect of auto-routing TRs with TRR rather than routing with humans. This corresponds to the $\beta_3$ regression parameter in the lognormal model.

\begin{figure}
    \centering
    \begin{subfigure}{6cm}
    \centering\includegraphics[width=6cm]{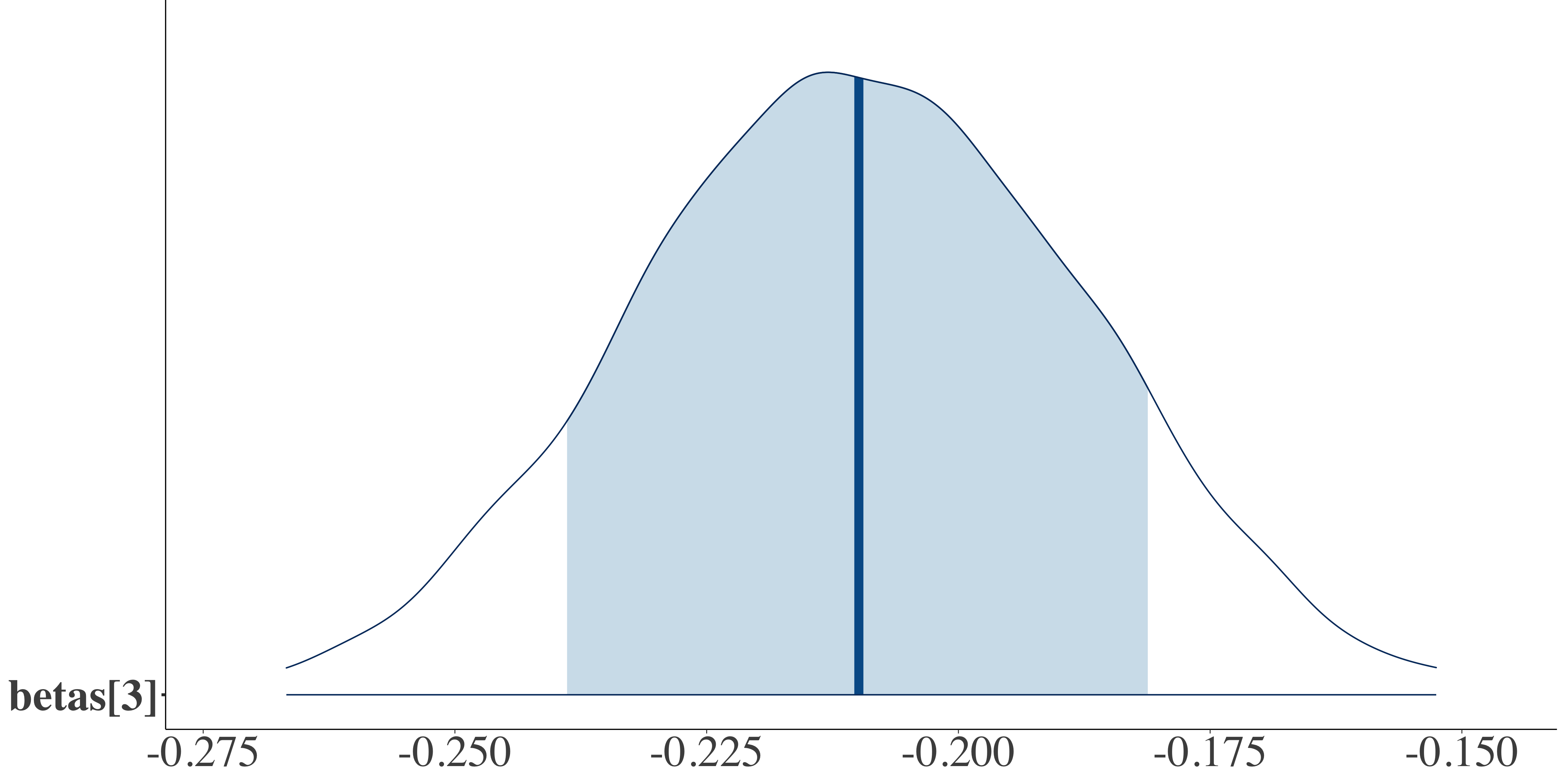}
    \caption{Fitted $\beta_3$ parameter.}\label{fig:fit}
    \end{subfigure}%
    \begin{subfigure}{6cm}
    \centering\includegraphics[width=4cm]{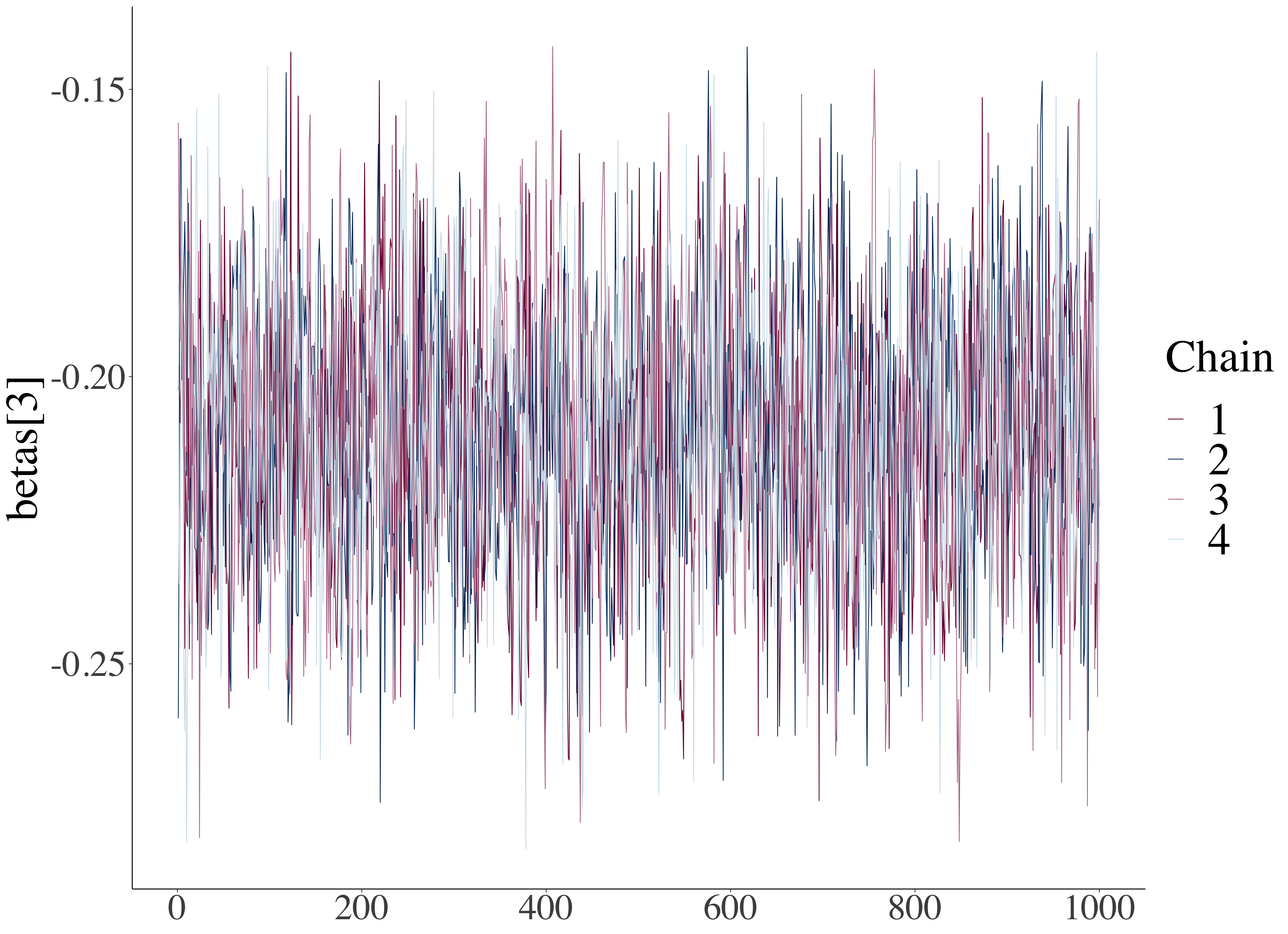}
    \caption{Traceplot of $\beta_3$ parameter}\label{fig:trace}
    \end{subfigure}\vspace{10pt}
    \caption{Fit and Traceplot of $\beta_3$ parameter}\label{fig:fit_trace}
\end{figure}

We inspect the posterior distribution of that parameter in Figure~\ref{fig:fit}. Since this parameter is on the log scale, the interpretation of the parameter is that auto-routed TRs on average require 21\% shorter total handling time. Note that this does not suggest that TRR should auto-route all TRs, since TRs with high uncertainty may be incorrectly routed and take longer time. Instead, the result shows that for the configured confidence thresholds, TRR results in time savings at Ericsson.

In Bayesian estimation the interpretation of uncertainty is very intuitive contrary to confidence intervals in (classical) frequentist estimates. Figure~\ref{fig:fit} is plotted with an 80\% \emph{high density region} (the light blue region) which symbolises the uncertainty of the estimate, and the interpretation is straightforward; there is an 80\% probability that the $\beta_3$ parameter resides in that region.

We conclude the Bayesian analysis by finally checking the \emph{traceplot} of the sampler to ensure that the sampling has gone well as well as sanity checking the sampling procedure using the sampler diagnostics of \emph{divergent transitions}, $\hat{R}$ and the \emph{effective sample size}.


Visual inspection of Figure~\ref{fig:trace} shows that the sampling has moved nicely in the sample space, and we see no trace of severe auto-correlation or abnormal behaviour. Divergent transitions are samples that have gone wrong and should ideally be zero, this is also what happens in both our models. $\hat{R}$ is 1 for all parameters and the \texttt{ess\_bulk} is 2,163 and \texttt{ess\_tail} is 2,773 out of 4,000 samples, both of which are sufficient. \citet{gelmanbda13} recommends $5m$ where $m$ is twice the number of chains. In our case that would lead to a recommended $5 \times 8 = 40$ samples, so samples in the thousands are plenty. 

\subsection{Sensitivity Analysis}\label{sec:sensitivity_analysis}

To examine how sensitive the model is to the specification of the priors is referred to as \emph{sensitivity analysis} in the Bayesian workflow. We run the model several times with different sets of priors to see how much the parameter estimates and fit are affected. We ran the lognormal model with several sets of different priors but the model is \emph{very} robust. We do not present any plots of the different fits since they are indistinguishable from our previous figures. This is true even with quite extreme values of the priors. The reason for this nice behaviour is that the lognormal model is a quite simple model that is usually easy to fit if there are no complex interactions with other hard-to-fit distributions. Another contributing factor is the very large amount of data that we have. It is a general rule in Bayesian inference that with large amounts of data it will overwhelm the priors and they will turn less important. But care must be taken, because for complex models it will require huge amounts of data which sometimes is not realistic.

\noindent\fbox{\begin{minipage}{0.97\textwidth}
Takeaways: 
\begin{enumerate}
    \item We present the first causal DAG of the time from issue report submission to completed implementation.
    \item TRs with an initial auto-routing by TRR are on average resolved 21\% faster compared to TRs routed by humans.
    \item BCA ensures unbiased estimates using domain knowledge encoded in causal DAGs and estimates have an intuitive measure of uncertainty.
\end{enumerate}
\end{minipage}}

\subsection{Perceived Benefits of TRR's Current Quality Level}
\begin{figure}
    \centering
    \includegraphics[width=0.75\textwidth]{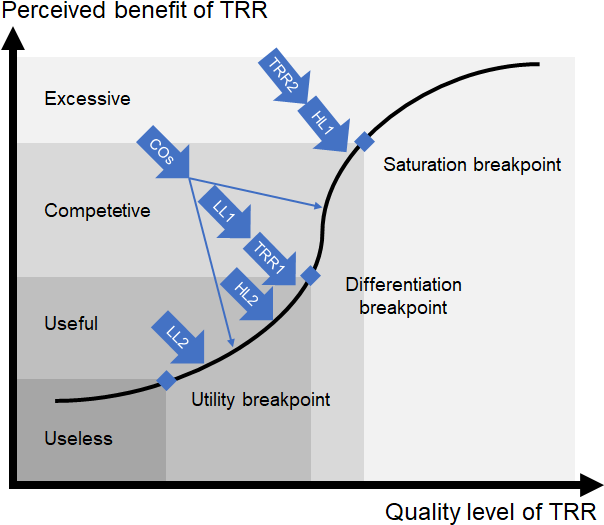}
    \caption{The interviewees' perception on the benefit of TRR given its current quality level.}
    \label{fig:quper}
\end{figure}

Figure~\ref{fig:quper} shows the relation between perceived benefit and quality levels according to the QUPER model~\citep{regnell_supporting_2008}. During the interviews, we presented the QUPER model and how it could be used to discuss the current benefit of the TRR adoption. Blue arrows depict the interviewees' impressions given TRR's current quality level. Using prediction accuracy as a proxy for TRR quality, all interviewees found that TRR had passed the utility breakpoint. However, we discovered contrasting viewpoints representing the entire spectrum from utility to saturation.

[TRR2] and [HL1] experienced the current version of TRR to be close to the saturation breakpoint. [TRR2] motivated his perspective from the tool development side as follows: \interviewquote{I would say [TRR] is near the saturation breakpoint. We see that we reached a level where [TRR] is quite enough and quite good. And we don't get any complaints about this tool. Or yeah, we got some but you know it is always [like that]... And now need to either drop something quite new in [TRR] or just maintain it as it is.}{[TRR2-365]} [TRR2] also clarified that TRR never reached the utility breakpoint when it comes to auto-routing all TRs, but when it comes to processing a fraction of TRs, saturation has almost been reached, i.e., additional quality improvements would be excessive and not get recognized by the users. [HL1] shared the same view: \interviewquote{I think we are at the saturation breakpoint. Everyone is quite happy in my module. Everyone understands that we are as good as we can be at the moment on predicting [TRs]. [TRR] can never be 100\% right.}{[HL1-328]}

[TRR1] and [LL1] believed that TRR is currently close to the differentiation breakpoint. This quality level implies that a slight accuracy improvement could turn the tool into a competitive solution that would be perceived as substantially better than the alternatives at Ericsson. Both interviewees argued that processing of TR attachments would be needed next, e.g., \interviewquote{Right now [TRR] is just under the differentiation breakpoint. A bit south of that and I think if actual log checking was introduced. [TRR] would definitely go deep into the competitive area. /.../ If [TRR] starts doing that tomorrow, it would be 95\% correct.}{[LL1-426]}

[HL2] positioned TRR in the useful range, but not as close to the differentiation breakpoint. In his view, the competitive range would represent making TR Coords. obsolete, and that would require TRR to go beyond auto-routing into also severity prediction etc. \interviewquote{Yeah, I would say we have definitely crossed the utility breakpoint /.../ [TRR] is providing a clear value in that 30\% [of auto-routed TRs] essentially shortcutting the process. And whether it has crossed full differentiation where we essentially conclude that a technical coordinator in Kista should perhaps do something else, that I wouldn't really say, but we are getting towards that at least. If it continues, if we get a couple of more percentages of automatic routing, if we get a little bit more adaptation for filling in a couple of the other TR fields, then we are probably going to be crossing over there.}{[HL2-453]}

In the QUPER discussion with the TR Coords., they preferred to split the perceived usefulness of TRR depending on feedback from the modules. [CO1-359] explained the viewpoint: \interviewquote{I would say for the modules that say `yes, we'll go for it. Let's do auto-routing.' We would be in the competitive [range]. They have identified that [TRR] saves them work, and it's good. And whereas the modules who say `Nah, we don't want auto-routing.' [TRR] is useful, absolutely, but they don't trust it enough.}{[CO1-359]}

[LL2] was the most critical interviewee, placing TRR just beyond the utility breakpoint. He finds that the current downsides of TRR outweights the upsides, but still recognized the value of the tool as a promising future approach for TR handling at Ericsson: \textit{``[TRR] is definitely not useless. Definitely not, because I see it more as a way forward. /.../ now [TRR] is more like maybe minus than plus, but it should not stay like that.''} [LL2-374] Moreover, he shared the view of [HL2] that TRR would have to make some human roles obsolete for the tool to qualify as competitive. However, instead of making the TR Coords. redundant, he talked about removing the need for pre-screeners on the module level --- the role whose work tasks he really defended earlier in the interview (cf. Section~\ref{sec:rq1_org_traits}). 

\noindent\fbox{\begin{minipage}{0.97\textwidth}
Takeaways: 
\begin{enumerate}
    \item Users' perceived benefit of TRR's accuracy spans the wide interval from barely useful to highly competitive. 
    \item Most interviewees expect more of TRR before considering it competitive.
    \item Two interviewees perceived TRR's accuracy almost as good as it can get, i.e., further improvements would be excessive.
\end{enumerate}
\end{minipage}}

\section{RQ4: TRR's Influence on the Way of Working}
\label{sec:rq4}

The adoption of TRR influenced the way of working at Ericsson in several ways, both positively and negatively. In the interviews we discussed both direct and indirect effects~\citep{engstrom2012indirect} of increasing the level of automation in TR routing. As discussed in the model by \citet{parasuraman2000model}, direct effects (i.e., the primary evaluative criteria for automation evaluation) refer to the human performance consequences of specific types and levels of automation. On the other hand, indirect effects (i.e., the secondary evaluative criteria) include automation reliability and the costs of action consequences. While \citet{parasuraman2000model} consider the automation effect on the individual in their model, we study the effect on the collective cognition (i.e., also including the effects on interactions and communication). In our analysis, direct effects refer to the intentional improvements of human performance while indirect effects refer to effects that were not the main intent with the automation. In \citet{parasuraman2000model}'s model, the secondary evaluative criteria aim to identify risks and costs. In our study, we also include positive side effects of the automation in the indirect effects.

\subsection{Direct Effects of TRR Adoption} \label{sec:rq4_direct}
Analogous to the deployment of IssueTag at IsBank reported by~\citet{aktas2020automated}, the deployment of TRR had direct effects on the issue assignment process at Ericsson. Table~\ref{tab:codes_direct} presents the corresponding four codes that emerged during the qualitative analysis of the interviews. The evolution of the codes is presented in the upper part of Figure~\ref{fig:code_evol_rq4} in Appendix~\ref{app:coding}.

\begin{table}[]
\caption{Codes used to describe direct effects of the TRR adoption.}
\begin{tabular}{|p{2.3cm}|p{8cm}|}
\hline
\textbf{Code}            & \textbf{Description} \\ \hline
Reduced leadtime                 & Shorter latency between the initiation and completion of the TR routing process.                 \\ \hline
Reduced manual work    & Less manual work is needed in the TR routing process.               \\ \hline
No defaulting to top-level & Before TRR, TRs were often assigned to the top-level module first. With TRR, more TRs are directly assigned to low-level modules.               \\ \hline
Shorter tossing chains      & Fewer reassignments of TRs, i.e., less ``bug tossing'' or ``TR ping-pong''.                 \\ \hline
\end{tabular}
\label{tab:codes_direct}
\end{table}

Interviewees reported \textit{shorter tossing chains}. This was not only an effect of the auto-routing (skipping the coordination step) but also mentioned by one interviewee (low level) as an effect on all TRs now being augmented with additional information, i.e., a ranked list of likely responsible modules with corresponding confidence levels. Instead of just returning TRs back to the highest level of the telecommunications stack, engineers can now assign TRs directly to another module guided by TRR's ranked list. This was described as a real gain by one of the [HL1-90]: \emph{``for us it was a real gain that [TRR] started to propose the [TR Coords.] to route things elsewhere because otherwise we got everything.''} We introduced the code ``No defaulting to top-level'' to describe this and Figure~\ref{fig:complex_process} illustrates the previously routine of ``top-down triaging.'' Note that this side-stepping of modules is not an entirely positive direct effect, as interviewees from LL-ModA reported an important downside. TRR sometimes auto-routed TRs to LL-ModA for which they could not initiate their analysis before the higher level modules had provided details from investigations on their levels of the stack --- [LL1] and [LL2] reported how this had caused considerable frustration in their teams. [HL1] confirmed the problem, which we will return to in Section~\ref{sec:rq4_ind_comm}

Interviewees from all three units of analysis except the TR Coords. estimated that the \textit{amount of manual work was reduced} thanks to TRR. [HL1] estimated that TRR roughly saves 50h/week for their team (due to less defaulting to the top level, i.e., fewer misrouted TRs to their high level module). On the lower level, [LL1] and [LL2] agreed that correctly routed TRs save manual work, but, they stressed the importance of accuracy --- an increased number of misrouted TRs would instead increase the manual workload for their teams. However, the guidance from TRR, i.e., the augmented information, helps in the manual pre-screening also for them. Currently around 30\% of the TRs are automatically routed and thus removed from the agenda of the pre-screening meetings. Many senior engineers are involved in these meetings, who can now be more focused and efficient. As also reported in Section~\ref{sec:rq1_improve}, [TRR1-382] envisioned a future where all TRs may be automatically routed \emph{``if 100\% of the TRs could be auto-routed, then you could really switch from one day to another, freeing up like hundreds of people from screening and save a lot of senior staff from this tedious job.''}

However, the same interviewees stressed that the most important gain from using TRR was the \textit{reduced lead times} for TR routing. [HL1-420] explained: \emph{``The TRR routing is faster. The initial step of TR routing from that it comes in and goes to a module that will actually start working with it compared to when it's on the top level and no one touches it.''} [HL1] continued: \emph{``Really, I mean that's a gain of on average /.../ like 12 hours for each TR, and that's a gain. It's hard to put money on it. For some TRs it doesn't matter. For some TRs that's gold.''}. The working day estimate is repeated by [TRR1], who explained it with an experienced use case: \interviewquote{We had a specific TR that was opened in North America between 2:00 and 3:00 AM CET. [TRR] auto-routed it at 4:00 AM and it was picked up in Asia within an hour. Most probably, according to the manual way of work, it would have waited until like 10 AM to be routed somewhere. Maybe it would have been pulled [by a pre-screener], but maybe it would have waited. And then on the next working day in Asia, the team would have picked up [the TR]. And that would mean a full working day [in reduced lead-time].}{[TRR1-142]}  With TRR, the dependency on the TC Coords. meeting at 10:30 AM every day is decreased and modules can pull TRs or work on auto-routed TRs when it suits them. [LL1] agreed: \emph{``Then we get the faster flow and everyone is happy.''}

\noindent\fbox{\begin{minipage}{0.97\textwidth}
Takeaways:
\begin{enumerate}
    \item Adopting automated bug assignment resulted in four direct effects: 1) reduced manual work, 2) reduced routing lead-time, 3) shorter tossing chains, and 4) less defaulting to top-level modules.
    \item Side-stepping high-level modules in the TR analysis sometimes cause frustration on lower levels due to missing vital clues from pre-analysis by higher-level modules.
\end{enumerate}

\end{minipage}}

\subsection{Indirect Effects of TRR Adoption} \label{sec:rq4_indirect}

Interviewees made several reflections about indirect effects of the TRR adoption. In general, we could see a pattern of changes in the TR handling process leading to changes in the internal communication about TRs, which in turn impacted the general awareness of the process itself but also of the products, the organization, and the customers' requests. In addition, changes along all these dimensions had a general impact on the work environment. Most of the identified indirect effects were considered positive and the few negative side effects were outweighed by the direct gains of deploying TRR: \emph{``And I mean [TRR] has given great improvements. So I think we just have to live with the other drawbacks and and mitigate those instead.''} [HL1-128]

Tables~\ref{tab:codes_indirect} presents the 20 codes that emerged in the analysis. The evolution of the codes is presented in Figure~\ref{fig:code_evol_rq4} in Appendix~\ref{app:coding}.

\begin{table}[]
\caption{Codes used to describe obstacles and enablers in the TRR adoption related to the high-level (HL) codes process, communication, awareness, environment.}
\begin{tabular}{|l|p{2.5cm}|p{7cm}|}
\hline
\textbf{HL Code}               & \textbf{Code}                      & \textbf{Description}                                                                                                                       \\ \hline
\multirow{4}{*}{Process}       & Misrouted TRs                      & Concerns views that TRR has not changed the issue assignment process much or at all.                                                       \\ \cline{2-3} 
                               & Prohibit questioning               & Using TRR might lead to TRs being assigned to incorrect modules.                                                                           \\ \cline{2-3} 
                               & Improved manual process            & When TRR auto-routes TRs, no human router gets questioned, i.e., TRR protects uncertain humans.                                            \\ \cline{2-3} 
                               & No differences                     & TRR has resulted in an improved manual issue assignment process.                                                                           \\ \hline
\multirow{5}{*}{Communication} & Effective communication            & TRR has resulted in more effective communication within the organization.                                                                  \\ \cline{2-3} 
                               & Increased documentation in TRs     & TRR triggers engineers to explicitly store more information in the TRs.                                                                    \\ \cline{2-3} 
                               & Sidestepping analysis              & TRR risks sidestepping human analysts on upper levels in the tech stack, resulting in less information available for low-level developers. \\ \cline{2-3} 
                               & Increased communication            & TRR output kickstarts conversations, e.g., "With 55\% confidence, Module X shall close this TR" will trigger human comments in the TR.     \\ \cline{2-3} 
                               & Decreased cross-team communication & When TRR correctly auto-routes TRs, different modules might not need to interact as much.                                                  \\ \hline
\multirow{6}{*}{Awareness}     & Increased process awareness        & TRR makes engineers more aware of how issue assignment works in the organization.                                                          \\ \cline{2-3} 
                               & Increased transparency             & TRR leads to a more transparent issue assignment process.                                                                                  \\ \cline{2-3} 
                               & Increased system knowledge         & Thanks to TRR, employees learn more about other modules.                                                                                   \\ \cline{2-3} 
                               & Decreased TR assessment skill      & As auto-routing increases. humans become less skilled at assessing TRs.                                                                    \\ \cline{2-3} 
                               & Decreased transparency             & TRR leads to a less transparent issue assignment process.                                                                                  \\ \cline{2-3} 
                               & Decreased TR overview              & As all TRs no longer pass through the TR Coords., they might lose an overview perspective.                                                 \\ \hline
\multirow{5}{*}{Environment}   & Job satisfaction                   & Engineers experience more satisfaction at work as TRR reduces the number of dull tasks.                                                    \\ \cline{2-3} 
                               & Fewer interruptions                & When simple TRs are auto-routed, individual developers get fewer questions from manual routers.                                            \\ \cline{2-3} 
                               & Dissatisfied employees             & Employees are put under pressure by TRR or get annoyed by its misclassifications.                                                          \\ \cline{2-3} 
                               & Increased workload                 & TRR increases the workload on individuals and teams.                                                                                       \\ \cline{2-3} 
                               & Decreased trust                    & TRs assigned by a machine are perceived as less trustworthy than human counterparts.                                                       \\ \hline
\end{tabular}
\label{tab:codes_indirect}
\end{table}

\subsubsection{Indirect Effects on the Process}
The automation effort itself had a positive effect on simplifying the process at Ericsson and aligning its inputs. By deploying TRR, the process had to be clarified and some complexity of the process could be reduced. By forcing people to communicate with a `dumb' tool, they adapted to better adhere to the process. [TRR2-317] elaborated: \emph{``I don't think [TRR] will be needed for a long time, so I think the main goal is to make the process and the description or the logs so clear that it is obvious without any tool or machine learning how to route [individual TRs].''} At the TR Coords. level, they stressed that the process learning originated in TRs that were not auto-routed, but augmented. \interviewquote{We have spent too much time discussing who should help, and that's why we wanted to have a tool to give us a good recommendation for who is the appropriate person [to contact] in the organization. Who should be going forward with the problems we're seeing, to give us [TR Coords.] more time to prioritize and less time to route.}{[CO1-37]}

On the negative side, interviewees discussed the risk of misrouted TRs. Especially [LL1] and [LL2] stressed the significance of this risk: \interviewquote{A TR is a very costly activity. We have a very limited budget for TRs and also this year we're talking about a reduction of that cost. It is extremely important for us that we are working in an efficient way so we don't get all the [TRs] to one module because there are tools proposing that. We need to have an accuracy which is very high. 80\% is really not good enough yet.}{[LL2-123]} 

However, although interviewees from all units of analysis brought up the risk, none of them considered it a current problem. One reason mentioned by [HL1-178] was the relative improvement: \emph{``[TRR] was smarter than the [TR Coords.] in a rush and that made it really a win situation for us. We could take the wrongly routed TRs because they were fewer than before.''} The TRR team pointed out that initially it was a problem that TRR made too many mistakes, but this was resolved by lowering the automation ambition to only auto-route if a high confidence threshold was met. At that initial state, [LL1] got negative response from the team confirming that it was a problem: \interviewquote{I got feedback immediately from the pre-screening groups. They were on me all the time. `What happened here?' Yeah [I explained] there is a transition period. We are going over to something more automatically I said. /.../ But [TRR] must be something we can trust because we are wasting time later on if [TRs] come wrong.}{[LL2-217]} As discussed in Section~\ref{sec:rq1_org_traits}, some TR Coords. felt relieved that they were no longer blamed for some of the misroutings --- they could instead blame TRR. Questioning of the TR assignments was prohibited by the automation. Before TRR, during a sensitive period, one administrator was dedicated to only communicate routing decisions, via the BTS, and deal with complaints.\\

\noindent\fbox{\begin{minipage}{0.97\textwidth}
Takeaways: 
\begin{enumerate}
    \item Increasing the level of automation can have a positive effect on process simplification and input alignment.
    \item The risk of misrouted TRs must be considered when deciding on the level of automation since the cost of misrouted TRs might be high. 
\end{enumerate}
\end{minipage}}

\subsubsection{Indirect Effects on Communication} \label{sec:rq4_ind_comm}

In most cases, the adoption of TRR increased and improved the communication at Ericsson. The increased communication was mainly noticed by the TRR team: \interviewquote{We have seen that [TRR] changed the way of work. When we introduced [TRR] the interactions on the TR tickets were really few. /.../ It changed because this machine started to act based on the information found in the ticket. And we see now that almost every module is adding comments. /.../ The other modules felt the pressure that `OK, we need to do something otherwise we will get it anyways from the [TR Coords.] if they have no clue.' So they started the mini-analysis and they started to document it in a standardized way. So now every module is [augmenting TRs].}{[TRR2-262]} One reason communication increased was that people were forced to explain why they believe TRR was wrong:
\interviewquote{People started to react to those messages like `oh oh oh, why do you say it was module A? It's obvious that it's not us. Send it to someone else.' /../ I think that this machinery challenges basically everyone on a daily basis, and that's good. That keeps the organization healthy /.../ These fortresses of organizations, who in this case [ANONYMOUS] that has a habit to be defensive and just push back everything.}{[TRR1-68,471,515]}
The TRR team also noted an increased documentation in TRs:
\interviewquote{I think if we would measure the number of notebook entries this pre-screening notebook entries. I think there are more notebook entries since 2017 then it was before. /.../ [Modules] started to document it in the notebook to have it next to the machine prediction. They state that `no, it's not that what that machine said'}{[TRR1-73]}

[TRR2] points out that a reason for more effective communication is that discussions start earlier after the TRR adoption. \interviewquote{I'm sure that there is some increase in the beginning in the discussions, but I think it pays off at the end because if we have the discussion before the actual work is started, it is easier to change and and find a better solution. If [a TR] is in the designer level where the design team tries to solve it, it is really hard to push it back to the higher layers and discuss it there and convince others that it is another problem. It costs more, so the earlier the communication, the better.}{[TRR2-303]} TC Coords. added more aspects of effective communication in the ability to avoid fruitless discussions about routing decisions. \interviewquote{There is another aspect of that routing, because sometimes modules are not so happy with TRs we send to them, so having [TRR] we can avoid the discussion, which is really sometimes endless and without any good value. So yeah, having [TRR] helps really. In that case it's really good.}{[CO3-58]}
\interviewquote{When we have [TRR], we have something that supports us. /.../ because normally we have what we call a ping-pong effect when every [module] says `This is not our problem. Someone else should start the analysis first' and then we need to put our foot down and say `No you should start.'}{[CO1-66]}

However, there was a reflection of possibly decreased communication as well, if TRR gets too good at predicting the final destination. As pointed out by [HL1], important intermediate steps of the analysis would then be omitted:

\interviewquote{The TR should be with us first to have the first investigation, and then later on be sent to another module, but we couldn't send the TR directly to the other module because then they were missing our analysis of what had actually been going on in the signaling. And this is the problem that [TRR] cannot cover. /.../ And this is pissing some of the modules off, to say the least. Because [LL-ModA] down in the very bottom of everything, they get a TR that they cannot do nothing with. They will solve it in the end, but if the modules on the way down to [LL-ModA] don't say what is going on on their layers, pointing to where the problem is... [Low-level modules] can never solve it.}{[HL1-100]}

\noindent\fbox{\begin{minipage}{0.97\textwidth}
Takeaways: 
\begin{enumerate}
    \item Adopting automated bug assignment can increase and improve communication.
    \item Automated bug assignment triggers earlier discussions of TRs, and involves more people, who tries to preempt misrouted TRs.
    \item If auto-routing gets too good at predicting the final destination, there is a risk that intermediate analysis steps will be omitted.
\end{enumerate}
\end{minipage}}

\subsubsection{Indirect Effects on Awareness}

The increased communication triggered by TRR led to increased awareness of some aspects such as the process. [HL2-306] explained: \emph{``[TRR] does kind of pave the way for additional work, since among other things it's a success case. [TRR] sets up basic data, awareness, infrastructure and whatnot, and it is also an organizational lesson in how you actually work and deploy these types of tools.''} [LL1-262] confirmed the benefit of increased awareness: \emph{``[TRR] kind of helps with the in between cases. It helps us to figure out what other modules do''}. 

Increased awareness was achieved by gathering all communication around TRs in one communication channel, instead of numerous informal channels: \interviewquote{I think this kind of [collected] communication was attempted before, but in other channels and the initiative failed /.../ Because of that we had almost nothing documented. So now it is documented in the same place. /.../ Not Teams or Skype chat or whatever, now it is the official [BTS].}{[TRR2-297]} 

However, there were also examples of decreased awareness. Lifting parts of the routing burden from the TR Coords, also results in coordinators losing some of the general overview of the TR inflow: \interviewquote{I think the people that would have been most affected are the [TR Coords.], because before they would actually have seen every heading of every TR, so they would have some kind of knowledge about the complete status. And they would have seen which customers that sent in TRs. But now when [TRs] are auto-routed, I think they miss a lot of TRs that they do not see.}{[HL1-258]}

Furthermore, the skill to manually route incoming TRs might decrease as automation increases. [LL2] expressed this risk and [TRR1] elaborated further: \interviewquote{It's really a complex product family, and let's say if from tomorrow on [all TRs] would be automated. [TR Coords.] would lose this really special competence to tell from the symptoms described to judge if it's this module or that module.}{[TRR1-455]} 


\noindent\fbox{\begin{minipage}{0.97\textwidth}
Takeaways: 
\begin{enumerate}
    \item Increased communication contributes to increased process awareness.
    \item Increased awareness was partly achieved by gathering all communication around TRs in one channel.
    \item Relieving humans from parts of the routing burden risks decreasing their overview of the product status on the market. 
\end{enumerate}
\end{minipage}}

\subsubsection{Indirect Effects on the Work Environment} \label{sec:rq4_ind_env}

There were mixed feelings regarding TRR's indirect effects on the work environment. The general reflection was that TRR had a positive effect on job satisfaction. Individual developers experience fewer interruptions due to questions about TRs, e.g., \emph{``I would suspect that the individual developers will be harassed slightly less by our poor middle project managers who do the routing otherwise.''} [HL2-302]. However, although the general manual workload decreased, some modules experienced an increase in manual TR work. Similarly, although the general trust in the decisions made by TRR is high, in parts of the organization the trust in TR assignments decreased as a result of the increased level of automation.

One example of positive impact on job satisfaction was improved relations between TC Coords. and the modules. As [CO1-254] expressed it: \emph{``Now \emph{[when decisions partly are attributed to TRR]} they don't hate us as much''}  Another example is the positive effect of generally reduced workload stressed by [HL2-228]: \emph{``Fundamentally it's also a matter of less work for literally everyone involved in the chain.''}, which is especially valuable for activities that traditionally are not in focus of process improvement. 

\interviewquote{In the TR world, there is often not that much done around optimizations and processes. There are quite a few organizations even today that literally have zero tool support for their work. So they work roughly the same way today as they did in the 90s. From any kind of business method, that's like halfway disastrous. Productivity is extremely low, morale is low, and so on. So anything we could do that would actually lessen the workload [of TR handling] or even just allow people to work slightly higher up in the food chain, that would be massively beneficial. Not just for like productivity and so on, but for working environments.}{[HL2-228]}

Both interviewees representing the low-level module believed their workload had increased as an effect of the automation. [LL1-243] had seen a higher expectation on analysis on their side also for misrouted TRs: \emph{``We heard the argument `OK, but this is your TR, TRR said that it's your TR, so you have to look into it more closely.'''} [LL2] elaborated on the increased cost for this extra analysis: 

\interviewquote{At the moment if [TRR] is making decisions, and the TR is routed to the wrong destination, it's going to be more costs. Because you can say that this is not in my module, but you can definitely not say where the root cause of the problem is, which module it is. And then you invoke a lot of front desks from different modules in the same discussion thread. To understand the root cause, many people get involved you know. It's good by the end. We come to a final conclusion, but it's more time consuming.}{[LL2-123]}

\citet{hoff2015trust} reviewed factors influencing trust in automation. They identified three layers of variability in human–automation trust (dispositional trust, situational trust, and learned trust). In our case, the variability in situational trust may be attributed to external variability, such as differences in placement in the telecommunications technology stack, task difficulty, workload, and perceived risks and benefits.\\

\noindent\fbox{\begin{minipage}{0.97\textwidth}
Takeaways: 
\begin{enumerate}
    \item Automated bug assignment can have a positive effect on the job satisfaction.
    \item Different roles are differently affected by increased automation. At Ericsson, the TR Coords. were impacted the most.
    \item The low-level module under study reported that TRR has caused extra work and they do not trust the auto-routing as much as the others.
\end{enumerate}
\end{minipage}}

\section{Quality Assurance and Validity} \label{sec:quality}
The value of design science research may be assessed from three different perspectives~\cite{runeson2020design}, i.e., its \emph{relevance}, its \emph{novelty} and its \emph{rigor}. The design knowledge gained from this research is \emph{relevant} for practitioners facing the challenge of manually assigning bugs to teams, and for researchers studying industrial adoption of ML approaches for automated bug assignment. Relevance is a subjective value and to support its assessment we identified and reported the context factors that affected the applicability and observed effects of the proposed intervention. Furthermore, the design knowledge is \emph{novel} in terms of increased maturity of the general technological rule and in proposing refined rules with respect to the scope of validity and the effects of adoption. \emph{Rigor} was achieved by following this pre-registered case study protocol~\citep{borg2021adopting} and by transparently reporting all steps of interpretation in the qualitative analysis. Furthermore, rigor may be assessed in terms of construct validity, internal validity, and reliability. As we design a single case study, pure statistical generalisation will not be possible. External validity is instead covered by the discussion on relevance above.

\textbf{Construct Validity.} Since we conducted an exploratory study, not all constructs were known upfront. Our high-level constructs such as ``value'' and ``ways of working'' were be refined in the qualitative analysis. The metrics proposed in Figure~\ref{fig:rqs} represent our initial assumptions of how to measure these aspects --- apart from M6 that was presented as ``TR assignment time'' in the registered report~\citep{borg2021adopting}. We had to adjust this to a relative metric for confidentiality reasons. To further increase the final construct validity, we asked the study participants to assess our interpretations. 

\textbf{Internal Validity.} As discussed in Section~\ref{sec:quan_anal}, we could not perform a controlled randomized trial to prove causal relationships within Ericsson --- we cannot disable HighAuto TRR for a random subset of teams. As we have to deal with the complexity of in vivo research, we conducted a BCA instead~\citep{pearl2009causal,hernan2020causal}. To increase the validity of the propositions, confounding factors were identified and all our assumptions can be scrutinized as our causal DAG is fully transparent (see Section~\ref{sec:critique}). Transparency is an important advantage of Bayesian analysis. 

We highlight two potential threats to the internal validity related to our interviews. First, the issue assignment process and TRR (see Section~\ref{sec:rq1_process}) and TRR (see Section~\ref{sec:rq1}) co-evolved. As the process and the tool are intertwined, it is possible that some interviewees did not clearly distinguish which of the two caused the indirect effects reported in Section~\ref{sec:rq4}. Second, interviewees with substantial experience have seen many tool adoptions at Ericsson. Some provided answers might be general and not only relate to TRR. We addressed both these threats by regularly reminding the interviewees to focus on TRR during the interviews.

\textbf{Reliability.} This aspect of rigor concerns to what extent the analysis depends on the specific researchers. We mitigated threats to reliability through researcher and method \textit{triangulation}~\cite{runeson2012case}. Additional measures included documentation of the evolving coding scheme (see Appendix~\ref{app:coding}), \textit{prolonged involvement}, i.e., the long-term relations that evolved during the study supported reliable interpretations, and \textit{member checking}, i.e., participants of the study validated both data collection. All transcripts were sent to the interviewees shortly after the interview sessions. Finally, our analysis, containing the main takeaways, were shared with all interviewees before we concluded the paper.



\section{Conclusions and Future Work} \label{sec:conc}
Ericsson's TRR adoption was successful and automated bug assignment is now an incorporated part of 4G/5G product development. In the next paragraphs, we answer the four research questions.

\textbf{RQ1: Evolution from prototype to tool.} Originating in academic research, Ericsson developed several proofs-of-concept for ML-based bug assignment between 2011 and 2017. Evolutionary prototyping followed and TRR recommended closing modules for all incoming bug reports during a year until an evaluation put the development on hold due to insufficient accuracy. In 2019, TRR was adjusted to only auto-route TRs to a module when highly confident and otherwise just augment recommendations. This version of TRR, embedding a substantially simpler ML model compared to the early research, has been in continuous operation at Ericsson since April 2019.


\textbf{RQ2: TRR's accuracy.}
As TRR operates at two levels of automation depending on prediction confidence, we report two figures. TRR's overall prediction accuracy is about 62.5\%. For automatically assigned TRs, the average accuracy is 75\%. Accuracy differences between modules are minor. Since April 2019, bug tossing chains are mostly short, i.e., 81\% of TRs are subject to 0--2 TR reassignments.

\textbf{RQ3: The value of TRR.}
On average, TRR auto-routes 30\% of all incoming bug reports. Compared to accuracy, the difference between modules is larger. The Bayesian causal analysis shows that TRs with an initial auto-routing by TRR are on average handled 21\% faster compared to TRs routed by humans. 

Moreover, several interviewees report that adopting TRR has saved highly seasoned engineers many hours of work at Ericsson. On the other hand, the value of TRR’s current accuracy level spans a wide interval of personal opinions from barely useful to highly competitive. Considering the abstraction levels of the telecommunications stack, high-level modules are more positive while low-level modules experienced some drawbacks.

\textbf{RQ4: TRR's influence on the way of working.}
Adopting automated bug assignment resulted in 1) reduced manual work, 2) reduced routing lead-time, 3) shorter bug tossing chains, and 4) less defaulting to top-level modules. Positive indirect effects of adopting TRR include 1) process improvements, 2) process awareness, 3) increased communication, and 4) higher job satisfaction. Negative effects include increased side-stepping of high-level modules in the TR analysis (when TRR immediately routes a TR to a low-level module) which can cause frustration on lower levels due to missing vital clues. Furthermore, when humans are less-involved in the bug assignment, there is a risk that a complete picture of the product status on the market is lost.

We conclude that TRR has saved time at Ericsson, but increasing the level of automation in the bus assignment was more intricate compared to similar endeavors reported from IsBank~\citep{aktas2020automated} and LG Electronics~\citep{oliveira2021issue}. We primarily attribute the difference to the very large size of the organization and the complex 4G/5G products. Key facilitators in the successful adoption included a gradual introduction, product champions, careful stakeholder analysis, and a mature internal tools team.

We propose three main directions to improve TRR within Ericsson. First, the accuracy could be improved by considering additional features in the ML model, most importantly extracted from attached logs. Moreover, for large modules with sufficient training data available, TRR could provide also sub-module predictions. Second, Ericsson could improve TRR's ``recommendation delivery'' by tailoring the amount of information depending on the user. A possible design could include information hiding, i.e., let interested users easily click a ``tell me more'' button to access the rationales behind predictions for better explainability.

Third, TRR could be adjusted to instead predict the module that should start a TR analysis. The current training data leads to predictions of the modules most likely to close a TR. We know that this is not necessarily the same as the module most suitable to initiate analysis into an issue. Potentially, this discrepancy leads to unsatisfied users, especially in the low-level modules of the telecommunications stack. Future work could either 1) annotate training data according to which module that historically initiated the analysis, or 2) introduce a two-step prediction process. The two-step process could involve a first prediction of whether a TR needs an investigation of multiple modules. If yes, assign it to the team most likely to start investigating the TR. If no, send it to the module most likely to close the TR.

\section*{Acknowledgment}
Our thanks go to all interviewees, managers, and product owners at Ericsson. We also want to thank the anonymous reviewers of the registered reports track at ESEM 2021 --- your comments improved this study. This work was partially conducted within the AIQ Meta-Testbed initiative funded by Kompetensfonden at Campus Helsingborg, Lund University, Sweden.

\section*{Conflict of Interests}
Ericsson representatives reviewed the manuscript to ensure that no sensitive information about products or processes are disclosed. Moreover, the second, fourth, and fifth authors are employed by Ericsson. The second author published the first conceptual paper about automated bug assignment at Ericsson. The first author joined shortly after to support development and empirical evaluations of the tool that is now TRR. Beyond this, the authors have no competing interests to declare that are relevant to the content of this article.

\section*{Data Availability Statement}
The raw interview data collected during the study are not publicly available to guarantee the interviewees’ anonymity (see Appendix~\ref{app:int_guide}). Data from Ericsson’s bug tracking system are not publicly available for confidentiality reasons. Reasonable requests by reviewers will be handled by the corresponding author on request.

\bibliography{trr}

\begin{thebibliography}{62}
\providecommand{\natexlab}[1]{#1}
\providecommand{\url}[1]{{#1}}
\providecommand{\urlprefix}{URL }
\providecommand{\doi}[1]{\url{https://doi.org/#1}}
\providecommand{\eprint}[2][]{\url{#2}}
 \bibcommenthead

\bibitem[{Aktas and Yilmaz(2020{\natexlab{a}})}]{aktas2020automated}
Aktas EU, Yilmaz C (2020{\natexlab{a}}) Automated issue assignment: results and
  insights from an industrial case. Empirical Software Engineering
  25(5):3544--3589

\bibitem[{Aktas and Yilmaz(2020{\natexlab{b}})}]{aktas2020exploratory}
Aktas EU, Yilmaz C (2020{\natexlab{b}}) An exploratory study on improving
  automated issue triage with attached screenshots. In: Proc. of the 42nd
  International Conference on Software Engineering: Companion Proceedings, pp
  292--293

\bibitem[{Anvik and Murphy(2011)}]{anvik2011reducing}
Anvik J, Murphy G (2011) Reducing the effort of bug report triage: Recommenders
  for development-oriented decisions. Transactions on Software Engineering and
  Methodology 20(3):1--35

\bibitem[{Baltes and Ralph(2020)}]{baltes2020sampling}
Baltes S, Ralph P (2020) Sampling in software engineering research: A critical
  review and guidelines. arXiv preprint arXiv:200207764

\bibitem[{Baysal et~al(2009)Baysal, Godfrey, and Cohen}]{baysal_bug_2009}
Baysal O, Godfrey M, Cohen R (2009) A {bug} {you} {like}: {A} {framework} for
  {automated} {assignment} of {bugs}. In: Proc. of the 17th {International}
  {Conference} on {Program} {Comprehension}, pp 297--298

\bibitem[{Bettenburg et~al(2008)Bettenburg, Premraj, Zimmermann, and
  Kim}]{bettenburg2008duplicate}
Bettenburg N, Premraj R, Zimmermann T, et~al (2008) Duplicate bug reports
  considered harmful… really? In: Proc. of the International Conference on
  Software Maintenance, pp 337--345

\bibitem[{Bhattacharya et~al(2012)Bhattacharya, Neamtiu, and
  Shelton}]{bhattacharya2012automated}
Bhattacharya P, Neamtiu I, Shelton C (2012) Automated, highly-accurate, bug
  assignment using machine learning and tossing graphs. Journal of Systems and
  Software 85(10):2275--2292

\bibitem[{Borg and Runeson(2014)}]{borg2014changes}
Borg M, Runeson P (2014) Changes, evolution, and bugs. In: Recommendation
  systems in software engineering. Springer, p 477--509

\bibitem[{Borg et~al(2016)Borg, Wnuk, Regnell, and
  Runeson}]{borg2016supporting}
Borg M, Wnuk K, Regnell B, et~al (2016) Supporting change impact analysis using
  a recommendation system: An industrial case study in a safety-critical
  context. Transactions on Software Engineering 43(7):675--700

\bibitem[{Borg et~al(2021)Borg, Jonsson, Engstr{\"o}m, Bartalos, and
  Szabo}]{borg2021adopting}
Borg M, Jonsson L, Engstr{\"o}m E, et~al (2021) Adopting automated bug
  assignment in practice: A registered report of an industrial case study.
  arXiv preprint arXiv:210913635

\bibitem[{Carver and Prikladnicki(2018)}]{carver2018industry}
Carver J, Prikladnicki R (2018) Industry--academia collaboration in software
  engineering. IEEE Software 35(5):120--124

\bibitem[{Chattamvelli and Shanmugam(2021)}]{chattamvelli2021continuous}
Chattamvelli R, Shanmugam R (2021) Continuous Distributions in Engineering and
  the Applied Sciences: Part I. Synthesis Lectures on Mathematics and
  Statistics Series, Morgan \& Claypool Publishers

\bibitem[{Crow and Shimizu(1988)}]{1988lognormal}
Crow E, Shimizu K (1988) Lognormal Distributions: Theory and Applications.
  Statistics: A Series of Textbooks and Monographs, Taylor \& Francis

\bibitem[{Cruzes and Dyb\aa(2011)}]{cruzes_recommended_2011}
Cruzes D, Dyb\aa T (2011) Recommended steps for thematic synthesis in software
  engineering. In: Proc. of the International Symposium on Empirical Software
  Engineering and Measurement, pp 275--284

\bibitem[{Davis(1989)}]{davis1989perceived}
Davis F (1989) Perceived usefulness, perceived ease of use, and user acceptance
  of information technology. MIS quarterly pp 319--340

\bibitem[{Engstr{\"o}m et~al(2012)Engstr{\"o}m, Feldt, and
  Torkar}]{engstrom2012indirect}
Engstr{\"o}m E, Feldt R, Torkar R (2012) Indirect effects in evidential
  assessment: a case study on regression test technology adoption. In:
  Proceedings of the 2nd international workshop on Evidential assessment of
  software technologies, pp 15--20

\bibitem[{Engstr{\"o}m et~al(2020)Engstr{\"o}m, Storey, Runeson, H{\"o}st, and
  Baldassarre}]{engstrom2020software}
Engstr{\"o}m E, Storey M, Runeson P, et~al (2020) How software engineering
  research aligns with design science: a review. Empirical Software Engineering
  25(4):2630--2660

\bibitem[{Favre et~al(2003)Favre, Estublier, and Sanlaville}]{favre2003tool}
Favre JM, Estublier J, Sanlaville A (2003) Tool adoption issues in a very large
  software company. In: Proceedings of 3rd International Workshop on
  Adoption-Centric Software Engineering (ACSE'03), Portland, Oregon, USA, pp
  81--89

\bibitem[{Flaounas(2017)}]{flaounas2017beyond}
Flaounas I (2017) Beyond the technical challenges for deploying machine
  learning solutions in a software company. In: Proc. of the Human in the Loop
  Machine Learning Workshop

\bibitem[{Furia et~al(2019)Furia, Feldt, and Torkar}]{furia2019bayesian}
Furia CA, Feldt R, Torkar R (2019) Bayesian data analysis in empirical software
  engineering research. IEEE Transactions on Software Engineering
  47(9):1786--1810

\bibitem[{Garousi et~al(2019)Garousi, Pfahl, Fernandes, Felderer,
  M{\"a}ntyl{\"a}, Shepherd, Arcuri, Co{\c{s}}kun{\c{c}}ay, and
  Tekinerdogan}]{garousi2019characterizing}
Garousi V, Pfahl D, Fernandes JM, et~al (2019) Characterizing industry-academia
  collaborations in software engineering: evidence from 101 projects. Empirical
  Software Engineering 24(4):2540--2602

\bibitem[{Garousi et~al(2020)Garousi, Borg, and Oivo}]{garousi2020practical}
Garousi V, Borg M, Oivo M (2020) Practical relevance of software engineering
  research: synthesizing the community’s voice. Empirical Software
  Engineering 25(3):1687--1754

\bibitem[{Gelman et~al(2013)Gelman, Carlin, Stern, and Rubin}]{gelmanbda13}
Gelman A, Carlin JB, Stern HS, et~al (2013) Bayesian Data Analysis, 3rd edn.
  Chapman and Hall/CRC

\bibitem[{Gelman et~al(2015)Gelman, Lee, and Guo}]{Gelman15stan:a}
Gelman A, Lee D, Guo J (2015) Stan: A probabilistic programming language for
  bayesian inference and optimization

\bibitem[{Hall et~al(2009)Hall, Frank, Holmes, Pfahringer, Reutemann, and
  Witten}]{hall_weka_2009}
Hall M, Frank E, Holmes G, et~al (2009) The {WEKA} {data} {mining} {software}:
  {An} {update}. SIGKDD Explorations Newsletter 11(1):10--18

\bibitem[{Hameed et~al(2012)Hameed, Counsell, and Swift}]{hameed2012conceptual}
Hameed MA, Counsell S, Swift S (2012) A conceptual model for the process of it
  innovation adoption in organizations. Journal of Engineering and Technology
  Management 29(3):358--390

\bibitem[{Hansen(2020)}]{hansen2020virtue}
Hansen KB (2020) The virtue of simplicity: On machine learning models in
  algorithmic trading. Big Data \& Society 7(1):2053951720926,558

\bibitem[{Hern{\'a}n and Robins(2020)}]{hernan2020causal}
Hern{\'a}n M, Robins J (2020) Causal inference: {What} if. Chapman \& Hall/CRC,
  Boca Raton, FL, USA

\bibitem[{Hoff and Bashir(2015)}]{hoff2015trust}
Hoff KA, Bashir M (2015) Trust in automation: Integrating empirical evidence on
  factors that influence trust. Human factors 57(3):407--434

\bibitem[{Jeong et~al(2009)Jeong, Kim, and Zimmermann}]{jeong2009improving}
Jeong G, Kim S, Zimmermann T (2009) Improving bug triage with bug tossing
  graphs. In: Proc. of the 7th Joint Meeting of the European Software
  Engineering Conference and the ACM SIGSOFT Symposium on the Foundations of
  Software Engineering, pp 111--120

\bibitem[{John et~al(2021)John, Olsson, and Bosch}]{john2021towards}
John MM, Olsson HH, Bosch J (2021) Towards mlops: A framework and maturity
  model. In: 2021 47th Euromicro Conference on Software Engineering and
  Advanced Applications (SEAA), IEEE, pp 1--8

\bibitem[{Jonsson(2013)}]{jonsson2013increasing}
Jonsson L (2013) Increasing anomaly handling efficiency in large organizations
  using applied machine learning. In: 2013 35th International Conference on
  Software Engineering (ICSE), IEEE, pp 1361--1364

\bibitem[{Jonsson et~al(2012)Jonsson, Broman, Sandahl, and
  Eldh}]{jonsson2012towards}
Jonsson L, Broman D, Sandahl K, et~al (2012) Towards automated anomaly report
  assignment in large complex systems using stacked generalization. In: Proc.
  of the 5th International Conference on Software Testing, Verification and
  Validation, pp 437--446

\bibitem[{Jonsson et~al(2016{\natexlab{a}})Jonsson, Borg, Broman, Sandahl,
  Eldh, and Runeson}]{jonsson2016automated}
Jonsson L, Borg M, Broman D, et~al (2016{\natexlab{a}}) Automated bug
  assignment: Ensemble-based machine learning in large scale industrial
  contexts. Empirical Software Engineering 21(4):1533--1578

\bibitem[{Jonsson et~al(2016{\natexlab{b}})Jonsson, Broman, Magnusson, Sandahl,
  Villani, and Eldh}]{jonsson2016automatic}
Jonsson L, Broman D, Magnusson M, et~al (2016{\natexlab{b}}) Automatic
  localization of bugs to faulty components in large scale software systems
  using bayesian classification. In: 2016 IEEE International Conference on
  Software Quality, Reliability and Security (QRS), IEEE, pp 423--430

\bibitem[{Just et~al(2008)Just, Premraj, and Zimmermann}]{just2008towards}
Just S, Premraj R, Zimmermann T (2008) Towards the next generation of bug
  tracking systems. In: Proc. of the Symposium on Visual Languages and
  Human-Centric Computing, pp 82--85

\bibitem[{Lee and See(2004)}]{lee2004trust}
Lee JD, See KA (2004) Trust in automation: Designing for appropriate reliance.
  Human factors 46(1):50--80

\bibitem[{McElreath(2020)}]{mcelreath2020statistical}
McElreath R (2020) Statistical rethinking: A Bayesian course with examples in R
  and Stan, 2nd edn. Chapman and Hall/CRC

\bibitem[{Murphy-Hill and Murphy(2014)}]{murphy-hill_recommendation_2014}
Murphy-Hill E, Murphy G (2014) Recommendation {Delivery}. In: Robillard M,
  Maalej W, Walker R, et~al (eds) Recommendation {Systems} in {Software}
  {Engineering}. Springer, p 223--242,
  \urlprefix\url{http://link.springer.com/chapter/10.1007/978-3-642-45135-5_9}

\bibitem[{Oliveira et~al(2021)Oliveira, Andrade, Nogueira, Barreto, and
  Bueno}]{oliveira2021issue}
Oliveira P, Andrade R, Nogueira T, et~al (2021) Issue auto-assignment in
  software projects with machine learning techniques. arXiv preprint
  arXiv:210401717

\bibitem[{Paleyes et~al(2020)Paleyes, Urma, and
  Lawrence}]{paleyes2020challenges}
Paleyes A, Urma RG, Lawrence ND (2020) Challenges in deploying machine
  learning: a survey of case studies. ACM Computing Surveys (CSUR)

\bibitem[{Parasuraman et~al(2000)Parasuraman, Sheridan, and
  Wickens}]{parasuraman2000model}
Parasuraman R, Sheridan T, Wickens C (2000) A model for types and levels of
  human interaction with automation. Transactions on Systems, Man, and
  Cybernetics-Part A: Systems and Humans 30(3):286--297

\bibitem[{Pearl(2009)}]{pearl2009causal}
Pearl J (2009) Causality: Models, reasoning and inference, 2nd edn. Cambridge
  University Press, Cambridge, UK

\bibitem[{Pearl and Mackenzie(2018)}]{pearl18why}
Pearl J, Mackenzie D (2018) The Book of Why: The New Science of Cause and
  Effect, 1st edn. Basic Books, Inc., USA

\bibitem[{Pearl et~al(2016)Pearl, Glymour, and Jewell}]{pearl2016causal}
Pearl J, Glymour M, Jewell N (2016) Causal Inference in Statistics: A Primer.
  Wiley

\bibitem[{Petersen and Wohlin(2009)}]{petersen_context_2009}
Petersen K, Wohlin C (2009) Context in {industrial} {software} {engineering}
  {research}. In: Proc. of the 3rd {International} {Symposium} on {Empirical}
  {Software} {Engineering} and {Measurement}, pp 401--404

\bibitem[{Premkumar and Potter(1995)}]{premkumar1995adoption}
Premkumar G, Potter M (1995) Adoption of computer aided software engineering
  (case) technology: an innovation adoption perspective. ACM SIGMIS Database:
  the DATABASE for Advances in Information Systems 26(2-3):105--124

\bibitem[{Regnell et~al(2008)Regnell, Berntsson~Svensson, and
  Olsson}]{regnell_supporting_2008}
Regnell B, Berntsson~Svensson R, Olsson T (2008) Supporting {roadmapping} of
  {quality} {requirements}. IEEE Software 25(2):42--47

\bibitem[{Rico et~al(2021)Rico, Bjarnason, Engstr{\"o}m, H{\"o}st, and
  Runeson}]{rico2021case}
Rico S, Bjarnason E, Engstr{\"o}m E, et~al (2021) A case study of
  industry-academia communication in a joint software engineering research
  project. Journal of software: Evolution and Process 33(10):e2372

\bibitem[{Runeson et~al(2012)Runeson, H\"ost, Rainer, and
  Regnell}]{runeson2012case}
Runeson P, H\"ost M, Rainer A, et~al (2012) Case study research in software
  engineering: Guidelines and examples. John Wiley \& Sons

\bibitem[{Runeson et~al(2020)Runeson, Engstr{\"o}m, and
  Storey}]{runeson2020design}
Runeson P, Engstr{\"o}m E, Storey M (2020) The design science paradigm as a
  frame for empirical software engineering. In: Contemporary Empirical Methods
  in Software Engineering. Springer, p 127--147

\bibitem[{Sajedi-Badashian and Stroulia(2020)}]{sajedi2020guidelines}
Sajedi-Badashian A, Stroulia E (2020) Guidelines for evaluating bug-assignment
  research. Journal of Software: Evolution and Process 32(9):e2250

\bibitem[{Sarkar et~al(2019)Sarkar, Rigby, and Bartalos}]{sarkar2019improving}
Sarkar A, Rigby P, Bartalos B (2019) Improving bug triaging with high
  confidence predictions at {Ericsson}. In: Proc. of the International
  Conference on Software Maintenance and Evolution, IEEE, pp 81--91

\bibitem[{Schroeder and Gibson(2009)}]{schroeder2009large}
Schroeder B, Gibson GA (2009) A large-scale study of failures in
  high-performance computing systems. IEEE transactions on Dependable and
  Secure Computing 7(4):337--350

\bibitem[{{Stan Development Team}(2022)}]{stan2022}
{Stan Development Team} (2022) Stan {Modeling} {Language} {User}'s {Guide} and
  {Reference} {Manual}, {Version} 2.30. \urlprefix\url{http://mc-stan.org/}

\bibitem[{Stefi(2015)}]{stefi2015developers}
Stefi A (2015) Do developers make unbiased decisions? {T}he effect of
  mindfulness and not-invented-here bias on the adoption of software
  components. In: Proc. of the 23rd European Conference on Information Systems,
  p Paper 175

\bibitem[{Tantithamthavorn and Jiarpakdee(2021)}]{xai4sebook}
Tantithamthavorn C, Jiarpakdee J (2021) Monash University,
  \doi{10.5281/zenodo.4769127}, \urlprefix\url{http://xai4se.github.io/},
  retrieved 2021-05-17

\bibitem[{Textor et~al(2016)Textor, Van~der Zander, Gilthorpe, Li{\'s}kiewicz,
  and Ellison}]{textor2016robust}
Textor J, Van~der Zander B, Gilthorpe MS, et~al (2016) Robust causal inference
  using directed acyclic graphs: the r package ‘dagitty’. International
  journal of epidemiology 45(6):1887--1894

\bibitem[{Vogelsang and Borg(2019)}]{vogelsang2019requirements}
Vogelsang A, Borg M (2019) Requirements engineering for machine learning:
  Perspectives from data scientists. In: 2019 IEEE 27th International
  Requirements Engineering Conference Workshops (REW), IEEE, pp 245--251

\bibitem[{Wirth and Hipp(2000)}]{wirth2000crisp}
Wirth R, Hipp J (2000) Crisp-dm: Towards a standard process model for data
  mining. In: Proceedings of the 4th international conference on the practical
  applications of knowledge discovery and data mining, Manchester, pp 29--39

\bibitem[{Wu et~al(2018)Wu, Liu, and Ma}]{wu2018empirical}
Wu H, Liu H, Ma Y (2018) Empirical study on developer factors affecting tossing
  path length of bug reports. IET Software 12(3):258--270

\bibitem[{Zhang et~al(2013)Zhang, Gong, and Versteeg}]{zhang2013predicting}
Zhang H, Gong L, Versteeg S (2013) Predicting bug-fixing time: an empirical
  study of commercial software projects. In: 2013 35th International Conference
  on Software Engineering (ICSE), IEEE, pp 1042--1051

\end{thebibliography}

\newpage

\begin{appendices}

\section{Interview Guide} \label{app:int_guide}
This section presents the key points in the interview guide. Subsections 4 and 5 contain different questions depending on which unit of analysis the interviewee represents, i.e., A) TRR Team, B) TR Coordinators or C) HighLevel/LowLevel Module.\\

\noindent \textbf{1. Formal introduction:}
\begin{itemize}
\item We are conducting a study to evaluate the adoption of TRR within its industrial context.
\item Ericsson, RISE and LU have signed a collaboration agreement including a non-disclosure clause.
\item You are guaranteed anonymity.
\item You do this interview voluntarily. You can skip answering any question, and you are free to stop the interview at any time.
\item The information collected during this interview will be securely stored.
\item Can we get your permission to record the interview?
\item (If yes) You will get an edited transcript of this interview soon. You will then have a chance to remove any information, or clarify yourself.
\end{itemize}

\noindent \textbf{2. Background (and warm-up)}

\begin{itemize}
\item Please describe your current role and your engineering background.
\item When did you graduate? When did you join this company?
\item What other roles have you had? For how many years?
\item For how long have you been in this role?
\item For how many years have you worked with this particular system?
\end{itemize}

\noindent \textbf{3. Open questions on TR assignment and the TRR intervention}

\begin{itemize}
\item Please describe the TR assignment task from your perspective.
\item How much do you know about the TRR tool? Do you use TRR today?
\item How do you recall the introduction of the TRR tool?
\item Please describe your impressions and lessons learned.
\end{itemize}

\noindent \textbf{4. Closed questions on TR assignment}\\
\noindent A) TRR Team:

\begin{itemize}
\item What were the biggest obstacles in turning TRR from a prototype to an Ericsson tool? (technology, people, process, organization, politics, \ldots)
\item How did you overcome the obstacles?
\item Can you compare the experience to other tools you have been involved in introducing?
\begin{itemize}
\item Did the fact that machine learning is involved in TRR  matter?
\end{itemize}
\item How accurate do you currently find the TRR assignments to be? Perception of bug tossing?
\end{itemize}

\noindent B) TR Coordinators:

\begin{itemize}
\item What is your general opinion about the TR assignment work task? Please explain your role in the task (module assignment, team assignment, person assignment, \ldots)
\item How accurate are your assignments? Perception of bug tossing?
\item Do you conduct TR assignments daily/weekly/monthly?
\item How much time do you spend on an individual TR assignment?
\begin{itemize}
\item Does it vary much?
\item Can you express it as minimum-normal-maximum time?
\end{itemize}
\item How do you make the assignment decision? 
\end{itemize}

\noindent C) HighLevel Module and LowLevel Module:

\begin{itemize}
\item How accurate do you find the assignments to be? Perception of bug tossing?
\item How frequently are you assigned a TR? (daily/weekly/monthly)
\item How much time does it take you to confirm that an individual TR belongs to you?
\begin{itemize}
\item Does it vary much?
\item Can you express it as minimum-normal-maximum time?
\end{itemize}
\end{itemize}

\noindent \textbf{5. Value of the TRR intervention and direct/indirect effects of increasing the level of automation}\\

\noindent A) TRR Team:

\begin{itemize}
\item How do you believe TRR will help Ericsson's TR coordinators do their work?
\item How will the TR assignment task evolve with the new level of automation?
\begin{itemize}
\item For TR assigners? For TR assignees?
\end{itemize}
\end{itemize}

\noindent B) TR Coordinators:

\begin{itemize}
\item How do you believe TRR will help you in your work?
\item How will your TR assignment task evolve with the new level of automation?
\begin{itemize}
\item For the engineers you assign TRs to?
\end{itemize}
\end{itemize}

\noindent C) HighLevel Module and LowLevel Module:

\begin{itemize}
\item How do you believe TRR will help you in your work?
\item How will your TR triage and resolution task evolve with the new level of automation?
\end{itemize}

\noindent Common for A), B) and C):

\begin{itemize}
\item What role does explainability play in the trustworthiness of TRR?
\item Will there be any effects on other work tasks?
\begin{itemize}
\item Possible downsides? (skill decrease, overtrust, lack of transparency, \ldots )
\item Other advantages that might appear? (new discussions, increased machine learning acceptance, increased job satisfaction, \ldots )
\end{itemize}
\item How do you regard TRR’s ease of use?
\begin{itemize}
\item Easy to understand and use the increased level of automation?
\item Please elaborate on TRR's current strengths and weaknesses? How will the TRR user experience evolve over time? (including requests for new features)
\end{itemize}
\end{itemize}

\noindent \textbf{6. Quality levels for the TRR intervention)}
\noindent Share screen and show the QUPER model. Explain the quality-benefit relation.

\begin{itemize}
\item How accurate does TRR have to be before you would recognize its value? (When does it go from useless to a solution that provides value)
\item Is there a TRR quality saturation point? (When would you stop bothering about the accuracy of the output)
\item Where on the quality scale would you position TRR today?
\end{itemize}

\noindent Share descriptive statistics about the accuracy of TRR in the field, preferably tailored for the interviewee.

\begin{itemize}
    \item Could you please comment on the figures?
\end{itemize}

\noindent \textbf{7. Concluding the interview}

\begin{itemize}
\item Do you have anything to add before we stop collecting data for this study?
\item Are there any other colleagues that you think we should interview?
\item Stop the recording and explain the next steps.
\end{itemize}


\newpage
\section{Evolution of Coding Scheme (TBD)}
\label{app:coding}
We combined deductive and inductive coding. In the interview material we extracted answers to three of the research questions as well as an in-depth case description. In the first iteration these questions alone guided the search for codes. 

\begin{itemize}
    \item Case description
    \item RQ1: How did TRR evolve from prototype to deployed tool? (A) Design decisions over time and B) obstacles/facilitators)
    \item RQ3: How much value does TRR provide in the organization?
    \item RQ4: How has the adoption TRR influenced the way of working? (Direct and indirect effects)
\end{itemize}

After each iteration new codes and themes were derived, which in turn were used as input for the next iteration. The interviews were assigned to iterations as follows: Iteration~1 = [HL1] and [TRR2], Iteration~2 = [TRR1] and [LL1], Iteration~3 = [HL2] and [LL2], and Iteration~4 = [TRR coords]. When all interviews had been coded once, the coding scheme was reworked, i.e., the codes were restructured and new themes identified. In Iteration~5 all transcripts were revisited to align the coding according to the final coding scheme.

This appendix shows how the coding scheme evolved in each iteration. The bold header in the boxes depict high-level codes and the itemized lists show low-level codes. Additions are highlighted in blue font and deletions in red font. Renamed or moved codes are depicted in green font. For moved codes, we add a strikethrough line for the origin that was effectively removed.

\begin{figure}
    \centering
    \includegraphics[width=1\textwidth]{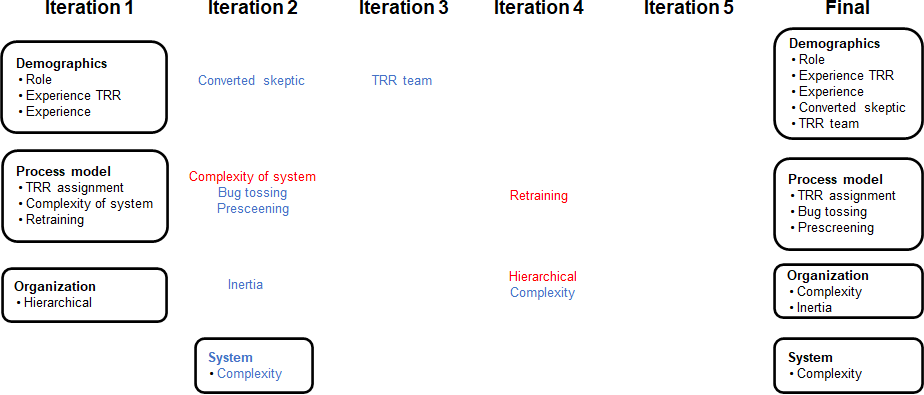}
    \caption{Evolution of codes for the case context.}
    \label{fig:code_evol_case}
\end{figure}

Figure~\ref{fig:code_evol_case} shows the evolution of codes for the case context. Six codes were added and three were removed during the iterations. The major change occurred in Iteration~2 as we added System as a high-level code.

\begin{figure}
    \centering
    \includegraphics[width=1\textwidth]{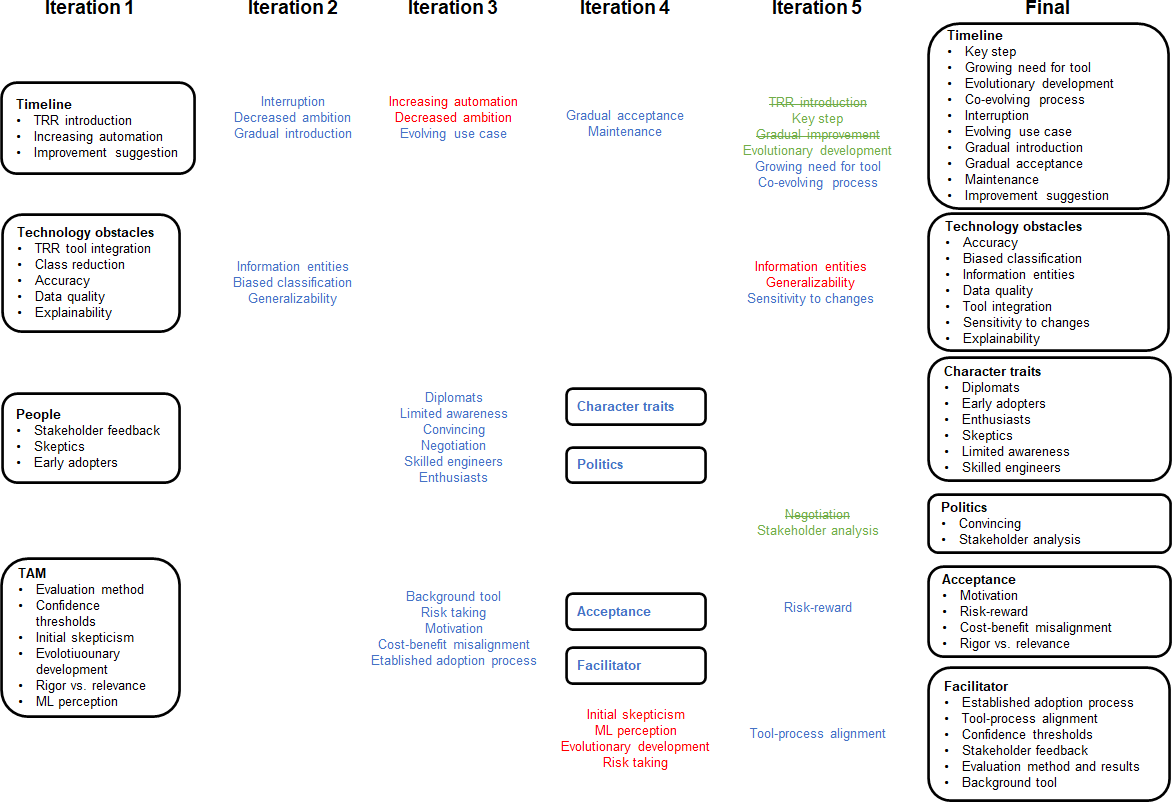}
    \caption{Evolution of codes for RQ1.}
    \label{fig:code_evol_rq1}
\end{figure}

Figure~\ref{fig:code_evol_rq1} presents the evolution of codes for RQ1. Several codes were added, removed, and renamed as our understanding increased. In Iteration~4 we reorganized the high-level codes from People and TAM (Technology Acceptance Model) to Character traits, Politics, Acceptance, and Facilitator.

\begin{figure}
    \centering
    \includegraphics[width=1\textwidth]{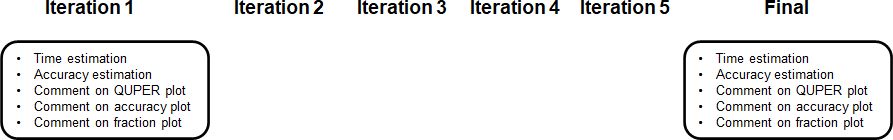}
    \caption{Evolution of codes for RQ3.}
    \label{fig:code_evol_rq3}
\end{figure}

Figure~\ref{fig:code_evol_rq3} shows that the codes for RQ3 remained stable over the five iterations.

\begin{figure}
    \centering
    \includegraphics[width=1\textwidth]{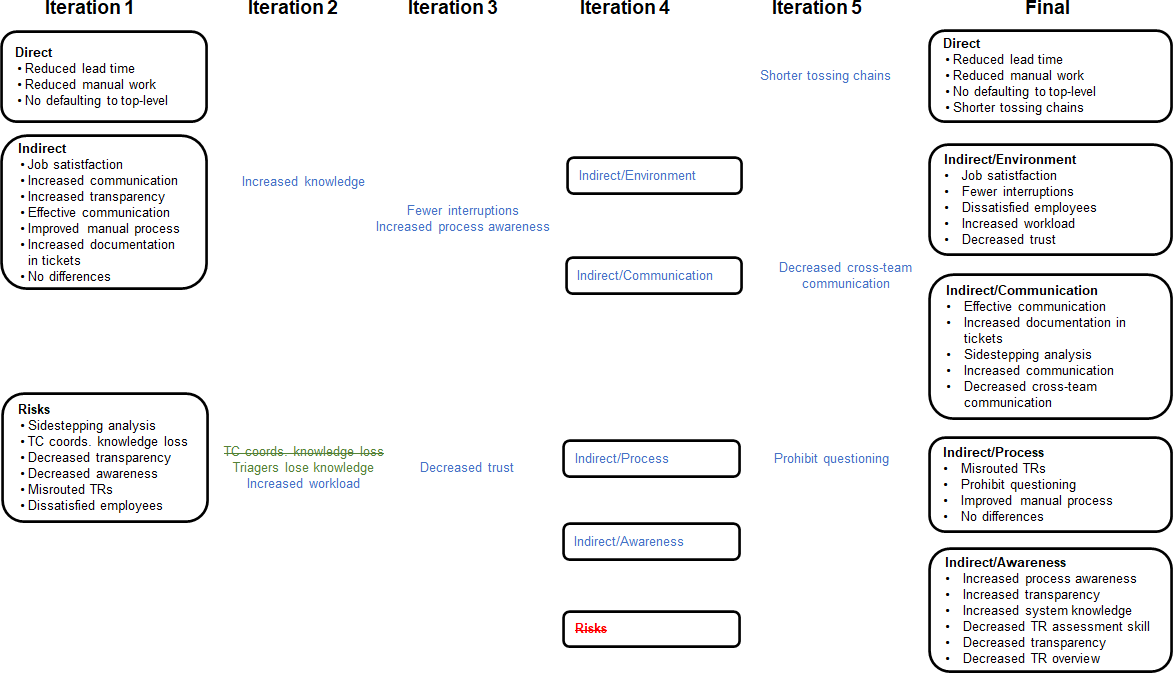}
    \caption{Evolution of codes for RQ4.}
    \label{fig:code_evol_rq4}
\end{figure}

Figure~\ref{fig:code_evol_rq4} presents the evolution of codes for RQ4. Codes were mostly added as we analyzed additional interviews. We made a major change in Iteration~4 as we introduced a mid-level code for indirect effects of the TRR introduction, i.e, we split the high-level code Indirect into Indirect/Environment, Indirect/Communication, Indirect/Process, and Indirect/Awareness. In this reorganization, we removed the high-level code Risks and merged its remaining low-level codes into the corresponding indirect effects.

\end{appendices}


\end{document}